\documentclass[a4paper,12pt]{article}
\usepackage{adjustbox}
\usepackage{comment}
\usepackage{jheppub}
\usepackage{placeins}
\usepackage[figuresright]{rotating}

\allowdisplaybreaks[4]

\title{
\Large \boldmath On sum rules for semi-leptonic $b \to c$ and $b \to u$ decays
}

\author[a]{Wen-Feng Duan,}
\author[b,c,d]{Syuhei Iguro,}
\author[a,e,1]{Xin-Qiang Li,\note{Corresponding author.}}
\author[a,2]{Ryoutaro Watanabe,\note{Corresponding author.}}
\author[a,f]{and Ya-Dong Yang}

\affiliation[a]{Institute of Particle Physics and Key Laboratory of Quark and Lepton Physics (MOE), Central China Normal University, Wuhan, Hubei 430079, China}
\affiliation[b]{Institute for Advanced Research, Nagoya University, Nagoya 464--8601, Japan}
\affiliation[c]{Kobayashi-Maskawa Institute for the Origin of Particles and the Universe, Nagoya University, Nagoya 464--8601, Japan}
\affiliation[d]{KEK Theory Center, IPNS, KEK, Tsukuba 305--0801, Japan}
\affiliation[e]{Center for High Energy Physics, Peking University, Beijing 100871, China}
\affiliation[f]{Institute of Particle and Nuclear Physics, Henan Normal University, Xinxiang 453007, China}

\emailAdd{dufewe@mails.ccnu.edu.cn}
\emailAdd{igurosyuhei@gmail.com}
\emailAdd{xqli@mail.ccnu.edu.cn}
\emailAdd{watanabe@ccnu.edu.cn}
\emailAdd{yangyd@ccnu.edu.cn}

\preprint{KEK-TH-2662}

\abstract{{\small
The semi-leptonic $b \to c l \nu$ processes are receiving a lot of attention, as the lepton flavor universality (LFU) violation has been hinted by the measured ratios $R_{D^{(*)}} = \Gamma(B \to D^{(*)} \tau\nu)/\Gamma(B \to D^{(*)} \ell\nu)$ for $\ell = e,\mu$.
Recently, it has also been pointed out that the baryonic counterpart, $R_{\Lambda_c} = \Gamma (\Lambda_b \to \Lambda_c \tau\nu)/\Gamma (\Lambda_b \to \Lambda_c \ell\nu)$, has a strong correlation with $R_{D^{(*)}}$, referred to as the $R$ ratio sum rule in this paper. 
The correlation is almost independent of the new physics (NP) implication and hence can predict $R_{\Lambda_c}$ from the measured $R_{D^{(*)}}$. 
On the other hand, we have fewer measurements and/or theoretical studies of the semi-leptonic $b \to u l \nu$ processes, although the same arguments can be applied to the ratios $R_\pi$, $R_\rho$, and $R_p$ as above. 
Since these processes are measurable at the ongoing LHCb run-3 and/or Belle~II experiments, precise studies on them are important as well. 
In this paper, we obtain the semi-analytic formulae for all the aforementioned $R_X$ ratios in the presence of NP contributions by using the available lattice QCD and/or light-cone sum rule fits to the form factors. 
Two novel points are highlighted: (i) We evaluate uncertainties of $R_X$ including both the Standard Model (SM) and NP terms, inherited from the form factor fits, and discuss how the uncertainties affect the $R$ ratio sum rules. 
(ii) We obtain the $R$ ratio sum rule among the semi-leptonic $b \to u l \nu$ processes for the first time, which provides a complementary motivation for observing these processes. 
In addition, based on our numerical results, we investigate how the different NP scenarios work in the $b\to c$ and $b\to u$ sectors and perform a combined study in the framework of SM effective field theory with specific flavor symmetries.
}
}

\begin{document}
\maketitle

\section{Introduction}
\label{sec:intro}

Testing the flavor structure of the Standard Model (SM) of particle physics has been performed with tremendous efforts since the CLEO, LEP, Tevatron, Belle, and BaBar experiments were established, 
and then it has been taken over by the LHCb and Belle~II experiments with the advantage of higher luminosities. 
One of their significant findings is that tensions between the experimental results and the SM predictions have been observed in the semi-tauonic $B$ meson decays. 
The observable is defined as the lepton flavor universality (LFU) ratios of $R_{D^{(*)}} = \Gamma(B \to D^{(*)} \tau\nu)/\Gamma(B \to D^{(*)} \ell \nu)$ for $\ell=e,\mu$, 
showing the tension at $3$--$4\sigma$ significance level~\cite{Iguro:2024hyk}.  
Its new physics (NP) implications have been studied from many point of views; see Ref.~\cite{Iguro:2024hyk} for a recent status and review.

Recently, a new direction regarding the NP study on the ratios has been discussed in Refs.~\cite{Blanke:2018yud,Blanke:2019qrx,Fedele:2022iib} 
that has found one significant relation among $R_D$, $R_{D^*}$ and $R_{\Lambda_c} = \Gamma (\Lambda_b \to \Lambda_c \tau\nu)/\Gamma (\Lambda_b \to \Lambda_c \mu\nu)$. 
In particular, 
\begin{align}
 \frac{R_{\Lambda_c}}{R_{\Lambda_c}^\text{SM}} \simeq 0.280 \frac{R_D}{R_D^\text{SM}} + 0.720 \frac{R_{D^*}}{R_{D^*}^\text{SM}} \,, 
 \label{eq:sumrulebc}
\end{align}
is obtained in the recent study of Ref.~\cite{Fedele:2022iib}, which will be referred to as the $R$ ratio sum rule for the semileptonic $b \to c$ decays in this paper. 
As pointed out in Ref.~\cite{Fedele:2022iib}, this relation holds unless the NP contribution is exceptionally large beyond the current experimental bounds. 
Thus, it gives a complementary check for the experimental results of the $R_X$ measurements, almost independent of the NP implication.  
In case the three experimental measurements of $R_D$, $R_{D^*}$ and $R_{\Lambda_c}$ do not follow Eq.~\eqref{eq:sumrulebc}, it would imply that some of these measurements are questionable. 
Therefore, the $R$ ratio sum rule gives a crucial feedback to the experimental measurements. 
This is an absolutely novel point, which cannot be provided by the individual $R_X$ study. 
It should be noted that the $R$ ratio sum rule of Eq.~\eqref{eq:sumrulebc} is empirically given since it could be violated if NP contributions are large to some extent. 
Very recently, Ref.~\cite{Endo:2025fke} has investigated a theoretical background for the condition holding the $R$ ratio sum rule, 
indicating that heavy quark symmetry enforces the exact sum rule and then actual inputs of the form factors are the source of its violation. 

In this paper, we further investigate two points that have not been addressed yet: 
one is the precise evaluation of uncertainties from form factor inputs, which is significant in the sense of the theoretical background as mentioned above, and another is the $R$ ratio sum rule for the $b \to u$ counterpart.  
Theoretical analyses of $R_X$ and the $R$ ratio sum rules rely on the modelling of the hadronic matrix elements. 
Thus, it is crucial to see how Eq.~\eqref{eq:sumrulebc} would be affected when theoretical uncertainties from the form factor inputs are taken into account. 
We will summarize the recent developments of the relevant form factors for $B \to D$, $B \to D^*$, and $\Lambda_b \to \Lambda_c$, 
with which we evaluate possible uncertainties on the $R_X$ formulae and the $R$ ratio sum rules, inherited from these inputs.

In a similar way, we will also construct the $R$ ratio sum rule for the semileptonic $b \to u$ decays. 
At present, the exclusive $B \to \pi \ell\nu$ process has been measured~\cite{ParticleDataGroup:2024cfk} to obtain $|V_{ub}|$ 
while the tauonic mode $B \to \pi \tau\nu$ was first measured by the Belle collaboration~\cite{Belle:2015qal} with a large uncertainty. 
Combining them, one may find $R_\pi^\text{exp} = \Gamma (B \to \pi \tau\nu)/\Gamma (B \to \pi \ell\nu) \approx 1.01 \pm 0.49$. 
On the other hand, although the exclusive processes of $B\to\rho \ell\nu$ and $\Lambda_b \to p \mu\nu$ have been observed~\cite{Belle-II:2024xwh,LHCb:2015eia}, 
their tauonic modes $B \to \rho \tau\nu$ and $B \to \omega \tau\nu$ have not been measured yet. 
However, it is expected that these decay processes are measurable at the LHCb run-3~\cite{LHCb:2018roe,LHCb:2022ine} and Belle~II~\cite{Belle-II:2018jsg,Belle-II:2022cgf} experiments. 
In light of this situation, it is remarkable to show the $R$ ratio sum rule among $R_\pi$, $R_\rho$, and $R_p$, with which we can make a useful feedback for the future experimental measurements of these missing tauonic modes. 
More detailed experimental status on the relevant semileptonic $b \to c$ and $b \to u$ processes of this paper are given in Appendix~\ref{app:status}.

With our updated numerical results, we then discuss whether the $R$ ratio sum rule is predictable and usable for the consistency check among the three measurements. 
It should be especially noted that we obtain the $R$ ratio sum rule in $b \to u l \nu$ decays for the first time. 
We also show the relations to the other observables, such as the inclusive ratio $R_{X_c}$ and the semi-leptonic $B_c$ decay ratio $R_{J/\psi}$. 
Based on our numerical results, we also investigate how the different NP scenarios work in the $b\to cl\nu$ and $b\to ul\nu$ transitions, and perform a combined study of these processes in the framework of SM effective field theory (SMEFT)~\cite{Buchmuller:1985jz,Grzadkowski:2010es,Brivio:2017vri,Isidori:2023pyp} with specific flavor symmetries~\cite{Faroughy:2020ina,Greljo:2022cah}.

This paper is organized as follows. In Sec.~\ref{sec:formula}, we will present the general formulae for $R_X$ and our procedure to get the $R$ ratio sum rules. Then, the recent form factor inputs are summarized, and the impacts of their uncertainties on $R_X$ and the $R$ ratio sum rules are discussed in Sec.~\ref{sec:error}. Phenomenological implications of the $R$ ratio sum rules are discussed in Sec.~\ref{sec:phen}, and then specific flavor structures are considered for a combined study in the SMEFT. Finally, our conclusion is made in Sec.~\ref{summary}. For convenience, details of the specific flavor symmetries we are considering are relegated in App.~\ref{app:U3andU2}, and the covariance tables for our results are presented in App.~\ref{app:covariancetable}.

\section{Sum rule formulae}
\label{sec:formula}

We describe the most general NP contributions to the quark-level $b \to q l \nu$ transitions for $q = c, u$ and $l = e, \mu, \tau$ in terms of the effective weak Hamiltonian,
\begin{equation} \label{eq:Hamiltonian}
 \mathcal{H}_{\mathrm{eff}}= 2 \sqrt2 G_FV_{qb}\biggl[ \left(1+C_{V_L}^{ql}\right) \mathcal O_{V_L}^{ql} + C_{V_R}^{ql} \mathcal O_{V_R}^{ql} + C_{S_L}^{ql} \mathcal O_{S_L}^{ql} + C_{S_R}^{ql} \mathcal O_{S_R}^{ql} + C_{T}^{ql} \mathcal O_{T}^{ql} \biggl]\,,
\end{equation}
with the four-fermion operators of our interest given by
\begin{equation} \label{eq:operator} 
\begin{aligned}
 \mathcal O_{V_L}^{ql} &= (\overline{q} \gamma^\mu P_Lb)(\overline{l} \gamma_\mu P_L \nu_{l})\,, \qquad & & 
 \mathcal O_{V_R}^{ql} = (\overline{q} \gamma^\mu P_Rb)(\overline{l} \gamma_\mu P_L \nu_{l})\,, \\[0.2cm]
 \mathcal O_{S_L}^{ql} &= (\overline{q}  P_Lb)(\overline{l} P_L \nu_{l})\,, \qquad & &
 \mathcal O_{S_R}^{ql} = (\overline{q}  P_Rb)(\overline{l} P_L \nu_{l})\,, \\[0.2cm]
 \mathcal O_{T}^{ql} &= (\overline{q}  \sigma^{\mu\nu}P_Lb)(\overline{l} \sigma_{\mu\nu} P_L \nu_{l}) \,,
\end{aligned}
\end{equation}
where $P_L=(1-\gamma_5)/2$ and $P_R=(1+\gamma_5)/2$. The NP contributions are encoded in the short-distance Wilson coefficients (WCs) $C_i^{ql}$ with $i=V_L, V_R, S_L, S_R, T$, which have been normalized by the SM factor of $2 \sqrt2 G_FV_{qb}$. We assume that the electronic and muonic processes are described by the SM, namely $C_i^{qe}= C_i^{q\mu} = 0$. It is also assumed that the light neutrino is always left-handed for simplicity. The case with right-handed neutrinos can be found, \textit{e.g.}, in Refs.~\cite{Iguro:2018qzf,Robinson:2018gza,Babu:2018vrl,Mandal:2020htr,Penalva:2021wye,Datta:2022czw}.

\subsection{General formula} 
\label{sec:formulaRD}

Depending on the spin properties of the hadrons involved in the semi-leptonic decays, we have specific forms of $R_X$ in terms of the WCs $C_i^{q\tau}$ for each NP effective operator. Explicitly, we can obtain the following general formulae:
\begin{align} \label{eq:RP}
 \frac{R_P}{R_{P}^\textrm{SM}} &=
  |1+C_{V_L}^{q\tau}+C_{V_R}^{q\tau}|^2  + a_{P}^{SS} |C_{S_L}^{q\tau}+C_{S_R}^{q\tau}|^2 + a_{P}^{TT} |C_{T}^{q\tau}|^2  \\
 & + a_{P}^{VS} \textrm{Re}\left[\left(1+C_{V_L}^{q\tau}+C_{V_R}^{q\tau}\right)\left(C_{S_L}^{q\tau*}+C_{S_R}^{q\tau*}\right)\right]  + a_{P}^{VT}\textrm{Re}\left[\left(1+C_{V_L}^{q\tau}+C_{V_R}^{q\tau}\right)C_{T}^{q\tau*}\right] \,, \notag \\[0.3cm] 
 \label{eq:RV}
 \frac{ R_{V}}{R_{V}^\textrm{SM}} &=
  |1+C_{V_L}^{q\tau}|^2 + |C_{V_R}^{q\tau}|^2  + a_V^{SS}|C_{S_L}^{q\tau}-C_{S_R}^{q\tau}|^2 + a_V^{TT} |C_{T}^{q\tau}|^2  \nonumber \\
 & +a_V^{V_LV_R} \textrm{Re}\left[\left(1+C_{V_L}^{q\tau}\right)C_{V_R}^{q\tau*}\right]  + a_V^{VS} \textrm{Re}\left[\left(1+C_{V_L}^{q\tau}-C_{V_R}^{q\tau}\right)\left(C_{S_L}^{q\tau*}-C_{S_R}^{q\tau*}\right)\right] \nonumber \\[0.6em]
 & +a_V^{V_LT} \textrm{Re}\left[\left(1+C_{V_L}^{q\tau}\right)C_{T}^{q\tau*}\right] + a_V^{V_RT} \textrm{Re}\left[C_{V_R}^{q\tau}C_{T}^{q\tau*}\right] \,, \\[0.3cm]
 \label{eq:RB}
\frac{R_{H}}{R^{\rm SM}_{H}} &= 
 |1+C_{V_L}^{q\tau}|^2 + |C_{V_R}^{q\tau}|^2 + a_{H}^{SS} \, \left[|C_{S_L}^{q\tau}|^2 + |C_{S_R}^{q\tau}|^2\right] + a_H^{TT} \, |C_T^{q\tau}|^2 \nonumber \\
 &+ a_H^{V_LV_R}\, \textrm{Re}\left[\left(1+C_{V_L}^{q\tau}\right)C_{V_R}^{q\tau*}\right] + a_H^{VS_1} \,\textrm{Re}\left[ \left(1 +C_{V_L}^{q\tau}\right) C_{S_L}^{q\tau\ast} + C_{V_R}^{q\tau} C_{S_R}^{q\tau\ast} \right] \notag\\[0.6em]
 &+ a_H^{VS_2} \,\textrm{Re}\left[ \left(1 +C_{V_L}^{q\tau}\right) C_{S_R}^{q\tau\ast} + C_{V_R}^{q\tau} C_{S_L}^{q\tau\ast} \right] + a_H^{S_LS_R} \, \textrm{Re}\left[ C_{S_L}^{q\tau} C_{S_R}^{q\tau\ast}  \right] \notag\\[0.6em]
 &+ a_H^{V_LT} \,\textrm{Re}\left[ \left(1 +C_{V_L}^{q\tau}\right)  C_T^{q\tau\ast}\right] + a_H^{V_RT} \textrm{Re}\left[ C_{V_R}^{q\tau}  C_T^{q\tau\ast}\right]  \,,
\end{align}
where $(P, V, H) = (D, D^*, \Lambda_c)$ for $q=c$ and $(P, V, H) = (\pi, \rho, p)$ for $q=u$. For each hadronic final state $X$, $a_X^{ij}$ is denoted as the numerical coefficient associated with the product of the WCs, where $ij$ represents the NP current structure of the corresponding term. Since some of these NP structures give identical coefficients, we have introduced the following abbreviations such that  
\begin{equation}
\begin{aligned}
 a_{P}^{VS} & \equiv a_{P}^{V_L S_L}=a_{P}^{V_L S_R}=a_{P}^{V_R S_L}=a_{P}^{V_R S_R} \,, \quad 
 a_{P}^{VT}  \equiv a_{P}^{V_L T}=a_{P}^{V_R T}\,,\\[0.2em] 
 a_{V}^{VS} & \equiv a_{V}^{V_L S_L}=-a_{V}^{V_L S_R}=-a_{V}^{V_R S_L}=a_{V}^{V_R S_R} \,, \\[0.2em] 
 a_{H}^{VS_1} & \equiv a_{H}^{V_L S_L}=a_{H}^{V_R S_R} \,, \quad 
 a_{H}^{VS_2} \equiv a_{H}^{V_L S_R}=a_{H}^{V_R S_L} \,,
\end{aligned}
\end{equation}
in Eqs.~\eqref{eq:RP}--\eqref{eq:RB}. In the next section, we will introduce our setup with a summary of the commonly adopted parametrizations of the form factors together with the latest fitting parameters, and obtain the numerical coefficients $a_X^{ij}$ with uncertainties inherited from the form factor inputs. Then, we find that Eqs.~\eqref{eq:RP}--\eqref{eq:RB} can be combined into a single relation 
\begin{align}\label{eq:Rsumrule-preliminary}
 \frac{R_{H}}{R^{\rm SM}_{H}} = b\, \frac{R_P}{R_{P}^\textrm{SM}} + c\, \frac{ R_{V}}{R_{V}^\textrm{SM}} + \delta_H \,. 
\end{align}
Here $\delta_H$ is some function of the WCs $C_i^{q\tau}$, and the two coefficients $b$ and $c$ are not unique but can be determined by setting a desired condition so that $\delta_H $ becomes small. In this paper, we follow Ref.~\cite{Fedele:2022iib} and set 
\begin{align}
 b + c = 1 \,, \quad\quad a_{P}^{VS}\, b + a_{V}^{VS}\, c = a_H^{VS_1} \,, \label{eq:SRset}
\end{align}
so that the terms $|1+C_{V_L}^{q\tau}|^2$ and $\textrm{Re}[(1+C_{V_L}^{q\tau})C_{S_L}^{q\tau*}]$ in Eq.~\eqref{eq:Rsumrule-preliminary} vanish, respectively. The other terms are all contained in $\delta_H$ that would shift the $R$ ratio sum rule if it is not negligible. For convenience, we call it the shift factor for the $R$ ratio sum rule throughout this paper. 

\subsection{Sum rule utility: a first look} 
\label{sec:centralSR}

Before showing details on the form factor inputs and evaluating the resulting uncertainties of the $R$ ratio sum rules, let us firstly summarize the current situation at the central values of the theoretical inputs. 
For the $b \to c$ case, we obtain 
\begin{align} \label{eq:Rsumrule_central}
 \frac{R_{\Lambda_c}}{R^{\rm SM}_{\Lambda_c}} 
 = 0.272\, \frac{R_D}{R^{\rm SM}_D} + 0.728\,\frac{R_{D^\ast}}{R^{\rm SM}_{D^\ast}} + \delta_{\Lambda_c} \,,
\end{align}
with 
\begin{align} \label{eq:deltalambdacdefault}
 \delta_{\Lambda_c} =
 & \,-0.001\,\left( |C_{S_L}^{c\tau}|^2 + |C_{S_R}^{c\tau}|^2\right) -0.007\,\textrm{Re}\left( C_{S_L}^{c\tau} C_{S_R}^{c\tau\ast} \right) 
  -2.681\,|C_T^{c\tau}|^2  \notag \\[0.5em]
 & \hspace{-0.2cm} \,+ \textrm{Re} \left[ \left(1 +C_{V_L}^{c\tau} \right) \left( 0.041 C_{V_R}^{c\tau\ast} + 0.594 C_T^{c\tau\ast} \right) \right]  -0.561\,\textrm{Re}\left( C_{V_R}^{c\tau} C_{T}^{c\tau\ast} \right) \notag\\[0.5em]
  & \hspace{-0.2cm} \,-0.002\, \textrm{Re} \left[ \left(1 +C_{V_L}^{c\tau} \right) C_{S_R}^{c\tau\ast} + C_{S_L}^{c\tau} C_{V_R}^{c\tau\ast} \right] \,.
\end{align}
This is consistent with the previous study where $C_{V_R}^{c\tau}$ has been omitted in Ref.~\cite{Fedele:2022iib}.\footnote{ \label{foonoteonCVR}
The reason for omitting $C_{V_R}^{c\tau}$ is motivated by the observation that the LFU violating operator $\mathcal O_{V_R}^{ql}$ is firstly generated by the $SU(2)_L\otimes U(1)_Y$ invariant operators at dimension eight, and thus the corresponding WC $C_{V_R}^{cl}$ will have an additional suppression compared with that of the other operators generated from the dimension-six operators~\cite{Jenkins:2017jig,Aebischer:2015fzz}.}
One can observe potentially large contributions from the terms $\textrm{Re}\left[\left(1 +C_{V_L}^{c\tau} \right) C_T^{c\tau\ast}\right]$, $\textrm{Re} \left( C_{V_R}^{c\tau} C_{T}^{c\tau\ast} \right)$ and $|C_T^{c\tau}|^2$, while all the other terms are completely negligible even for $|C_i^{c\tau}| \sim \mathcal O(1)$. Their effects can be further checked by looking at the collider bounds on $C_i^{c\tau}$ through the analysis of the $\tau + \text{missing}$ searches at ATLAS and CMS, as demonstrated in Refs.~\cite{Iguro:2020keo,Greljo:2018tzh}. 
It has turned out that the allowed range depends on the mediator mass, which is given as~\cite{Iguro:2020keo}
\begin{align}
 |C_T^{c\tau}| \lesssim 0.3\,\,\text{--}\,\, 0.2 \,, 
\end{align}
for the tensor-type interaction, where $0.3$ corresponds to a $t$-channel NP mediator with mass of $2\,\text{TeV}$ and $0.2$ indicates an effective-field-theory-limit case. 
Given the above bound, we find that the $R$ ratio sum rule could be shifted by $\delta_{\Lambda_c} \approx -0.4\,\,\text{--}\,-0.2$ at most. If this were the case, such a shift factor would be not negligible. 

On the other hand, if we try to find a NP solution to the $R_D$ and $R_{D^*}$ anomalies by the tensor-type interaction, 
the best-fit point to the current experimental data is obtained as $C_T^{c\tau} \approx 0.02 \pm i\,0.13$~\cite{Iguro:2024hyk}. 
Taking this fit point as input leads to $\delta_{\Lambda_c} \approx -0.035$, 
which is sufficiently smaller than the current experimental resolutions for the $R_X$ measurements, especially for the ratio $R_{\Lambda_c}$~\cite{LHCb:2022piu,Bernlochner:2022hyz}. 
Thus, the current situation of the $R_{D^{(*)}}$ anomalies tells us that $\delta_{\Lambda_c}$ is ``empirically small'', that is, 
the $R$ ratio sum rule should be satisfied without the shift factor $\delta_{\Lambda_c}$.
The world average of the $R_{D^{(*)}}$ measurements, reported by the Heavy Flavor Averaging Group (HFLAV) for Moriond 2024~\cite{HFLAV2024winter}, reads
\begin{align} \label{eq:RDRD*data}
 R_D^\text{exp} = 0.342\pm 0.026\,, \qquad R_{D^\ast}^\text{exp} = 0.287\pm 0.012 \,. 
\end{align}
If we take the $R$ ratio sum rule by neglecting the small shift factor $\delta_{\Lambda_c}$ that has been verified by the above discussion, we may \textit{predict} a value of $R_{\Lambda_c}$ from Eq.~\eqref{eq:Rsumrule_central} as
\begin{align}
 R_{\Lambda_c}^\text{SR} = 0.372 \pm 0.017 \,, \label{eq:RlamcSR}
\end{align}
where $R^{\rm SM}_X$ is presented in Eq.~\eqref{eq:SMprediciton}, as we will evaluate later, and Eq.~\eqref{eq:RDRD*data} is taken. 
This should be compared with the current LHCb measurement~\cite{LHCb:2022piu}
\begin{align}
 R_{\Lambda_c}^\text{LHCb} = 0.242 \pm 0.076\,. \label{eq:RlamcLHCb} 
\end{align}
Therefore, the $R$ ratio sum rule looks violated for the moment, implying that
\begin{align}
 \delta_{\Lambda_c}^\text{SR/LHCb} = - 0.39 \pm 0.23 \,. \label{eq:deviation_SRLHCb}
\end{align}
This is, however, inconsistent with the shift of $\delta_{\Lambda_c} \approx -0.035$ (indicated by the $R_{D^{(*)}}$ best-fit point as demonstrated above) beyond the above error. 
We will get back to this point once we take the form factor uncertainties into account. 

For the $b \to u$ case, we collect the recent updates for the form factor inputs (which will be detailed in the next section), and then obtain for the first time the following $R$ ratio sum rule: 
\begin{equation}
\frac{R_p}{R_p^{\rm SM}} = 0.284 \frac{R_\pi}{R_\pi^{\rm SM}} + 0.716 \frac{R_\rho}{R_\rho^{\rm SM}} + \delta_p \,, 
\end{equation}
with
\begin{align} \label{eq:deltapdefault}
\delta_p =
 & \,
 -0.090\, \left( |C_{S_L}^{u \tau}|^2 + |C_{S_R}^{u \tau}|^2 \right) 
 -0.185\, \operatorname{Re}\left[C_{S_L}^{u \tau} C_{S_R}^{u \tau *}\right]  
 -0.913\, |C_T^{u \tau}|^2 \notag \\[0.5em]
 & \hspace{-0.2cm} \, 
 +\operatorname{Re}\left[ \left( 1 + C_{V_{L}}^{u \tau} \right) \left(0.169 C_{V_R}^{u \tau *} + 0.370 C_T^{u \tau *}\right)\right] -0.203\, \operatorname{Re}\left(C_{V_R}^{u \tau}C_T^{u \tau *}\right) \notag\\[0.5em]
 &\hspace{-0.2cm} \, -0.079\, \operatorname{Re}\left[ \left(1 + C_{V_{L}}^{u \tau} \right) C_{S_R}^{u \tau *} + C_{S_L}^{u \tau} C_{V_R}^{u \tau *} \right] \,. 
\end{align}
From Eqs.~\eqref{eq:deltalambdacdefault} and \eqref{eq:deltapdefault}, we see that the suppression of the numerical factors in $\delta_p$ is milder than those in $\delta_{\Lambda_c}$ for the scalar and vector terms. 
On the other hand, the terms involving the tensor contribution in $\delta_p$ are more suppressed than those in $\delta_{\Lambda_c}$. 
Note that, to obtain the numerical coefficients involving $C_T^{u \tau}$, we need a prescription for the $\Lambda_b \to p$ tensor form factors, as will be discussed in the next section.

The collider bound on $|C_T^{u\tau}|$ obtained in Ref.~\cite{Iguro:2020keo} is roughly of $|C_T^{u\tau}| \lesssim 0.4$, which is rather milder than on $C_T^{c\tau}$. This is based on the observation that the net effect due to $V_{ub}$ suppression and parton-distribution-function enhancement of $b\bar{u} \to \tau^- \bar{\nu}$, compared with the $b\bar{c} \to \tau^- \bar{\nu}$ case, results in a reduction of the production rate and hence less statistics of the process $pp \to \tau^\pm +\text{missing}$ from $b\bar{u} \to \tau^- \bar{\nu}$. Therefore, we obtain a milder upper bound on the NP contribution, and a shift factor of $\delta_p \approx -0.29$ may be possible for the moment. 
In any case, the $R$ ratio sum rule provides us with a good motivation for measuring the semi-tauonic decays of $B \to \pi \tau\nu$, $B \to \rho \tau\nu$ and $\Lambda_b \to p \tau\nu$ at the ongoing LHCb run-3~\cite{LHCb:2018roe,LHCb:2022ine} and/or Belle~II~\cite{Belle-II:2018jsg,Belle-II:2022cgf} experiments. 

\section{Uncertainties from the form factor inputs}
\label{sec:error}

The numerical coefficients $a_X^{ij}$ in Eqs.~\eqref{eq:RP}--\eqref{eq:RB} rely on the form factor inputs that can be evaluated by several theoretical methods, such as the lattice QCD calculations and the LCSR evaluations. 
Our numerical results for $a_X^{ij}$ will be shown in Sec.~\ref{sec:analysis}, but here we begin with summarizing recent developments on the form factor evaluations, and then give the explicit inputs we will use in our analysis.

\subsection{Decay rate descriptions} 
\label{sec:DecayRate}

Let us start with providing a general description for the decay rates of the semi-leptonic $B \to P l\nu$, $B \to V l\nu$, and $\Lambda_b \to H l\nu$ processes for $P = (D,\pi)$, $V=(D^*,\rho)$, and $H=(\Lambda_c,p)$~\cite{Chen:2006nua,Celis:2012dk,Tanaka:2012nw,Fajfer:2012vx,Sakaki:2013bfa,Bernlochner:2015mya,Tanaka:2016ijq,Li:2016pdv,Hu:2018veh,Hu:2020axt,Bernlochner:2021rel,Biswas:2021cyd}. By taking the usual notations of 
\begin{align} \label{eq:NBdefinition}
 N_B \equiv \frac{G_F^2 \left|V_{qb}\right|^2}{192\pi^3m_B^3} \sqrt{Q_+^X Q_-^X} \left(1-\frac{m_l^2}{q^2}\right)^2 \,, \quad \quad
 Q_\pm^X \equiv (m_B\pm m_X)^2-q^2 \,,
\end{align}
with $q^2$ being the invariant mass squared of the lepton-neutrino system and $X$ the final-state hadron, we obtain the formulae for the 
differential decay rates as follows. 
\begin{itemize}
\item For $B \to P l\nu$ decays, 
\begin{align}
 & \frac{d\Gamma(B\to P l\nu)}{dq^2} 
 = \frac{N_B}{2} \bigg\{|1+C_{V_L}^{q l} + C_{V_R}^{q l}|^2 \left[\left(2q^2 + m_l^2\right) (H_{V_+}^{P})^2 + 3m_l^2 (H_{V_0}^{P})^2\right] 
 \notag \\[0.5em]
 &\hspace{8.5em} + 3 |C_{S_L}^{q l} + C_{S_R}^{q l} |^2 q^2 (H_S^{P})^2 + 16 |C_T^{q l}|^2 \left(q^2 + 2m_l^2\right) (H_T^{P})^2  \notag\\[0.7em]
 &\hspace{8.5em} + 6 {\rm Re} \left[(1+C_{V_L}^{q l}+C_{V_R}^{q l}) (C_{S_L}^{q l\ast} + C_{S_R}^{q l\ast})\right] m_l \sqrt{q^2} H_{V_0}^{P} H_S^{P} \notag\\[0.5em]
 &\hspace{8.5em} - 24 {\rm Re} \left[(1+C_{V_L}^{q l}+C_{V_R}^{q l}) C_T^{q l\ast}\right] m_l \sqrt{q^2} H_{V_+}^P H_T^P \bigg\} \,,
\end{align}
with the helicity amplitudes given by 
\begin{align}
H_{V_+}^P &\equiv \frac{\sqrt{Q_+^PQ_-^P}}{\sqrt{q^2}}f_+^P (q^2) \,, & H_{V_0}^P &\equiv \frac{m_B^2-m_P^2}{\sqrt{q^2}} f_0^P (q^2) \,,  \\[0.5em]
H_S^P &\equiv \frac{m_B^2-m_P^2}{m_b-m_q} f_0^P (q^2) \,, & H_T^P &\equiv -\frac{\sqrt{Q_+^PQ_-^P}}{m_B+m_P} f_T^P (q^2) \,,
\end{align}
where $f_{+,0}^P$ and $f_T^P$ are the $B \to P$ transition form factors for the vector and tensor currents, respectively. Here we follow their definitions as given in Refs.~\cite{Tanaka:2012nw,Du:2015tda,Tanaka:2016ijq}.

\item For $B \to V l\nu$ decays,  
\begin{align}
 &\frac{d\Gamma(B\to V l\nu)}{dq^2} 
 = \frac{N_B}{2} \bigg\{ \left(|1+C_{V_L}^{q l}|^2 + |C_{V_R}^{q l}|^2\right) \left[\left(2q^2+m_l^2\right)\left( (H_{V_+}^{V})^2+(H_{V_-}^{V})^2 \right.\right. \notag\\[0.5em]
 &\hspace{2.5em} \left.\left. +(H_{V_0}^{V})^2\right) + 3m_l^2 (H_{V_t}^{V})^2\right] + 3|C_{S_L}^{q l} - C_{S_R}^{q l}|^2 q^2 (H_S^{V})^2 \notag\\[0.7em]
 &\hspace{2.5em} +16|C_T^{q l}|^2 \left(q^2+2m_l^2\right) \left((H_{T_+}^{V})^2 + (H_{T_-}^{V})^2 + (H_{T_0}^{V})^2\right)  \notag\\[0.7em]
 &\hspace{2.5em} -2 {\rm Re}\left[(1+C_{V_L}^{q l})C_{V_R}^{q l\ast}\right] \left[\left(2q^2+m_l^2\right)\left(2H_{V_+}^V H_{V_-}^V + (H_{V_0}^{V})^2\right) + 3m_l^2 (H_{V_t}^{V})^2\right] \notag\\[0.7em]
 &\hspace{2.5em} -6 {\rm Re}\left[(1+C_{V_L}^{q l}-C_{V_R}^{q l})(C_{S_L}^{q l\ast} - C_{S_R}^{q l\ast})\right] m_l \sqrt{q^2} H_{V_t}^V H_S^V \notag \\[0.7em]
 &\hspace{2.5em} -24 {\rm Re}\left[(1+C_{V_L}^{q l})C_T^{q l\ast}\right] m_l \sqrt{q^2} \left(H_{T_0}^V H_{V_0}^V + H_{T_+}^V H_{V_+}^V - H_{T_-}^V H_{V_-}^V\right) \notag\\[0.5em]
 &\hspace{2.5em} +24 {\rm Re}\left[C_{V_R}^{q l}C_T^{q l\ast}\right] m_l \sqrt{q^2} \left(H_{T_0}^V H_{V_0}^V + H_{T_+}^V H_{V_-}^V - H_{T_-}^V H_{V_+}^V \right)  \bigg\} \,,
\end{align}
with the helicity amplitudes given by 
\begin{align}
H_{V_\pm}^V &\equiv (m_B + m_V) A_1^V (q^2) \mp \frac{\sqrt{Q_+^VQ_-^V}}{m_B + m_V} V^V(q^2) \,,  \\[0.5em]
H_{V_0}^V &\equiv \frac{m_B+m_V}{2m_V \sqrt{q^2}} \left[\frac{Q_+^VQ_-^V}{(m_B+m_V)^2} A_2^V(q^2) - (m_B^2-m_V^2-q^2) A_1^V(q^2)\right], \\[0.5em]
H_{V_t}^V &\equiv - \frac{\sqrt{Q_+^VQ_-^V}}{\sqrt{q^2}} A_0^V (q^2)\,, \quad\quad\quad\quad H_S^V \equiv - \frac{\sqrt{Q_+^VQ_-^V}}{m_b+m_q} A_0^V (q^2) \,, \\[0.5em]
H_{T_\pm}^V &\equiv \frac{1}{\sqrt{q^2}} \left[\sqrt{Q_+^VQ_-^V}\, T_1^V (q^2) \pm (m_B^2-m_V^2) T_2^V(q^2)\right], \\[0.5em]
H_{T_0}^V &\equiv \frac{1}{2m_V} \left[-(m_B^2+3m_V^2-q^2) T_2^V(q^2) + \frac{Q_+^VQ_-^V}{m_B^2-m_V^2} T_3^V(q^2)\right] \,,
\end{align}
where $V^V$, $A_{0,1,2}^V$, and $T_{1,2,3}^V$ are the $B \to V$ transition form factors for the vector, axial-vector, and tensor currents, respectively. We refer to Refs.~\cite{Ball:2004rg,Tanaka:2012nw,Bharucha:2015bzk} for their explicit definitions.

\item For $\Lambda_b \to H l\nu$ decays,  
\begin{align}
 &\frac{d\Gamma\left(\Lambda_b\to H l\nu\right)}{dq^2}  = \frac{N_{\Lambda_b}}{4} \bigg\{ \left(|1+C_{V_L}^{q l}|^2 + |C_{V_R}^{q l}|^2\right) \Big[\left(m_l^2 + 2q^2\right) \left( (H_{V_+}^{H+})^2 \right. \notag \\[0.5em]
 & \hspace{1.3em} \left. + (H_{V_+}^{H-})^2 + 2(H_{V_\perp}^{H+})^2 + 2(H_{V_\perp}^{H-})^2\right) + 3m_l^2 \left( (H_{V_0}^{H+})^2 + (H_{V_0}^{H-})^2\right) \Big] \notag\\[0.5em]
 & \hspace{1.3em} + 3q^2 \Big[|C_{S_L}^{q l} H_S^{H+} + C_{S_R}^{q l} H_S^{H-}|^2 + |C_{S_L}^{q l} H_S^{H-} + C_{S_R}^{q l} H_S^{H+}|^2\Big] \notag\\[0.5em]
 & \hspace{1.3em} + 16 |C_T^{q l}|^2 \left(q^2 + 2m_l^2\right) \Big( (H_{T_+}^{H+})^2 + (H_{T_+}^{H-})^2 + 2(H_{T_\perp}^{H+})^2 + 2(H_{T_\perp}^{H-})^2\Big) \notag \\[0.5em]
 & \hspace{1.3em} + 4 {\rm Re} \left[(1+C_{V_L}^{q l}) C_{V_R}^{q l\ast}\right] \Bigl[\left(m_l^2 + 2q^2\right) \left(H_{V_+}^{H+} H_{V_+}^{H-} + 2H_{V_\perp}^{H+} H_{V_\perp}^{H-}\right)  \notag\\[0.5em]
 & \hspace{1.3em}  + 3m_l^2 H_{V_0}^{H+} H_{V_0}^{H-}\Bigr] + 6 {\rm Re} \left[(1+C_{V_L}^{q l}) C_{S_L}^{q l\ast} + C_{V_R}^{q l} C_{S_R}^{q l\ast}\right] m_l \sqrt{q^2} \notag \\[0.5em]
 & \hspace{1.3em} \times \Big(H_{V_0}^{H+} H_S^{H+} + H_{V_0}^{H-} H_S^{H-}\Big) + 6 {\rm Re} \left[(1+C_{V_L}^{q l}) C_{S_R}^{q l\ast} + C_{V_R}^{q l} C_{S_L}^{q l\ast} \right] \notag \\[0.5em]
 & \hspace{1.3em} \times m_l \sqrt{q^2} \Big(H_{V_0}^{H+} H_S^{H-} + H_{V_0}^{H-} H_S^{H+}\Big) + 24 {\rm Re} \left[(1+C_{V_L}^{q l}) C_T^{q l\ast}\right] m_ l \sqrt{q^2} \notag \\[0.5em]
 & \hspace{1.3em} \times \left(H_{V_+}^{H+} H_{T_+}^{H+} + H_{V_+}^{H-} H_{T_+}^{H-} + 2H_{V_\perp}^{H+} H_{T_\perp}^{H+} + 2H_{V_\perp}^{H-} H_{T_\perp}^{H-}\right) + 24 {\rm Re} \left[C_{V_R}^{q l} C_T^{q l\ast}\right]  \notag\\[0.5em]
 & \hspace{1.3em} \times m_l \sqrt{q^2} \left(H_{V_+}^{H+} H_{T_+}^{H-} + H_{V_+}^{H-} H_{T_+}^{H+} + 2H_{V_\perp}^{H+} H_{T_\perp}^{H-} + 2H_{V_\perp}^{H-} H_{T_\perp}^{H+}\right) \bigg\}\,,
\end{align}
with the helicity amplitudes given by 
\begin{align}
H_{V_+}^{H\pm} &\equiv \frac{1}{\sqrt{q^2}} \left[\left(m_{\Lambda_b} + m_H\right) F_+^H \sqrt{Q_-^H} \mp \left(m_{\Lambda_b} - m_H\right) G_+^H \sqrt{Q_+^H}\right], \\[0.5em]
H_{V_0}^{H\pm} &\equiv \frac{1}{\sqrt{q^2}} \left[\left(m_{\Lambda_b} - m_H\right) F_0^H \sqrt{Q_+^H} \mp \left(m_{\Lambda_b} + m_H\right) G_0^H \sqrt{Q_-^H}\right], \\[0.5em]
H_{V_\perp}^{H\pm} &\equiv F_\perp^H \sqrt{Q_-^H} \mp G_\perp^H \sqrt{Q_+^H}, \\[0.5em]
H_S^{H\pm} &\equiv \frac{m_{\Lambda_b} - m_H}{m_b - m_q} F_0^H \sqrt{Q_+^H} \pm \frac{m_{\Lambda_b} + m_H}{m_b + m_q} G_0^H \sqrt{Q_-^H}, \\[0.5em] 
H_{T_+}^{H\pm} &\equiv h_+^H \sqrt{Q_-^H} \pm \Tilde{h}_+^H \sqrt{Q_+^H}, \\[0.5em]
H_{T_\perp}^{H\pm} &\equiv \frac{1}{\sqrt{q^2}} \left[\left(m_{\Lambda_b} + m_H\right) h_\perp^H \sqrt{Q_-^H} \pm \left(m_{\Lambda_b} - m_H\right) \Tilde{h}_\perp^H \sqrt{Q_+^H}\right],
\end{align}
where the expression of $N_{\Lambda_b}$ is the same as that of $N_B$ defined by Eq.~\eqref{eq:NBdefinition}, but now with the meson mass $m_B$ replaced by the baryon mass $m_{\Lambda_b}$. 
The form factors $F_{+,0,\perp}^H$, $G_{+,0,\perp}^H$, $h_{+,\perp}^H$ and $\Tilde{h}_{+,\perp}^H$ for the vector, axial-vector, and tensor currents are defined as in Refs.~\cite{Chen:2001zc,Feldmann:2011xf,Detmold:2015aaa,Detmold:2016pkz}.
\end{itemize}

The form factors introduced as above need to be extracted from experimental data and/or evaluated by non-perturbative methods like lattice QCD and LCSR. 
Concerning the $q^2$ dependence of these form factors, it is now standard to use the $z$ parametrization that satisfies unitarity, analyticity and perturbativity~\cite{Caprini:1997mu,Boyd:1995cf,Bourrely:2008za}. 
To this end, it is convenient to introduce the conformal mapping variable
\begin{align}
 z(q^2) \equiv \frac{\sqrt{t_+ - q^2} - \sqrt{t_+ - t_0}}{\sqrt{t_+ - q^2} + \sqrt{t_+ - t_0}} \,, \label{eq:zq2}
\end{align}
which maps the $q^2$-plane cut for $q^2>t_+$ onto the disk $|z(q^2)|<1$ in the $z$ complex plane, with $z(t_+)=-1$, $z(\infty)=1$, and $z(t_0)=0$, 
where $t_+$ is the threshold parameter and $t_0$ determines the point $q^2$ mapped onto the origin in the $z$ plane. 
The values of $t_+$ and $t_0$ are fixed for each form factor depending on the individual setup of the analysis. 
In this way, the $q^2$ dependence of the form factors is given as a Taylor expansion in the variable $z(q^2)$, where the expansion coefficients are free parameters to be fitted. 
Explicit definitions are different for each form factor analysis, which will be described in the next two subsections. 
We will summarize recent updates of the form factor inputs and indicate what will be used in our analysis.

Let us briefly mention QED corrections to the (semi-)leptonic $B$ decays. 
Photon emissions from the pseudo-scalar mesons (scalar QED corrections) have been taken into account in the experimental setup (via PHOTOS~\cite{Golonka:2005pn}). 
On the other hand, virtual QED corrections that are not covered by PHOTOS have also been studied in the literature. 
In Ref.~\cite{deBoer:2018ipi}, a Coulomb-type interaction (soft-photon correction) was evaluated, which produces a $\sim4\%$ overall correction to $R_D$. 
Furthermore, structure-dependent QED corrections that affect the meson transition have been investigated in leptonic decays~\cite{Beneke:2017vpq,Beneke:2019slt,Cornella:2022ubo,Boer:2023vsg,Rowe:2024jml,Rowe:2024pfs}.  
This study has been extended to the semi-leptonic decay $B \to K \ell^+\ell^-$ in Ref.~\cite{Isidori:2020acz}, which has found a $\alpha \log(m_\ell / m_B)$ type correction. 
It could be relevant for $B^+ \to \bar D^0 \ell^+\nu$ in our case. 
See also Ref.~\cite{Zwicky:2021olr} for $B^+ \to \pi^0 \ell^+\nu$. 
In the present study, we do not consider such kinds of QED effects. 
%

\subsection[\texorpdfstring{$B \to D$}{B2D}, \texorpdfstring{$B \to D^*$}{B2Dstar}, and \texorpdfstring{$\Lambda_b \to \Lambda_c$}{Lambdab2Lambdac} form factors]{\boldmath \texorpdfstring{$B \to D$}{B2D}, \texorpdfstring{$B \to D^*$}{B2Dstar}, and \texorpdfstring{$\Lambda_b \to \Lambda_c$}{Lambdab2Lambdac} form factors}
\label{sec:FFbc}

The $B \to D$ and $B \to D^*$ form factors have been studied for various purposes, as they affect both the $|V_{cb}|$ determinations and the $R_{D^{(\ast)}}$ evaluations. 
Theoretical studies have been done mainly by the lattice QCD and LCSR approaches. 
The lattice results are available for $B \to D$ in Refs.~\cite{MILC:2015uhg,Na:2015kha} while for $B \to D^*$ in Refs.~\cite{FermilabLattice:2021cdg,Aoki:2023qpa,Harrison:2023dzh}. In particular, the latter has been recently reported for the first time but, for the moment, we do not have a conclusive combined fit for these results. 
Our purpose in this paper is to highlight the significance of uncertainties from the theoretical inputs, which is useful once a conclusive form factor evaluation (or fit) is available in the future. 
Here let us also point out that the lattice result of the tensor form factor is only available for $B \to D^{*}$ by HPQCD~\cite{Harrison:2023dzh}, which is however based on the heavy quark symmetry. 
Hence, it can be concluded that we have no direct lattice point for the tensor form factor in the so-called Boyd-Grinstein-Lebed (BGL) parametrization~\cite{Boyd:1995cf}. 
The LCSR calculation has been done in Ref.~\cite{Gubernari:2018wyi}, and then updated by another group in Ref.~\cite{Cui:2023jiw} by including the next-to-leading-order QCD corrections and the various power-suppressed contributions. 
It should be clarified that the former study includes the tensor form factor, 
while the latter study provides a combined fit to their LCSR result and the lattice data point from Ref.~\cite{FermilabLattice:2021cdg} for the SM currents. 

\begin{table}[t]
\centering
\renewcommand{\arraystretch}{1.5}
\begin{adjustbox}{width=0.99\columnwidth,center}
\begin{tabular}{c|cc|cc|c}
  \hline\hline
    & \multicolumn{2}{c|}{Lattice} & \multicolumn{2}{c|}{LCSR} & Lattice + LCSR \\
   & SM & Tensor & SM & Tensor & SM   \\
  \hline
  $B \to D$ & Refs.~\cite{MILC:2015uhg,Na:2015kha} & no data & Ref.~\cite{Gubernari:2018wyi,Cui:2023jiw} & Ref.~\cite{Gubernari:2018wyi} &  Ref.~\cite{Cui:2023jiw}$\,^{(**)}$  \\
  \hline
  $B \to D^*$ & Refs.~\cite{FermilabLattice:2021cdg,Aoki:2023qpa,Harrison:2023dzh} & no data$\,^{(*)}$ & Ref.~\cite{Gubernari:2018wyi,Cui:2023jiw} & Ref.~\cite{Gubernari:2018wyi} &   Ref.~\cite{Cui:2023jiw}$\,^{(**)}$   \\
  \hline
  $\Lambda_b \to \Lambda_c$ & Ref.~\cite{Detmold:2015aaa} & Ref.~\cite{Datta:2017aue} & no data & no data & -- \\
  \hline\hline
\end{tabular}
\end{adjustbox}
\caption{
Summary of recent theoretical evaluations of the $B \to D$, $B \to D^*$, and $\Lambda_b \to \Lambda_c$ form factors from lattice QCD and/or LCSR approaches. 
$\,^{(*)}$ The $B \to D^*$ tensor form factor is obtained in Ref.~\cite{Harrison:2023dzh} only under the heavy quark symmetry. 
$\,^{(**)}$ The lattice QCD results of Refs.~\cite{MILC:2015uhg,Na:2015kha} and \cite{FermilabLattice:2021cdg} are taken to provide their Lattice + LCSR combined fit~\cite{Cui:2023jiw} for the $B \to D$ and $B \to D^*$ form factors, respectively. \label{tab:FFsummaryBD}
}
\end{table}

We list the current situation on the form factor evaluations in Table~\ref{tab:FFsummaryBD}, where the columns labelled by ``SM'' indicate the vector and axial-vector form factors. 
As can be seen from the above summary table, the form factor inputs for $B \to D$ and $B \to D^*$ transitions have different sources and the lattice data points are still incoherent for the moment. 
Concerning our purpose, however, it is sufficient to simply choose the results 
from Ref.~\cite{Cui:2023jiw} for the vector ($f_{+,0}^D$, $V^{D^\ast}$) and axial-vector ($A_{0,1,2}^{D^\ast}$) and from Ref.~\cite{Gubernari:2018wyi} for the tensor ($f_T^D$ and $T_{1,2,3}^{D^\ast}$) form factors. It is also noted that the BGL form factors for $B \to D^*$ transition are usually given, instead of $V^{D^\ast}$ and $A_{0,1,2}^{D^\ast}$ themselves, by~\cite{Cui:2023jiw,Kapoor:2024ufg}
\begin{align}
 & f(q^2) = (m_B + m_{D^*}) A_1^{D^*}(q^2) \,, \quad g(q^2) = \frac{2V^{D^*}(q^2)}{m_B + m_{D^*}} \,, \quad F_2(q^2) = 2 A_0^{D^*}(q^2) \,, \\[0.5em]
 & F_1(q^2) = \frac{m_B+m_{D^*}}{2m_{D^*}} \left[(m_B^2-m_{D^*}^2-q^2) A_1^{ D^*}(q^2) - \frac{Q_+^{D^\ast}Q_-^{D^\ast}}{(m_B+m_{D^*})^2} A_2^{D^*}(q^2) \right] \,. 
\end{align}
These form factors can be schematically parametrized as~\cite{Boyd:1995cf,Boyd:1997kz}
\begin{align}
 F(z) = \frac{1}{P_F(z) \phi_F(z)} \sum_{n=0}^{N_F} a^{F}_n z(q^2)^n \,, 
\end{align}
in terms of the BGL series coefficients $a^{F}_n$, where $N_F$ stands for the truncation order and the $z$ expansion variable is defined as in Eq.~\eqref{eq:zq2} with the choice of 
\begin{align}
 t_+ = (m_B + m_{D^{(*)}})^2 \,, \qquad t_0 = (m_B - m_{D^{(*)}})^2 \,. 
\label{eq:zq2DSM}
\end{align}
The prefactors $P_F(z)$ and $\phi_F$ are introduced to control the convergence of the form factors; $P_F(z)$ are the Blaschke factors served to account for the explicit poles in the variable $q^2$ that are associated with the on-shell productions of $B_c^*$ bound states for $q^2<t_+$, and $\phi_F(z)$ are the outer functions that guarantee the form factors to satisfy the unitarity bounds. 
The BGL series coefficients $a_n^F$ have been fitted to the lattice QCD and LCSR evaluations with $N_F =2$ for the SM currents in Ref.~\cite{Cui:2023jiw}. 
Explicit forms of the functions and numerical inputs of $P_F(z)$, $\phi_F(z)$, and $a_n^F$ are given in Appendix~\ref{app:FFinputs}.

For the $B \to D^{(*)}$ tensor form factors, we follow Ref.~\cite{Gubernari:2018wyi} and adopt their fitted expansion parameters obtained by incorporating the LCSR data points. 
Here the form factors are parametrized in the Bharucha-Straub-Zwicky (BSZ) pattern~\cite{Bharucha:2015bzk} as
\begin{align}
 F(q^2) = \frac{1}{1 -q^2/M_{F}^2} \sum_{n=0}^{N_F} a_n^F \big[z(q^2) - z(0) \big]^n \,, \label{BSZ}
\end{align}
where the $z$ variable takes the same form as in Eq.~\eqref{eq:zq2} but now with the choice of 
\begin{align}
 t_+ \equiv \left(m_B + m_{D^{(\ast)}}\right)^2 \,, \quad\quad t_0 \equiv \left(m_B + m_{D^{(\ast)}}\right) \left(\sqrt{m_B} - \sqrt{m_{D^{(\ast)}}}\right)^2 \,.
\end{align}
We employ the numerical inputs for $a_n^F$ given in Ref.~\cite{Gubernari:2018wyi}. 
See Appendix~\ref{app:FFinputs} for the precise expression.

On the other hand, the $\Lambda_b \to \Lambda_c$ form factors have been evaluated by lattice QCD in Ref.~\cite{Detmold:2015aaa} for the vector and axial-vector currents, and then in Ref.~\cite{Datta:2017aue} for the tensor current. Their $q^2$ dependence is now parametrized as~\cite{Bourrely:2008za}
\begin{align}
 F (q^2) = \frac{1}{1 -q^2/M_{F}^2} \sum_{n=0}^{N_F} a_n^F  z(q^2)^n \,, \label{FF_bar}
\end{align}
for $F= F_+^{\Lambda_c}$, $F_0^{\Lambda_c}$, $F_\perp^{\Lambda_c}$, $G_+^{\Lambda_c}$, $G_0^{\Lambda_c}$, $G_\perp^{\Lambda_c}$, $h_+^{\Lambda_c}$, $h_\perp^{\Lambda_c}$, $\tilde{h}_+^{\Lambda_c}$, and $\tilde{h}_\perp^{\Lambda_c}$, where the $z$ variable is also defined as in Eq.~\eqref{eq:zq2} with
\begin{align}
  t_+ \equiv M_F^2 \,, \quad\quad  t_0 \equiv (m_{\Lambda_b} - m_{\Lambda_c})^2 \,, 
\end{align}
and the numerical descriptions for $M_F$ and $a_n^F$ are shown in Appendix~\ref{app:FFinputs}. 
In Refs.~\cite{Detmold:2015aaa,Datta:2017aue}, the authors have also provided a procedure to calculate the ``systematic'' uncertainty associated with a different choice of the expansion order $N_F$ (the so-called higher-order fit). In our study, we follow exactly the same procedure proposed in Ref.~\cite{Detmold:2015aaa} to calculate this additional uncertainty, and the total uncertainty is then obtained by adding the statistical and systematic ones in quadrature. For further details, we refer the readers to Eqs.~(82)--(84) in Ref.~\cite{Detmold:2015aaa}.

\subsection[\texorpdfstring{$B \to \pi$}{B2pi}, \texorpdfstring{$B \to \rho$}{B2rho}, and \texorpdfstring{$\Lambda_b \to p$}{Lambdab2proton} form factors]{\boldmath \texorpdfstring{$B \to \pi$}{B2pi}, \texorpdfstring{$B \to \rho$}{B2rho}, and \texorpdfstring{$\Lambda_b \to p$}{Lambdab2proton} form factors} 
\label{sec:FFbu}

In contrast to the extensively investigated $b\to c$ case, the current explorations of the semi-leptonic $b\to u l \nu$ decays are less studied both experimentally and theoretically. Regarding the experimental measurements, only the light-lepton modes have been observed for the moment, as mentioned in Sec.~\ref{sec:intro}. On the other hand, several theoretical evaluations are available for the $B\to\pi$, $B\to\rho$, and $\Lambda_b\to p$ form factors, but not for all kinds of methods and currents. For example, the $B\to\pi$ form factors have been widely studied using both the lattice QCD~\cite{Flynn:2015mha,FermilabLattice:2015mwy,Colquhoun:2022atw,FermilabLattice:2015cdh} and LCSR~\cite{Ball:2004ye,Duplancic:2008ix,Wang:2015vgv,Khodjamirian:2017fxg,Lu:2018cfc,Gubernari:2018wyi,Leljak:2021vte,Cui:2022zwm} approaches, with which the semi-leptonic $B\to\pi l \nu$ decay has been used to determine the CKM matrix element $|V_{ub}|$ exclusively. The $B\to\rho$ form factors are provided only by the LCSR calculations~\cite{Bharucha:2015bzk,Gubernari:2018wyi,Gao:2019lta} for both the SM (vector and axial-vector) and NP (tensor) currents. 

\begin{table}[t]
\centering
\renewcommand{\arraystretch}{1.5}
\begin{adjustbox}{width=0.99\columnwidth,center}
\begin{tabular}{c|cc|cc|c}
	\hline\hline
	& \multicolumn{2}{c|}{Lattice} & \multicolumn{2}{c|}{LCSR} & Lattice + LCSR \\
	& SM & Tensor & SM & Tensor & SM + Tensor \\
	\hline
	$B \to \pi$ & Refs.~\cite{Flynn:2015mha,FermilabLattice:2015mwy,Colquhoun:2022atw} & Ref.~\cite{FermilabLattice:2015cdh} & \multicolumn{2}{c|}{Refs.~\cite{Ball:2004ye,Duplancic:2008ix,Wang:2015vgv,Khodjamirian:2017fxg,Lu:2018cfc,Gubernari:2018wyi,Leljak:2021vte}}  & Ref.~\cite{Cui:2022zwm}  \\
	\hline
	$B \to \rho$ & no data & no data & \multicolumn{2}{c|}{Refs.~\cite{Bharucha:2015bzk,Gubernari:2018wyi,Gao:2019lta}} & -- \\
        \hline
        $\Lambda_b\to p$ & Ref.~\cite{Detmold:2015aaa} & no data & Ref.~\cite{Wang:2009hra} & no data & -- \\
	\hline\hline
\end{tabular}
\end{adjustbox}
\caption{
Summary of recent theoretical estimations of the $B \to \pi$, $B \to \rho$, and $\Lambda_b \to p$ form factors from lattice QCD and/or LCSR approaches. 
Note that Ref.~\cite{Wang:2015vgv} provides the LCSR fit result for $B \to \pi$ as well, but only for the SM currents. 
\label{tab:FFsummaryBpirho}
}
\end{table}

In Table~\ref{tab:FFsummaryBpirho}, we list the recent theoretical evaluations of the form factors.
At first, we follow Ref.~\cite{Cui:2022zwm} to parametrize the $q^2$ dependence of the $B\to\pi$ form factors through a modified BGL scenario, the so-called Bourrely-Caprini-Lellouch~(BCL) form~\cite{Bourrely:2008za} that combines the pole factorization with an expansion in powers of the conformal mapping variable and simplifies the BGL unitary requirement. The BCL parametrization of the vector and tensor form factors reads~\cite{Bourrely:2008za}
\begin{equation}
F(q^2) = \frac{1}{1-q^2/M_{F}^2} \sum_{n=0}^{N_F-1} b_n^F \left[z(q^2)^n - (-1)^{n-N_F} \frac{n}{N_F} z(q^2)^{N_F}\right] \,, 
\end{equation}
with the sub-threshold resonance masses given by~\cite{ParticleDataGroup:2024cfk}
\begin{equation}
M_{f_{+}^\pi} = M_{f_{T}^\pi} = 5.325\,\text{GeV} \,, 
\end{equation}
while for the scalar form factor we have 
\begin{equation}
f_0^\pi(q^2) = \sum_{n=0}^{N_{f_0^\pi}-1} b_n^{f_0^\pi} z(q^2)^n \,,
\end{equation}
where the disappearance of the pole factor is due to the fact that the lowest-lying resonance in the $J^P = 0^+$ channel is located above the $B\pi$ production threshold. 
The conformal mapping variable $z(q^2)$ is given as in Eq.~\eqref{eq:zq2}, with 
\begin{align}
 t_+ \equiv \left(m_B + m_\pi\right)^2 \,, \quad\quad t_0 \equiv \left(m_B + m_\pi\right) \left(\sqrt{m_B} - \sqrt{m_\pi}\right)^2 \,.
\end{align}
The fit results for the BCL series coefficients $b_n^F$ from Ref.~\cite{Cui:2022zwm} are exhibited in Appendix~\ref{app:FFinputs}.

Regarding the $B \to \rho$ form factors, we employ the LCSR results provided in Ref.~\cite{Bharucha:2015bzk}, although Refs.~\cite{Gubernari:2018wyi,Gao:2019lta} have also provided the fit results.
This is a reasonable choice because, compared to the results obtained with the $B$-meson light-cone distribution amplitude, the form factors can be currently evaluated with better accuracy by using the light-meson light-cone distribution amplitude~\cite{Bharucha:2015bzk}, as analyzed in Ref.~\cite{Gubernari:2018wyi}.\footnote{Notice that the modern evaluations of the $B \to D^*$ form factors in the LCSR approach are only available with the former setup~\cite{Gubernari:2018wyi,Cui:2023jiw,Faller:2008tr}.}
This choice results in smaller uncertainties in the LCSR form factor fits of the $B \to \rho l\nu$ decay amplitudes. 
Instead, a large decay width of the $\rho$ meson can be a source of uncertainty. 
It is briefly mentioned in Appendix~\ref{app:FFinputs}. 
The form factors of our concern can be parametrized in the BSZ form as  
\begin{align}
 F(q^2) = \frac{1}{1 -q^2/M_{F}^2} \sum_{n=0}^{N_F} a_n^F \big( z(q^2) - z(0) \big)^n \,, 
\end{align}
with the $z$ variable given by Eq.~\eqref{eq:zq2} and
\begin{align}
 t_+ \equiv \left(m_B + m_{\rho}\right)^2 \,, \quad\quad t_0 \equiv \left(m_B + m_{\rho}\right) \left(\sqrt{m_B} - \sqrt{m_{\rho}}\right)^2 \,.
\end{align}
See Appendix~\ref{app:FFinputs} for the explicit values of the form factor inputs~\cite{Bharucha:2015bzk}.

For the baryonic $\Lambda_b\to p l \nu$ process, detailed computations of the relevant form factors have been carried out by both the lattice QCD~\cite{Detmold:2015aaa} and LCSR~\cite{Wang:2009hra,Huang:2022lfr} approaches for the vector and axial-vector currents. Here we simply take the lattice QCD results, in which the $q^2$ dependence of the form factors is represented as~\cite{Detmold:2015aaa}
\begin{align}
 F (q^2) = \frac{1}{1 -q^2/M_{F}^2} \sum_{n=0}^{N_F} a_n^F  z(q^2)^n \,, 
\end{align}
for $F= F_+^p$, $F_0^p$, $F_\perp^p$, $G_+^p$, $G_0^p$, and $G_\perp^p$. 
It is noted that $t_+$ and $t_0$ involved in the $z(q^2)$ variable defined by Eq.~\eqref{eq:zq2} is now given by~\cite{Detmold:2015aaa}
\begin{align}
  t_+ \equiv (m_B + m_\pi)^2 \,, \quad\quad  t_0 \equiv (m_{\Lambda_b} - m_{p})^2 \,.
\end{align}
See Appendix~\ref{app:FFinputs} for the numerical inputs.

At present, we have no lattice QCD nor LCSR results for the $\Lambda_b \to p$ tensor form factors $h_{+,\perp}^p$ and $\tilde{h}_{+,\perp}^p$. 
However, we can infer the following approximations when taking into account the $J^P$ properties of the $\Lambda_b \to p$ transition:
\begin{equation}
 h_{+,\perp}^p \equiv R_{+,\perp} F_{+,\perp}^p\,, \quad\quad 
 \tilde{h}_{+,\perp}^p \equiv \tilde{R}_{+,\perp} G_{+,\perp}^p\,,
\end{equation}
where $R_{+,\perp}$ and $\tilde{R}_{+,\perp}$ indicate some real constant parameters that need to be estimated. 
According to the preliminary analysis made in Ref.~\cite{StefanComment}, 
we have $R_+ \approx 1.2$, $R_\perp \approx 0.7$, $\tilde{R}_{+} \approx 1.0$, and $\tilde{R}_{\perp} \approx 1.0$, with an about $10\%$ accuracy. 
Based on this observation, we will take these estimations as input, with a conservative uncertainty of $20\%$ for each tensor form factor. The status of ongoing next-generation calculations of these form factors can be found in Ref.~\cite{Meinel:2023wyg}.

\subsection{Numerical results} 
\label{sec:analysis}

We are now ready to evaluate the appropriate propagation of the form factor uncertainties on the general $R_X$ formulae. 
Taking into account all the form factor inputs summarized in Sec.~\ref{sec:FFbc}, we obtain the following values of the coefficients $a_{X}^{ij}$ defined in Eqs.~\eqref{eq:RP}--\eqref{eq:RB}:
\begin{align}
 & \frac{R_D}{R_{D}^\textrm{SM}}: \quad & 
 & a_{D}^{SS} = 1.070 \pm 0.006 \,,
 & &a_{D}^{TT} = 0.721 \pm 0.341 \,, \notag \\
 & & 
 &a_{D}^{VS} = 1.528 \pm 0.006 \,,&  
 &a_{D}^{VT} = 1.015 \pm 0.233 \,,
 \label{eq:RDform} \\[1em]
 &\frac{R_{D^*}}{R_{D^*}^\textrm{SM}}: \quad &
 &a_{D^*}^{SS} = 0.043 \pm 0.002 \,,& 
 &a_{D^*}^{TT} = 17.76 \pm 9.37 \,, \notag \\
 & &
 &a_{D^*}^{V_LV_R} = -1.797 \pm 0.015 \,,&  
 &a_{D^*}^{VS} = -0.113 \pm 0.004 \,, \notag \\[0.5em]
 & &
 &a_{D^*}^{V_LT} = -5.470 \pm 1.717 \,,& 
 &a_{D^*}^{V_RT} = 7.098 \pm 1.944 \,, 
 \label{eq:RDstform} \\[1em]
 &\frac{R_{\Lambda_c}}{R_{\Lambda_c}^\textrm{SM}}: \quad &
 &a_{\Lambda_c}^{SS} = 0.321 \pm 0.005 \pm 0.013  \,,& 
 &a_{\Lambda_c}^{TT} = 10.44 \pm 0.21 \pm 0.91  \,, \notag \\
 & &
 &a_{\Lambda_c}^{V_LV_R} = -0.722 \pm 0.027 \pm 0.060 \,,& 
 &a_{\Lambda_c}^{S_LS_R} = 0.513 \pm 0.011 \pm 0.029 \,, \notag \\[0.5em]
 & &
 &a_{\Lambda_c}^{VS_1} = 0.334 \pm 0.009 \pm 0.022  \,,& 
 &a_{\Lambda_c}^{VS_2} = 0.497 \pm 0.008 \pm 0.019 \,,& \notag \\[0.5em]
 & &
 &a_{\Lambda_c}^{V_LT} = -3.110 \pm 0.065 \pm 0.211  \,,&
 &a_{\Lambda_c}^{V_RT} = 4.880 \pm 0.053 \pm 0.215  \,, 
 \label{eq:RLcform}
\end{align} 
for the semi-leptonic $b \to c$ decays, where the third terms in $a_{\Lambda_c}^{ij}$ account for the systematic uncertainties obtained by following the same procedure as in Ref.~\cite{Detmold:2015aaa}.
Combining all these results, we finally obtain the $R$ ratio sum rule,
\begin{align}
 \frac{R_{\Lambda_c}}{R^{\rm SM}_{\Lambda_c}} 
 = (0.272 \pm 0.015)\, \frac{R_D}{R^{\rm SM}_D} 
 + (0.728 \mp 0.015)\,\frac{R_{D^\ast}}{R^{\rm SM}_{D^\ast}} 
 + \delta_{\Lambda_c}\,, \label{eq:RLc_sum_rule}
\end{align} 
with the shift factor given by
\begin{align} 
 \delta_{\Lambda_c} = 
 & \,(-0.001 \pm 0.005)\,\left( |C_{S_L}^{c\tau}|^2 + |C_{S_R}^{c\tau}|^2\right) 
 + (-0.007 \pm 0.005) \,\textrm{Re}\left( C_{S_L}^{c\tau} C_{S_R}^{c\tau\ast} \right) \notag \\[0.3em]
 & + (-2.681\pm 6.907) \,|C_T^{c\tau}|^2  
 + (-0.561\pm 1.439) \,\textrm{Re}\left( C_{V_R}^{c\tau} C_{T}^{c\tau\ast} \right) \notag \\[0.3em]
 & + \textrm{Re} \left[ \left(1 +C_{V_L}^{c\tau} \right) \left\{ 
 (0.041\pm 0.034) C_{V_R}^{c\tau\ast} 
 + (0.594\pm 1.274) C_T^{c\tau\ast} \right\}\right] \notag \\[0.3em]
 & + (-0.002 \pm 0.009) \textrm{Re} \left[ \left(1 +C_{V_L}^{c\tau} \right) C_{S_R}^{c\tau\ast} + C_{S_L}^{c\tau} C_{V_R}^{c\tau\ast} \right] \,.  \label{eq:delta_Lc}
\end{align} 
We observe a fluctuation of $\pm 0.015$ from the central values for the $R_X$ ratio coefficients. 
It can also be seen that the vector and scalar terms do not generally have considerable uncertainties in $\delta_{\Lambda_c}$, 
while the terms involving the tensor WC $C_{T}^{c\tau}$ receive non-negligible uncertainties, which stem from the LCSR results for the $B \to D^{(*)}$ tensor form factors~\cite{Gubernari:2018wyi}. 
Once future precise lattice QCD results are confirmed with a good consistency and a combined fit is available, 
it is expected that these terms can be evaluated at a similar level of accuracy with that of the vector and scalar terms. 
Let us remind that the recent lattice QCD results of Refs.~\cite{FermilabLattice:2021cdg,Aoki:2023qpa,Harrison:2023dzh} are not conclusive yet, and making a private combined fit with detailed correlations taken into account is beyond the scope of our work. 
Since our purpose is to exhibit how the $R$ ratio sum rule is violated by the uncertainties of the theoretical inputs, our setup is rather informative.

It is also noted that the $R$ ratio sum rule and the shift factor $\delta_{\Lambda_c}$ in Ref.~\cite{Fedele:2022iib} were obtained by taking their combined fits to the theoretical/experimental inputs and a unique form factor parametrization based on the heavy quark effective theory. This causes a difference from our results that are based on the well-known $z$-expansion parametrizations of the $q^2$ dependence of the form factors, although they are well consistent with each other within the uncertainty. The suppression of the numerical coefficients in $\delta_{\Lambda_c}$ can be obviously confirmed in both cases.  

Similarly, we obtain the following numerical results for the $b \to u$ case, where our inputs are summarized in Sec.~\ref{sec:FFbu} 
and the additional conservative errors are imposed on the tensor form factors, $R_+ \simeq 1.20 \pm 0.24$, $R_\perp \simeq 0.70\pm0.14$, and $\tilde R_{+,\perp} \simeq 1.00\pm0.20$:
\begin{align}
 \label{RXRXSM_1} 
 &\frac{R_\pi}{R_{\pi}^\textrm{SM}}: \quad & 
 &a_{\pi}^{SS} = 1.497 \pm 0.094 \,,& 
 &a_{\pi}^{TT} = 3.804 \pm 0.238 \,,& \notag \\
 & &
 &a_{\pi}^{VS} = 1.239 \pm 0.073 \,,& 
 &a_{\pi}^{VT} = 2.511 \pm 0.088 \,, \\[1.0em] 
\label{RXRXSM_2} 
 &\frac{R_{\rho}}{R_{\rho}^\textrm{SM}}: \quad &
 &a_{\rho}^{SS} = 0.276 \pm 0.030 \,,& 
 &a_{\rho}^{TT} = 12.75 \pm 1.32 \,,& \notag \\
 & &
 &a_{\rho}^{V_LV_R} = -1.294 \pm 0.081 \,,& 
 &a_{\rho}^{VS} = -0.349 \pm 0.037 \,,& \notag \\[0.5em]
 & &
 &a_{\rho}^{V_LT} = -2.327 \pm 0.189 \,,& 
 &a_{\rho}^{V_RT} = 5.172 \pm 0.332 \,, \\[1.0em]
 \label{RXRXSM_7}
 &\frac{R_{p}}{R_{p}^\textrm{SM}}: \quad &
 &a_{p}^{SS} = 0.533 \pm 0.029 \pm 0.022 \,,& 
 &a_{p}^{TT} = 9.304 \pm 2.108 \pm 0.544 \,,& \notag \\
 & &
 &a_{p}^{V_LV_R} = -0.190 \pm 0.095 \pm 0.130 \,,& 
 &a_{p}^{S_LS_R} = 0.266 \pm 0.056 \pm 0.065 \,,&  \notag \\[0.5em]
 & &
 &a_{p}^{VS_1} = 0.101 \pm 0.033 \pm 0.037 \,,& 
 &a_{p}^{VS_2} = 0.523 \pm 0.029 \pm 0.024 \,,& \notag \\[0.5em]
 & &
 &a_{p}^{V_LT} = -0.583 \pm 0.528 \pm 0.306 \,,&
 &a_p^{V_RT} = 4.214 \pm 0.453 \pm 0.124 \,,
\end{align}
where the systematic errors shown by the third terms of $a_p^{ij}$ are obtained by following Ref.~\cite{Detmold:2015aaa} as well as $a_{\Lambda_c}^{ij}$.
In the end, we arrive at the following $R$ ratio sum rule: 
\begin{equation}
\frac{R_{p}}{R^{\rm SM}_{p}} 
= (0.284 \pm 0.037)\, \frac{R_\pi}{R^{\rm SM}_\pi} + (0.716 \mp 0.037)\,\frac{R_{\rho}}{R^{\rm SM}_{\rho}} + \delta_{p}\,, \label{eq:Rp_sum_rule}
\end{equation}
with
\begin{align}
 \delta_{p} = 
 & \,(-0.090 \pm 0.059)\,\left( |C_{S_L}^{u\tau}|^2 + |C_{S_R}^{u\tau}|^2\right) + (-0.185 \pm 0.038) \,\textrm{Re}\left( C_{S_L}^{u\tau} C_{S_R}^{u\tau\ast} \right) \notag  \\[0.3em]
 & + (-0.913 \pm 2.403)\,|C_T^{u\tau}|^2 + (-0.203 \pm 0.538) \,\textrm{Re}\left( C_{V_R}^{u\tau} C_{T}^{u\tau\ast} \right) \notag \\[0.3em]
 & + \textrm{Re} \left[ 
 \left(1 +C_{V_L}^{u\tau} \right) \left\{ (0.169 \pm 0.158) C_{V_R}^{u\tau\ast} + (0.370 \pm 0.632) C_T^{u\tau\ast} \right\}
 \right] \notag \\[0.3em]
 &
 + (-0.079 \pm 0.056)\, \textrm{Re} \left[ \left(1 +C_{V_L}^{u\tau} \right) C_{S_R}^{u\tau\ast} + C_{S_L}^{u\tau} C_{V_R}^{u\tau\ast} \right]\,. \label{eq:delta_p}
\end{align} 
We can find that, compared with $\delta_{\Lambda_c}$, the shift factor $\delta_{p}$ has larger (smaller) numerical coefficients for the scalar (tensor) terms. 
This could indicate that (i) the $R$ ratio sum rule for $b \to u$ is more (less) sensitive to the scalar (tensor) NP contribution, or 
(ii) there is another different setup so that the NP effect on $\delta_p$ is changed. However, such a setup can be checked only when the evaluation of the $\Lambda_b \to p$ tensor form factors is well-developed. 
As pointed out in Sec.~\ref{sec:FFbu}, the tensor terms in $B \to \rho\tau\nu$ are evaluated with a better accuracy than in $B \to D^*\tau\nu$, thanks to the precise light-meson inputs in the LCSR study. 
For the tensor terms in $\Lambda_b \to p\tau\nu$, on the other hand, they are currently obtained based on the estimate of the tensor form factors as detailed in Sec.~\ref{sec:FFbu}. 
Although a direct lattice QCD calculation of the $\Lambda_b \to p$ tensor form factors is still missing, our result is already well informative for the moment: 
one can see that the tensor (scalar) NP contribution is less (more) sensitive in the $b \to u$ than in the $b \to c$ case. 

For complementary, the SM predictions of the LFU ratios $R_X^\text{SM}$ with the use of our form factor setup are summarized as
\begin{align}
 &R_D^\text{SM} = 0.302 \pm 0.008 \,,&
 &R_{D^*}^\text{SM} = 0.257 \pm 0.005 \,,&
 &R_{\Lambda_c}^\text{SM} = 0.332 \pm 0.010 \,, \notag \\[0.5em]
 &R_\pi^\text{SM} = 0.719 \pm 0.028 \,,&
 &R_\rho^\text{SM} = 0.532 \pm 0.011 \,,&
 &R_p^\text{SM} = 0.688 \pm 0.064 \,. \label{eq:SMprediciton}
\end{align}
These values are well consistent with that found in the literature at the $1\sigma$ level; 
see, \textit{e.g.}, Refs.~\cite{HFLAV2024winter,Iguro:2024hyk,Bernlochner:2018bfn} and \cite{Bernlochner:2021rel,Leljak:2021vte,Dutta:2015ueb,Biswas:2021cyd} for the semi-leptonic $b\to c$ and $b\to u$ decays, respectively. 
As suggested in Refs.~\cite{Tanaka:1994ay,Freytsis:2015qca}, it would also be advantageous to define the ratio
\begin{align} \label{eq:RtildeM}
 \widetilde{R}_M = {\int_{m_\tau^2}^{q^2_\text{max}} \frac{d \Gamma\left(B,\Lambda_b \rightarrow M \tau \nu\right)}{d q^2} d q^2} \Bigg/
 {\int_{m_\tau^2}^{q^2_\text{max}} \frac{d \Gamma\left(B,\Lambda_b \rightarrow M \ell \nu\right)}{d q^2} d q^2} \,, 
\end{align} 
in which the range of $q^2$ integration in the numerator is the same as in the denominator. This observable has a nice advantage for the lattice inputs, because the lattice study of the form factors has to be performed at around the zero recoil point ($q^2 \sim q^2_\text{max}=(m_{B,\Lambda_b}-m_M)^2$), and the resulting form factors extrapolated to the large recoil region ($q^2 \sim 0$) usually contain large uncertainties. Thus, the uncertainties of the theoretical predictions can be further reduced if we employ the ratio $\widetilde{R}_M$. We can apply the same description of the $R$ ratio sum rule shown above to the ratio $\widetilde{R}_M$ by the replacement $R_M \to \widetilde{R}_M$. Compared to the $R_M/R_M^\text{SM}$ case, the NP contributions to $\widetilde{R}_M/\widetilde{R}_M^\text{SM}$ do not change, while the SM predictions are obtained as 
\begin{align}
&\widetilde{R}_D^\text{SM} = 0.575 \pm 0.005\,, & 
&\widetilde{R}_{D^*}^\text{SM} = 0.339 \pm 0.003\,, &  
&\widetilde{R}_{\Lambda_c}^\text{SM} = 0.424 \pm 0.006\,, \notag \\[0.5em]
&\widetilde{R}_\pi^\text{SM} = 0.814 \pm 0.023\,, & 
&\widetilde{R}_\rho^\text{SM} = 0.605\pm 0.007\,, & 
&\widetilde{R}_p^\text{SM} = 0.728 \pm 0.038\,.
\end{align}
One can see the efficient reduction of the uncertainties compared to the numerical results given in Eq.~\eqref{eq:SMprediciton}. 
In the future, the lattice studies could provide the form factor inputs with much better precision than the LCSR evaluations, which is the reason that the ratio $\widetilde{R}_X$ can be significant for testing the LFU in these processes. 

\section{Phenomenological implications}
\label{sec:phen}

Based on our numerical results, we will discuss phenomenological implications of the $R$ ratio sum rules and make some NP investigations in light of the current experimental measurements. 

\subsection[The \texorpdfstring{$b \to c$}{b2c} mode]{\boldmath The \texorpdfstring{$b \to c$}{b2c} mode}
\label{sec:b2cstudy}

In this paper, we do not proceed with the NP investigation in the $b \to c$ decays, since it has been extensively studied in both theoretical and experimental aspects in the literature (see Refs.~\cite{Capdevila:2023yhq,London:2021lfn,Bernlochner:2021vlv,Bifani:2018zmi} for recent reviews). 
Instead, we simply employ the recent fit results for the $R_{D^{(*)}}$ measurements as well as the constraints from the $B_c$ lifetime and the collider searches, analyzed in Ref.~\cite{Iguro:2024hyk}. 

\subsubsection{Sum rule predictions and shift factor}
\label{sec:SRprediction}

The $R$ ratio sum rule introduced in Eq.~\eqref{eq:Rsumrule-preliminary} is valid in any tau-philic NP scenario described by the effective weak Hamiltonian in Eq.~\eqref{eq:Hamiltonian}. 
One of its significant points is that we can obtain a unique prediction of the baryonic counterpart from the two semi-leptonic mesonic decays, which is free from any NP implications. 
This provides a critical consistency check of the experimental measurements among the relevant decay processes, with the theoretical estimations encoded in the sum rule coefficients $b$ and $c$, as well as the shift factor $\delta_H$.
For the present case of the semi-leptonic $b \to c$ decays, assuming a negligible $\delta_{\Lambda_c}$ and taking the latest world averages of $R_{D}^\text{exp}$ and $R_{D^*}^\text{exp}$ summarized in Sec.~\ref{sec:centralSR}, we find
\begin{align}
 R_{\Lambda_c}^\text{SR} = 0.372 \pm 0.017 \big|_{R_X^\text{SM, exp}} \pm (< 0.001) \big|_{\text{SR}} \,, \quad\quad (\text{precise}) \label{eq:SRfitprecise}
\end{align}
where the first uncertainty comes from the $R_{D,D^*,\Lambda_c}^\text{SM, exp}$ inputs, 
whereas the second one from the sum rule coefficients and ``$(<0.001)$'' means that the uncertainty is less than $0.001$. 
Thus, we can see that the current form factor uncertainties are actually negligible with respect to the sum rule prediction for $R_{\Lambda_c}$. 
Such a suppression is caused by the condition $b+c=1$ of Eq.~\eqref{eq:SRset}, which leads to a perfectly negative correlation between the coefficients $b$ and $c$. 
This means that the sum rule prediction relies mainly on the central values of the theoretical inputs. 
For instance, even if we put a $100\%$ uncertainty in the sum rule coefficients by hand, it still gives a sufficiently small uncertainty of $\pm 0.001 \big|_{\text{SR}}$, which may sound incongruous. 
Just to be sure, we also take the conservative error estimation without considering the correlation between $b$ and $c$, which results in 
\begin{align}
 R_{\Lambda_c}^\text{SR} = 0.372 \pm 0.017 \big|_{R_X^\text{SM, exp}} \pm 0.008 \big|_{\text{SR}} \,. \quad\quad (\text{conservative}) 
\end{align}
This exercise demonstrates, besides trying to improve the theoretical calculations of the form factors, importance of properly considering the correlation among the form factor parameters when calculating the $R$ ratio sum rule.

It is well-known that the current experimental measurements of $R_D$ and $R_{D^*}$ have discrepancies from the corresponding SM predictions, which is usually referred to as the $R_{D^{(*)}}$ anomalies. 
A recent NP fit study by two of us in Ref.~\cite{Iguro:2024hyk} points out some NP solutions. 
For instance, a tensor NP scenario with $C_{T, \text{sol}}^{c\tau} \approx 0.02 \pm i \,0.13$ can address the anomalies, which is also allowed by the collider search~\cite{Iguro:2020keo}. 
If we take this NP scenario, it induces a non-zero shift factor for the $R$ ratio sum rule such as 
\begin{align}
 \delta_{\Lambda_c} (C_{T, \text{sol}}^{c\tau}) = -0.035 \pm 0.096 \,.  \label{eq:deltaLc}
\end{align}
The previous study made in Ref.~\cite{Fedele:2022iib}, based just on the central values, concludes that this effect is negligible. 
As one can see, however, it includes a large uncertainty for the moment and thus the $R$ ratio sum rule relation can be shifted by $\delta_{\Lambda_c} \approx -0.1$ at the $1\sigma$ level. 
This can be compared with $\delta_{\Lambda_c}^\text{SR/LHCb} = - 0.39 \pm 0.23$ as shown in Eq.~\eqref{eq:deviation_SRLHCb}, indicating a discrepancy between the sum rule fit $R_{\Lambda_c}^\text{SR}$ and the measured $R_{\Lambda_c}^\text{LHCb}$. 
It also implies that this discrepancy may be compensated with the tensor NP contribution, consistent with the NP solution to $R_{D^{(*)}}^\text{exp}$, within the current uncertainty. 
Therefore, reducing further the uncertainties from the form factor inputs is crucial to test the $R_{\Lambda_c}$ measurement in comparison with the $R_{\Lambda_c}^\text{SR}$ prediction.

\subsubsection[Relations to \texorpdfstring{$R_{J/\psi}$}{RJpsi} and \texorpdfstring{$R_{X_c}$}{RXc}]{\boldmath Relations to \texorpdfstring{$R_{J/\psi}$}{RJpsi} and \texorpdfstring{$R_{X_c}$}{RXc}}

The quark-level $b \to c l \nu$ transition also gives rise to other hadronic processes that can be measured. 
In particular, the $B_c \to J/\psi l \nu$ decays have been observed~\cite{ParticleDataGroup:2024cfk}, and the first measurement of $R_{J/\psi}$ has been reported by the LHCb~\cite{Aaij:2017tyk} and CMS~\cite{CMS:2024seh,CMSRJpsi2} collaborations. 
Indeed, it is pointed out that the general NP contribution to $R_{J/\psi}/R_{J/\psi}^\text{SM}$ and $R_{D^*}/R_{D^*}^\text{SM}$ described by Eq.~\eqref{eq:Hamiltonian} has an interesting relation~\cite{Yasmeen:2024cki} 
\begin{align}
 \frac{R_{J/\psi}}{R_{J/\psi}^\text{SM}} \simeq \frac{R_{D^*}}{R_{D^*}^\text{SM}} \,. \label{eq:RJpsi}
\end{align} 
This can also be regarded as a sum rule with $b \simeq 0$ and $c \simeq 1$. 
By taking the recent experimental measurements as introduced in Sec.~\ref{sec:intro}, we see that
\begin{align}
 \frac{R_{J/\psi}^\text{exp}}{R_{J/\psi}^\text{SM}} - \frac{R_{D^*}^\text{exp}}{R_{D^*}^\text{SM}} = 1.2 \pm 0.7 \,, 
\end{align}
where we refer to $R_{J/\psi}^\text{SM} = 0.258 \pm 0.004$ from Ref.~\cite{Harrison:2020nrv} whereas $R_{D^*}^\text{SM}$ from our result of Eq.~\eqref{eq:SMprediciton}.
Although the data of $R_{J/\psi}^\text{exp}$ still includes a large uncertainty, the relation of Eq.~\eqref{eq:RJpsi} is satisfied within $2\sigma$ for the moment. 
We do not go into further detail on this relation, but this is also a good point to be checked and will become significant once the $R_{J/\psi}$ measurement is improved. 

There is another counterpart for the exclusive semi-leptonic $b \to c$ decay, \textit{i.e.} the inclusive process $B \to X_c l \nu$. 
The NP contribution to $R_{X_c}/R_{X_c}^\text{SM}$ has been studied in Refs.~\cite{Grossman:1994ax,Goldberger:1999yh,Colangelo:2016ymy,Celis:2016azn,Mannel:2017jfk,Kamali:2018fhr,Bhattacharya:2018kig,Hu:2018veh,Kamali:2018bdp,Fael:2022wfc,ToAppearXQL}. 
Since the numerical analysis of the inclusive decay depends on the expansion order and the scheme for the bottom-quark mass $m_b$, 
evaluating the uncertainties of the numerical coefficients associated with the NP WCs is not straightforward and is beyond the scope of this paper. 
Thus, we simply refer to the numerical result obtained with the $1S$ scheme for $m_b$ in Ref.~\cite{ToAppearXQL}, which reads 
\begin{align}
 \frac{R_{X_c}}{R_{X_c}^\text{SM}} 
 &\simeq\,
 \left| 1+C_{V_L}^{c\tau} \right|^2 + \left| C_{V_R}^{c\tau} \right|^2 + 0.354 (\left| C_{S_L}^{c\tau} \right|^2 + \left| C_{S_R}^{c\tau} \right|^2) + 11.194 \left| C_{T}^{c\tau} \right|^2  \notag \\
 & \, + 0.360\, \text{Re}\left[ (1+C_{V_L}^{c\tau}) C_{S_L}^{c\tau*} + C_{V_R}^{c\tau} C_{S_R}^{c\tau*} \right] - 0.511\, \text{Re}\left[ (1+C_{V_L}^{c\tau}) C_{V_R}^{c\tau*} \right]  \notag \\[0.5em]
 & \, + 0.553\, \text{Re}\left[ C_{S_L}^{c\tau} C_{S_R}^{c\tau*} \right]  + 0.564\, \text{Re}\left[(1+C_{V_L}^{c\tau}) C_{S_R}^{c\tau*} + C_{V_R}^{c\tau} C_{S_L}^{c\tau*} \right] \notag\\[0.5em]
 & \, - 2.705\, \text{Re}\left[ (1+C_{V_L}^{c\tau}) C_{T}^{c\tau*} \right] + 1.939\, \text{Re}\left[ C_{V_R}^{c\tau} C_{T}^{c\tau*} \right]\,,
\end{align} 
where the $\mathcal{O}(1/m_b^2)$ power and the $\mathcal{O}(\alpha_s)$ perturbative corrections to both the SM and NP terms have been properly taken into account. Furthermore, the SM prediction of $R_{X_c}^\text{SM}$ including the theoretical uncertainty has been already given in the literature (see, \textit{e.g.}, Refs.~\cite{Ligeti:2014kia,Rahimi:2022vlv}), and we refer to $R_{X_c}^\text{SM}=0.220 \pm 0.001$ again from Ref.~\cite{ToAppearXQL}. 
One can find that the structure of the NP terms for $R_{X_c}/R_{X_c}^\text{SM}$ is the same as that for the baryonic decay $R_{\Lambda_c}/R_{\Lambda_c}^\text{SM}$ given by Eq.~\eqref{eq:RB}.
Therefore, it is possible to construct the $R$ ratio sum rule among $R_D$, $R_{D^*}$, and $R_{X_c}$. 
As an illustration with the central values of the inputs, we have 
\begin{align} \label{eq:RXcsumrule}
 \frac{R_{X_c}}{R^{\rm SM}_{X_c}} 
 \simeq 0.288\, \frac{R_D}{R^{\rm SM}_D} 
 + 0.712\,\frac{R_{D^\ast}}{R^{\rm SM}_{D^\ast}} 
 + \delta_{X_c}\,,
\end{align}
with 
\begin{align}
 \delta_{X_c} \simeq
 & \,0.015\,\left( |C_{S_L}^{c\tau}|^2 + |C_{S_R}^{c\tau}|^2\right) -0.003\,\textrm{Re}\left( C_{S_L}^{c\tau} C_{S_R}^{c\tau\ast} \right) 
  -1.655\,|C_T^{c\tau}|^2  \notag \\[0.3em]
 & + \textrm{Re} \left[ \left(1 +C_{V_L}^{c\tau} \right) \left\{ 0.192 C_{V_R}^{c\tau\ast} + 0.896 C_T^{c\tau\ast} \right\}\right]  
  -3.405\,\textrm{Re}\left( C_{V_R}^{c\tau} C_{T}^{c\tau\ast} \right) \notag\\[0.3em]
  &+0.043 \textrm{Re} \left[ \left(1 +C_{V_L}^{c\tau} \right) C_{S_R}^{c\tau\ast} + C_{S_L}^{c\tau} C_{V_R}^{c\tau\ast} \right] \,.
\end{align}
Then, following the same argument that gives the prediction of $R_{\Lambda_c}^\text{SR}$ from the $R$ ratio sum rule as in the previous subsection, we immediately obtain  
\begin{align}
 R_{X_c}^\text{SR} \simeq 0.247 \pm 0.008 \big|_{R_X^\text{SM, exp}} \,, 
\end{align}
where only the uncertainties from the SM predictions and the experimental measurements are taken into account, while those from the sum rule coefficients are ignored. It can be seen that the sum rule prediction is consistent with the current experimental measurement, $R_{X_c}^\text{exp}=0.228 \pm 0.039$~\cite{Belle-II:2023aih}, after taking into account the large experimental uncertainty. 
In a similar way, we also find that
\begin{align}
 \delta_{X_c} (C_{T, \text{sol}}^{c\tau}) \simeq -0.011 \,, 
\end{align}
where $C_{T, \text{sol}}^{c\tau} \approx 0.02 \pm i \,0.13$ denotes one of the NP solutions to the $R_{D^{(*)}}$ anomalies, as already introduced. 
Then, we can see that $\delta_{X_c} (C_{T, \text{sol}}^{c\tau})$ has a similar (central) value with that of Eq.~\eqref{eq:deltaLc}. 
This can be understood from the fact that the numerical formulae of 
${R_{X_c}}/{R^{\rm SM}_{X_c}}$ and ${R_{\Lambda_c}}/{R^{\rm SM}_{\Lambda_c}}$
have similar values and signs. To obtain a more conclusive result, however, we need to evaluate the uncertainties of the NP contributions in the inclusive decay, which is beyond the scope of the present work and will be explored in Ref.~\cite{ToAppearXQL}. 

As the inclusive branching fraction corresponds to a sum over that of all the possible exclusive final states, the sum over the decay rates of all these exclusive processes should saturate the inclusive rate. Specific to the $b\to c$ case, the degree of saturation can be explored by comparing the inclusive branching ratio of $B \to X_c l \nu$ with the sum over that of $B \to D^{(\ast)} l \nu$ and $B \to D^{\ast\ast} l \nu$~\cite{Freytsis:2015qca,Bernlochner:2021vlv,Rahimi:2022vlv}, where $D^{\ast\ast}$ denote the lightest orbitally excited
$D$ states. Our sum rule relation given by Eq.~\eqref{eq:RXcsumrule} could provide another complementary test of the dynamics behind these decays.

\subsubsection[Predictions of $D^*$ and $\tau$ polarizations]{\boldmath Predictions of \texorpdfstring{$D^*$}{Dst} and \texorpdfstring{$\tau$}{tau} polarizations}

The $D^*$ and $\tau$ longitudinal polarizations $F_L^{D^*}$ and $P_\tau^{D^*}$ in $B \to D^* \tau\nu$ are good observables to test the NP implication observed from $R_X$, which will be measured more precisely at the ongoing Belle~II experiment. 
See Ref.~\cite{Tanaka:2012nw} and Ref.~\cite{Iguro:2024hyk} for their definitions and status, respectively. 
From our evaluation, we show the semi-numerical formulae for them in the same way such that  
\begin{align}
 \frac{F_{L}^{D^\ast}}{F_{L,{\rm SM}}^{D^\ast}} \frac{R_{D^\ast}}{R_{D^\ast}^{\rm SM}} 
 = 
 & \left|1+C_{V_L}^{c\tau} - C_{V_R}^{c\tau}\right|^2 + (9.554 \pm 6.755) \left|C_T^{c\tau}\right|^2 + (0.099 \pm 0.003) \left|C_{S_L}^{c\tau} - C_{S_R}^{c\tau}\right|^2 \notag \\
 & + (-0.263\pm0.007) \,\text{Re}\! \left[\left(1+C_{V_L}^{c\tau}-C_{V_R}^{c\tau}\right)\left(C_{S_L}^{c\tau}-C_{S_R}^{c\tau}\right)^\ast\right] \notag\\[1em]
 & + (-5.116\pm1.762) \,\text{Re}\!\left[\left(1+C_{V_L}^{c\tau} - C_{V_R}^{c\tau}\right) C_T^{c\tau\ast}\right], \label{eq:FL_NP_SM}
\end{align}
and
\begin{align}
 \frac{P_\tau^{D^\ast}}{P_{\tau,{\rm SM}}^{D^\ast}} \frac{R_{D^\ast}}{R_{D^\ast}^{\rm SM}} 
 = 
 & \left|1+C_{V_L}^{c\tau}\right|^2 + \left|C_{V_R}^{c\tau}\right|^2 + (-1.837\pm1.544) \left|C_T^{c\tau}\right|^2  \notag \\
 & + (-0.082\pm0.004) \left| C_{S_L}^{c\tau} - C_{S_R}^{c\tau}\right|^2 + (-1.755\pm0.018) \,\text{Re}\!\left[\left(1+C_{V_L}^{c\tau}\right) C_{V_R}^{c\tau}\right]  \notag\\[1em]
 & + (0.218\pm0.011) \,\text{Re}\!\left[\left(1+C_{V_L}^{c\tau}-C_{V_R}^{c\tau}\right) \left(C_{S_L}^{c\tau} - C_{S_R}^{c\tau}\right)^\ast\right] \notag \\[1em]
 & + (-3.521\pm1.105) \,\text{Re}\!\left[\left(1+C_{V_L}^{c\tau}\right) C_T^{c\tau\ast}\right] \notag \\[1em]
 & + (4.569\pm1.250) \,\text{Re}\!\left[C_{V_R}^{c\tau} C_T^{c\tau\ast}\right] \,, \label{eq:Ptau_NP_SM}
\end{align}
where $R_{D^*}/R_{D^*}^\text{SM}$ is taken from the formula with our result of Eqs.~\eqref{eq:RDstform}. 
The SM predictions are obtained as 
\begin{align}
 F_{L,{\rm SM}}^{D^\ast} = 0.430 \pm 0.007 \,, \quad P_{\tau,{\rm SM}}^{D^\ast} = -0.518 \pm 0.006 \,.
\end{align}

\subsection[The \texorpdfstring{$b \to u$}{b2u} mode]{\boldmath The \texorpdfstring{$b \to u$}{b2u} mode}
\label{sec:b2ustudy}

Regarding the semi-leptonic $b \to u$ decays, we start with the NP investigations. At first, it is noted that parts of the NP WCs $C_i^{u\tau}$ relevant to the $b \to u\tau\nu$ transition can be constrained from the measured branching ratio of the purely leptonic decay, $\mathcal{B}(B \to \tau \nu)^\text{exp} = (1.09 \pm 0.24)\times10^{-4}$~\cite{ParticleDataGroup:2024cfk}. 
The theoretical prediction including both the SM and NP contributions can be written as 
\begin{align}
 \mathcal{B}(B \to \tau \nu) =
 \frac{\tau_{B^{-}} G_F^2\left|V_{u b}\right|^2 f_{B^{-}}^2}{8 \pi} m_{B^-} m_\tau^2\left(1-\frac{m_\tau^2}{m_{B^-}^2}\right)^2 \left|1+r^{u\tau}_{\mathrm{NP}}\right|^2\,,
\end{align} 
where $\tau_{B^{-}}$ and $f_{B^{-}}$ are the lifetime and decay constant of the charged $B$ meson, and $r^{u\tau}_{\mathrm{NP}}$ represents the NP effect given by
\begin{align}
r^{u\tau}_{\mathrm{NP}}=C_{V_L}^{u\tau}-C_{V_R}^{u\tau}+\frac{m_{B^-}^2}{m_b m_\tau}\left(C_{S_R}^{u\tau}-C_{S_L}^{u\tau}\right) \,. 
\end{align} 
Note that the tensor-type NP operator does not contribute to this decay at all. 
Taking the experimental value of $\mathcal{B}(B \to \tau \nu)^\text{exp} = (1.09 \pm 0.24)\times10^{-4}$~\cite{ParticleDataGroup:2024cfk}, together 
with the theoretical inputs of $\tau_{B^{-}}=(1.638 \pm 0.004)\,\text{ps}$, $f_{B^{-}} = (190.0 \pm 1.3)\,\text{MeV}$ and $|V_{ub}|=(3.82 \pm 0.20) \times 10^{-3}$ (the average over the direct determinations of $|V_{ub}|$ from exclusive and inclusive $B$ decays)~\cite{ParticleDataGroup:2024cfk}, we have 
\begin{align}
 \left|1+r_{\mathrm{NP}}^{u\tau}\right| = 1.08 \pm 0.13 \,, \label{eq:b2u_pure_lep}
\end{align} 
which immediately leads to\footnote{Here we keep only those solutions with $|C_{i}^{u\tau}|<1$, as generally expected from the perturbativity requirement.}
\begin{align}
 &C_{V_L}^{u\tau} \in [-0.05,0.21]\,,&  
 &C_{V_R}^{u\tau} \in [-0.21,0.05] \,, \label{eq:CVallowed}  \\[0.5em]
 &C_{S_L}^{u\tau} \in [-0.06,0.01] \cup [0.52,0.59]\,,& 
 &C_{S_R}^{u\tau} \in [-0.59,-0.52] \cup [-0.01,0.06]\,, \label{eq:CSallowed}
\end{align}
at the $1\sigma$ level by assuming the presence of a single and real WC. 
Although the $R_\pi$ measurement is currently available~\cite{Belle:2015qal} as well, it gives mild bounds on $C_i^{u\tau}$ at present due to the large experimental uncertainty. Thus, we do not combine it in the following discussions but check the consistency below.

\subsubsection[Correlations among \texorpdfstring{$R_X$}{RX} in the presence of NP]{\boldmath Correlations among \texorpdfstring{$R_X$}{RX} in the presence of NP}

\begin{figure}[t]
\begin{center}
 \includegraphics[viewport=0 0 1000 543, width=0.99\textwidth]{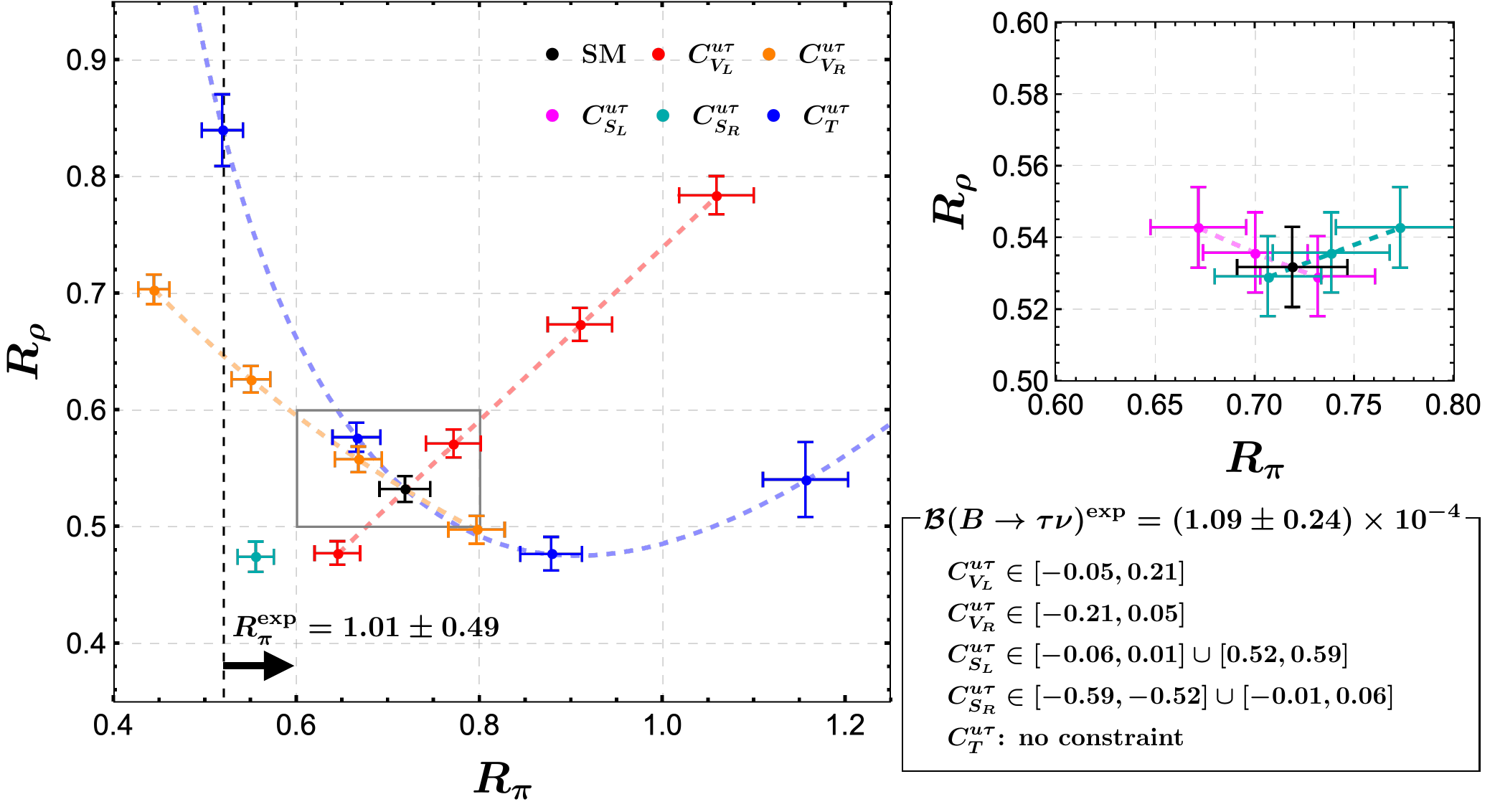} 
 \caption{
 The correlation between $R_\pi$ and $R_\rho$ in the presence of a single NP contribution indicated on the plot. 
 The SM prediction together with its uncertainty is indicated by the black cross bar, while the NP benchmark points by the cross bars in color, which give the uncertainties as illustrations. See the main text for the other descriptions in detail. 
 \label{fig:Rpi_Rrho}
 }
\end{center} 
\end{figure}

\begin{figure}[b]
\begin{center}
 \includegraphics[viewport=0 0 1000 543, width=0.99\textwidth]{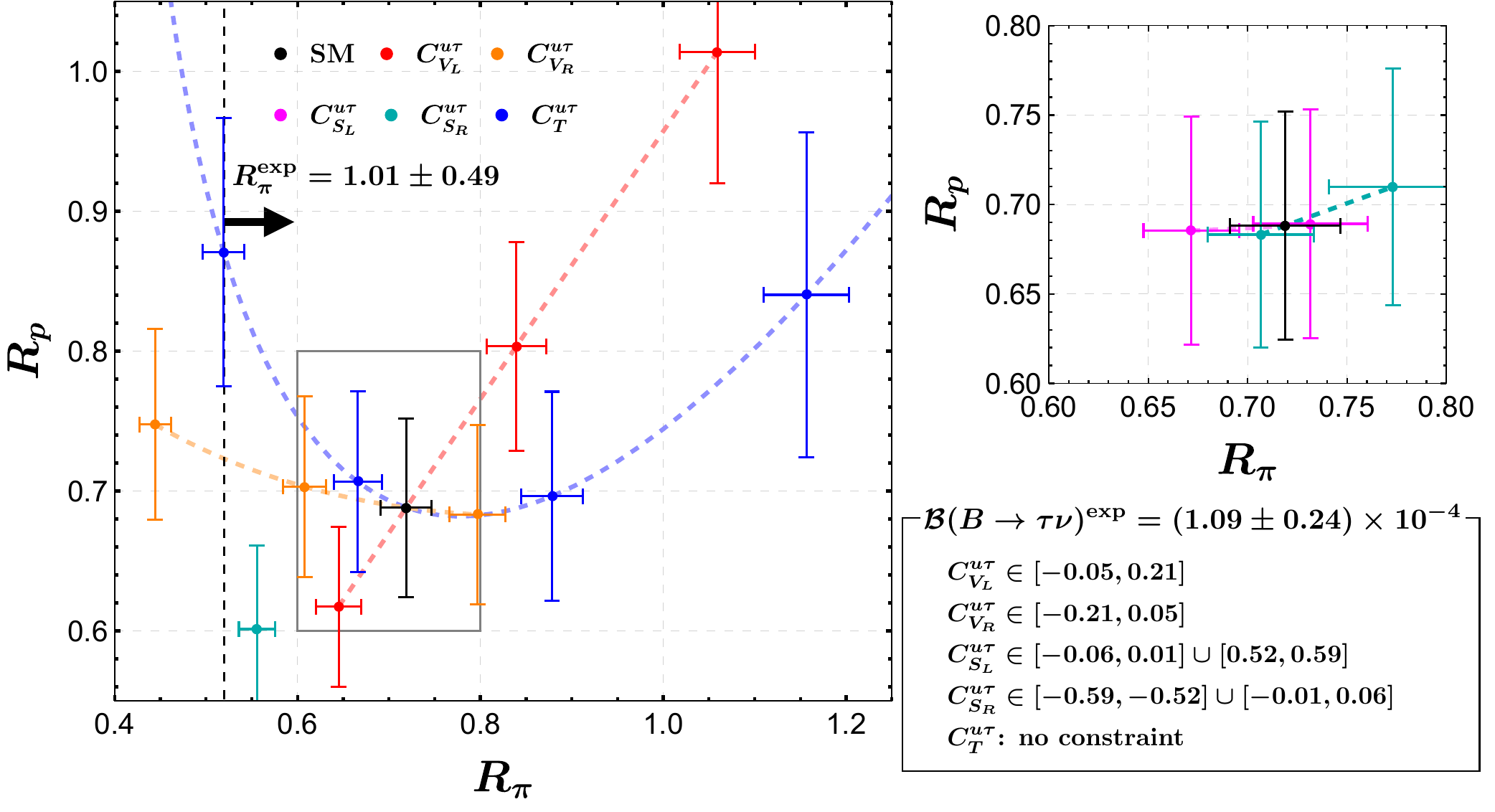}
 \caption{
 The correlation between $R_\pi$ and $R_p$ in the presence of a single NP contribution.
 The other captions are the same as in Fig.~\ref{fig:Rpi_Rrho}. 
 \label{fig:Rpi_Rp}
 }
\end{center}
\end{figure}

In Fig.~\ref{fig:Rpi_Rrho}, we show the correlation between $R_\pi$ and $R_\rho$ in the presence of a single NP contribution within its allowed range as specified in Eqs.~\eqref{eq:CVallowed} and \eqref{eq:CSallowed}. Regarding the tensor contribution, we just assume a real $C_{T}^{u\tau}$.
The dashed curves represent the trajectories for the central values of $R_\pi$ and $R_\rho$, while the cross bars indicate the uncertainties at the benchmark points as illustrations. 
The black cross bar shows the SM prediction together with its uncertainty. 
The $1\sigma$ lower value of $R_\pi^\text{exp}$ is put as the black dashed line, and the upper one is out of the plot range. 
The effects of the scalar contributions in the vicinity of $C_{S_{L,R}}^{u\tau}\approx0$ (corresponding to the gray rectangle in the left panel) are separately shown in the upper-right panel for better visibility.  
The $S_R$ ($S_L$) effect in the other allowed range of Eq.~\eqref{eq:CSallowed} is in the very limited range shown as the cross bar (out of the plot range) in the left panel. 
Furthermore, we show in Fig.~\ref{fig:Rpi_Rp} the correlation between $R_\pi$ and $R_p$, where the prescriptions of the lines, cross bars, and colors are the same as in Fig.~\ref{fig:Rpi_Rrho}. 
From these figures, we can see that (i) the NP effects projected onto the $R_\pi$--$R_\rho$ and $R_\pi$--$R_p$ planes look similar, (ii) the NP contributions to $R_\pi$ and $R_\rho$ are well distinctive from the corresponding SM predictions with the present accuracy of the form factor inputs, and (iii) $R_p$ is less predictive due to the large uncertainty from the form factor inputs for the moment. One can also realize that the $R$ ratio sum rule for the $b\to u$ case is not explicitly readable from these figures. 

Apart from the less predictive $R_p$, the NP constraints as given by Eqs.~\eqref{eq:CVallowed} and \eqref{eq:CSallowed} also lead to the predictions on $R_\pi$ and $R_\rho$ in the following ranges:  
\begin{align}
 V_L~:\qquad & 0.62 \lesssim R_\pi \lesssim 1.10 \,, \quad 0.47 \lesssim R_\rho \lesssim 0.80 \,, \\[0.5em]
 V_R~:\qquad & 0.43 \lesssim R_\pi \lesssim 0.82 \,, \quad 0.72 \gtrsim R_\rho \gtrsim 0.49 \,, \\[0.5em]
 S_L~:\qquad & 0.65 \lesssim R_\pi \lesssim 0.76 \,, \quad 0.55 \gtrsim R_\rho \gtrsim 0.52 \,, \\[0.3em]
 \, &  \!\!(1.38 \lesssim R_\pi \lesssim 1.73 \,, \quad 0.49 \gtrsim R_\rho \gtrsim 0.46) \,, \\[0.5em]
 S_R~:\qquad & 0.68 \lesssim R_\pi \lesssim 0.81 \,, \quad 0.52 \lesssim R_\rho \lesssim 0.55 \,, \\[0.3em]
 \, &  0.59 \gtrsim R_\pi \gtrsim 0.53 \,, \quad 0.46 \lesssim R_\rho \lesssim 0.49 \,, 
\end{align}
where the correlation is aligned with the direction of the inequality signs, and the range in the parenthesis means that this region is out of the plot in Fig.~\ref{fig:Rpi_Rrho}.  
Regarding the NP prediction from the tensor contribution, we need the $R_X$ measurement to study its effect. 
The bound from the present $R_\pi^\text{exp}$ measurement~\cite{Belle:2015qal} gives a very loose allowed range of $C_T^{u\tau} \in [-0.14, 0.30]$,\footnote{
Another range of $C_T^{u\tau} \in [-0.96, -0.52]$ is also allowed by $R_\pi^\text{exp}$, but excluded by the collider search, which gives $|C_{T}^{u\tau}|<0.42$; see Ref.~\cite{Iguro:2020keo} for further details. 
Note that the allowed ranges for the other NP scenarios shown in the main text satisfy the collider bound. 
} and it leads to 
\begin{align}
 T~:\qquad  0.50 \lesssim R_\pi \lesssim 1.57 \,, \quad 0.46 \lesssim R_\rho \lesssim 0.87 \,. 
\end{align}

\subsubsection{Sum rule predictions and shift factor}
\label{sec:sumruleandshiftfactor}

Since the $b \to u$ case has two missing measurements of $B \to \rho\tau\nu$ and $\Lambda_b \to p \tau\nu$, the sum rule prediction is only available for the combination of $R_\rho$ and $R_p$. 
This can be represented with a vanishing $\delta_p$ as 
\begin{equation}
 R_{p}^\text{SR} - (0.926 \pm 0.088 \big|_{R_X^\text{SM, exp}} \mp 0.048 \big|_\text{SR}) R_{\rho}^\text{SR} = 0.274 \pm 0.136 \big|_{R_X^\text{SM, exp}} \pm 0.036 \big|_\text{SR} \,. 
\end{equation} 
Due to the currently large experimental uncertainty of $R_\pi^\text{exp}$, the sum rule is less predictive at present. 
Within the SM, the branching ratios of the two missing decay modes, $\bar B^0 \to \rho^+\tau^-\nu$ and $\Lambda_b^0 \to p \tau^-\nu$, are both predicted to be of $\mathcal{O}(10^{-4})$~\cite{Bernlochner:2021rel,Dutta:2015ueb}.
Explicitly, with the updated form factor inputs, we obtain the SM predictions
\begin{align}
\mathcal{B}(\bar B^0 \to \rho^+\tau^-\nu)_\text{SM} &= \left(2.12\pm 0.43\right) \times10^{-4}\,, \\[0.3em]
\mathcal{B}(\Lambda_b^0 \to p \tau^-\nu)_\text{SM} &= \left(3.80 \pm 0.44\right) \times10^{-4}\,.
\end{align} 
They should be measurable at the ongoing LHCb run-3~\cite{LHCb:2018roe,LHCb:2022ine} and/or Belle~II~\cite{Belle-II:2018jsg,Belle-II:2022cgf} experiments. In addition, when $R_{p}$ and $R_{\rho}$ are measured for the first time, the ratio $R_\pi$ is also expected to be measured more precisely. Therefore, we believe that our above study offers another good motivation to perform precise measurements of these decays at the future experiments. 

\begin{figure}[t]
\begin{center}
 \includegraphics[viewport=0 0 600 532, width=24em]{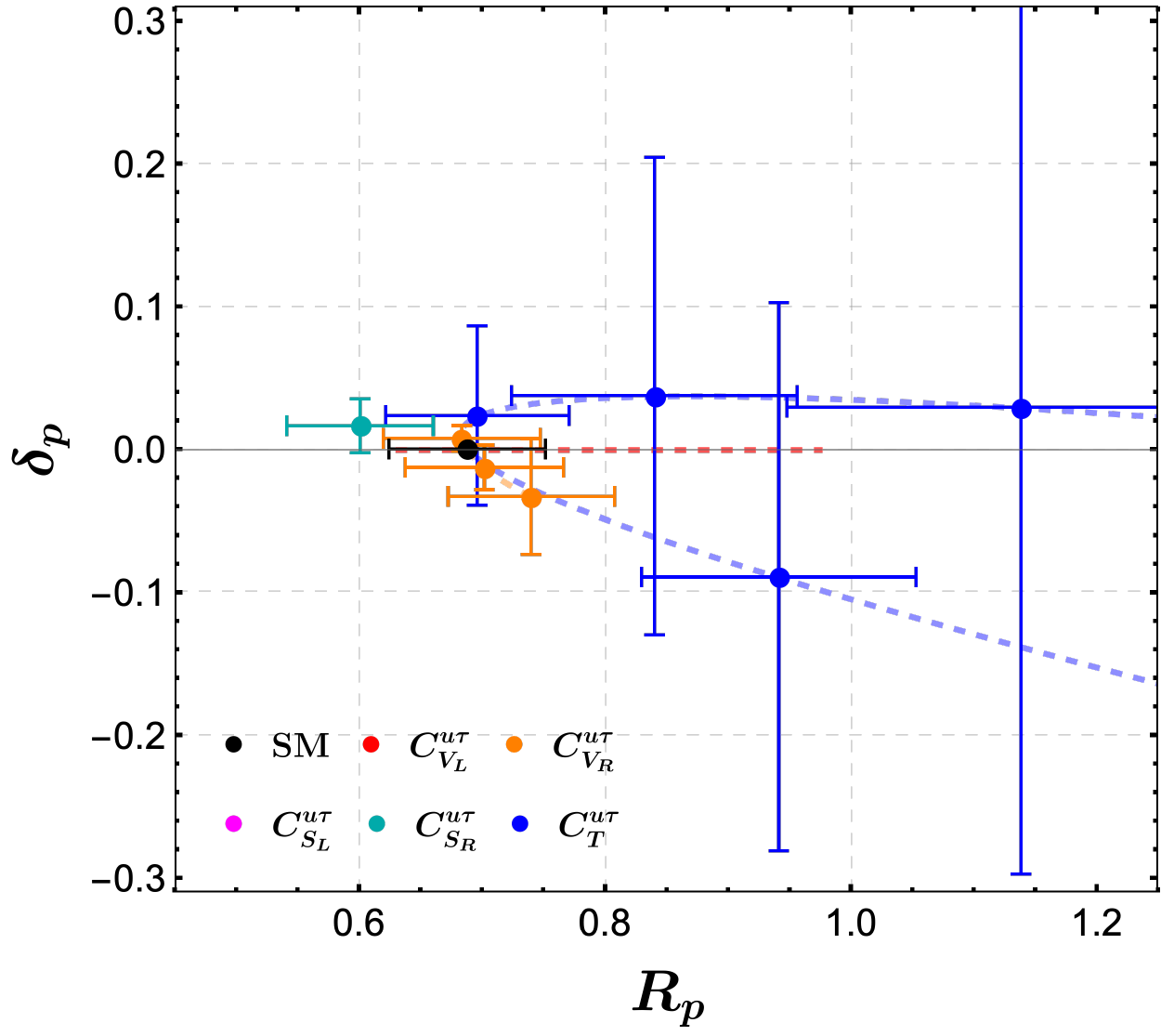}
 \caption{
 The relation between $R_p$ and $\delta_p$ in the presence of a single NP contribution. 
 The $V_L$ scenario gives $\delta_p = 0$ by definition. 
 The $S_{L,R}$ scenarios with $C_{S_L}^{u\tau} \in [-0.06,0.01]$, $C_{S_R}^{u\tau} \in [-0.01,0.06]$ give $\delta_p \approx 0$, and thus they are both omitted in the plot. 
 The color prescription for the figure is the same as in Fig.~\ref{fig:Rpi_Rrho}. \label{fig:Rp_Delta}
}
\end{center}
\end{figure}

In Fig.~\ref{fig:Rp_Delta}, we show the correlation between $R_p$ and $\delta_p$ in the presence of a single NP contribution, where the prescription is again the same as in Fig.~\ref{fig:Rpi_Rrho}. Note that the $V_L$ scenario gives $\delta_p =0$ by definition. The $S_{L,R}$ scenarios with $C_{S_L}^{u\tau} \in [-0.06,0.01]$ and $C_{S_R}^{u\tau} \in [-0.01,0.06]$ are also omitted in the figure, since they give a negligible value of $\delta_p$ even with the form factor uncertainties taken into account. 
We can see from the figure that the $V_R$ ($S_R$) scenario could have $\delta_p \approx -0.07$ ($+0.03$) at most, while the $T$ scenario could reach $\delta_p \approx \pm0.3$ due to the large uncertainty. 
If a future measurement of $R_p$ has a different value from the corresponding SM prediction, we can then check how large the shift factor $\delta_p$ could be from this figure for each NP scenario. 
As an illustration, we take the $T$ scenario as an example and represent the numerical values of the shift factor $\delta_{p}$ at some representative points of $C_{T}^{u\tau}$: 
\begin{align} 
 & \delta_{p} (C_{T}^{u\tau}=-0.14) = -0.07 \pm 0.11 \,, \\[0.5em]
 & \delta_{p} (C_{T}^{u\tau}=+0.30) = 0.03 \pm 0.25 \,, \\[0.5em]
 & \delta_{p} (C_{T}^{u\tau}=-0.41) = -0.31 \pm 0.53 \,, 
\end{align}
where the first two correspond to the $1\sigma$ allowed range of $C_{T}^{u\tau}$ obtained from the measured $R_\pi^\text{exp}$, while the last one to the upper bound set by the collider search~\cite{Iguro:2020keo}, as mentioned in the last subsection. 
Concerning the scale of the branching ratio and the future statistics of the measurement of this baryonic process, we expect that a shift factor of $|\delta_p| \approx O(0.1)$ can be tested. 
Recalling that the $\Lambda_b \to p$ tensor form factor inputs in our analysis are only preliminary~\cite{StefanComment}, a precise evaluation of the tensor form factors, especially with the aid of the next-generation lattice QCD techniques~\cite{Meinel:2023wyg}, is also necessary. 

\subsection{Combined study: SMEFT explanation}
\label{sec:SMEFT_analysis}

Finally, we perform a combined study of the $b \to c l \nu$ and $b \to u l \nu$ decays in the framework of SMEFT~\cite{Buchmuller:1985jz,Grzadkowski:2010es,Brivio:2017vri,Isidori:2023pyp} with specific flavor symmetries~\cite{Faroughy:2020ina,Greljo:2022cah}. We begin with the so-called minimal flavor violation (MFV) hypothesis~\cite{Chivukula:1987py,DAmbrosio:2002vsn,Cirigliano:2005ck} and discuss its validity and possible extensions.

\subsubsection[{Brief description of SMEFT and \texorpdfstring{$U(3)^5$}{U(3)5} symmetry}]{\boldmath Brief description of SMEFT and \texorpdfstring{$U(3)^5$}{U(3)5} symmetry}

In the case where the NP particles are much heavier than the electroweak scale and the electroweak symmetry breaking is realized linearly, the SMEFT is known to provide a model-independent framework to systematically study such kinds of NP effects, which are encoded in the short-distance WCs of a tower of higher-dimensional operators invariant under the SM gauge group. Following the same conventions as in Ref.~\cite{Grzadkowski:2010es}, we can write the SMEFT Lagrangian as
\begin{equation}
\mathcal{L}_{\rm SMEFT} = \mathcal{L}_{\rm SM} + \frac{1}{\Lambda} \sum_i c_i^{(5)} \mathcal{Q}_i^{(5)} + \frac{1}{\Lambda^2} \sum_i c_i^{(6)} \mathcal{Q}_i^{(6)} + \cdots, \label{eq:SMEFT_Lag}
\end{equation}
where $\mathcal{L}_{\rm SM}$ is the renormalizable SM Lagrangian before the electroweak spontaneous symmetry breaking, 
whereas $\mathcal{Q}_i^{(n)}$ are the dimension-$n$ operators induced by integrating out the heavy new particles and constructed solely in terms of the SM fields, with $c_i^{(n)}$ the corresponding dimensionless WCs normalized by the NP scale $\Lambda$. The ellipsis represents the terms with dimension higher than six. 

Below the electroweak scale, the heavy SM particles of the top quark and the $W^\pm$, $Z$, Higgs bosons are all decoupled. 
The resulting theory is then described by the low-energy effective field theory (LEFT)~\cite{Jenkins:2017jig,Aebischer:2015fzz} invariant under the strong and electromagnetic gauge groups, which is our starting point of consideration as in Eq.~\eqref{eq:Hamiltonian} for the semi-leptonic $b \to q l \nu$ transitions. The WCs $C_i^{ql}$ of the LEFT operators in Eq.~\eqref{eq:operator} encode all the physics related to the heavy SM and NP degrees of freedom. In principle, these low-energy WCs are independent of $c_i^{(n)}$ in Eq.~\eqref{eq:SMEFT_Lag}. However, if we start with the SMEFT, we can fix the former in terms of the latter by performing a matching between these two effective theories at the electroweak scale~\cite{Jenkins:2017jig,Dekens:2019ept}. This is our motivation to discuss the SMEFT framework. In practice, the WCs $C_i^{ql}$ are usually evaluated at a low-energy scale $\mu_\text{low}$ characteristic of the process considered (\textit{e.g.}, $\mu_\text{low} \simeq m_b$ for $b \to q l \nu$ decays) to avoid additional large corrections to the matrix elements of the LEFT operators. The connection between the electroweak and the low-energy scale in LEFT is realized by the renormalization group~\cite{Jenkins:2017dyc,Naterop:2023dek}. Such a standard procedure will be followed in our study. 

In a bottom-up approach of the SMEFT, where the underlying ultraviolet-complete NP model is unknown, the flavor structure of the operators $\mathcal{Q}_i^{(n)}$ in Eq.~\eqref{eq:SMEFT_Lag} is generally not specified. In this case, the corresponding WCs $c_i^{(n)}$ will be flavor dependent, and must be taken as independent free parameters for different flavors. This causes a significant increase in the number of independent terms in the SMEFT Lagrangian, when all the possible flavors are taken into account. The flavor structures of the SMEFT operators can be probed on the basis of the flavor symmetries satisfied by the SM gauge sector. In particular, from the known structure of the SM Yukawa couplings indicated by the observed hierarchies among the SM fermion masses and mixings, we have already known that flavor is highly non-generic. 
It is therefore natural to employ specific hypotheses about the flavor symmetries on the whole SMEFT, such that the number of independent parameters in $c_i^{(n)}$ can be significantly reduced (see Refs.~\cite{Brivio:2017vri,Isidori:2023pyp} for reviews). In this paper, we consider first the $U(3)^5$ symmetry~\cite{Chivukula:1987py,Gerard:1982mm}, which is the maximal flavor symmetry compatible with the SM gauge group. It also allows us to implement the most restrictive hypothesis of MFV that can be utilized to suppress non-standard contributions to the various flavor violating observables~\cite{Isidori:2023pyp,Faroughy:2020ina}. 
To this end, let us start with the $U(3)^5$ global symmetry transformation of the SM fermions~\cite{Chivukula:1987py,Gerard:1982mm}
\begin{equation} \label{eq:U(3)_symmetry}
 \mathcal{G}_F \equiv U(3)^5 = U(3)_Q \otimes U(3)_U \otimes U(3)_D \otimes U(3)_L \otimes U(3)_E = SU(3)^5 \otimes U(1)^5\,,
\end{equation}
where the lepton and quark $SU(2)_L$ doublets are written as $L$ and $Q$ respectively, while the right-handed $SU(2)_L$ singlets are denoted by $U$ (up-type quark), $D$ (down-type quark), and $E$ (charged lepton). 
Here, each $SU(3)$ subgroup acts on the flavor indices of the corresponding fermion field characterized by the subscript. Regarding the $U(1)$ transformations, each fundamental/anti-fundamental representation has a charge $\pm 1$ under the associated Abelian subgroup.

\subsubsection[Application to the semi-leptonic \texorpdfstring{$b\to ql\nu$}{b2qlnu} processes]{\boldmath Application to the semi-leptonic $b\to ql\nu$ processes}

Here, let us apply the above formalism to the case of semi-leptonic $b\to ql\nu$ transitions. 
We assume that the NP preserves both the baryon and lepton numbers, which is the present case and thus the dimension-five operators $\mathcal{Q}_i^{(5)}$ will be irrelevant. 
Then, the most relevant dimension-six terms for $b\to ql\nu$ can be written as~\cite{Grzadkowski:2010es,Aebischer:2015fzz} 
\begin{align}
 \sum_i c_i^{(6)} \mathcal{Q}_i^{(6)} \biggr|_{b \to ql\nu} = 
 &~ c_{H_\ell}^{ij} \left(H^\dagger i\overleftrightarrow{D}_\mu^I H\right) \left(\Bar{L}^i \gamma^\mu \tau^I L^j\right) 
 + c_{H_q}^{mn} \left(H^\dagger i\overleftrightarrow{D}_\mu^I H\right) \left(\Bar{Q}^m \gamma^\mu \tau^I Q^n\right) \notag \\
 & + c_{V}^{mnij} \left(\Bar{Q}^m \gamma^\mu \tau^I Q^n\right) \left(\Bar{L}^i \gamma_\mu \tau^I L^j\right) 
 + \bigg\{ {c}_{H_{\tilde q}}^{mn} \left(\tilde{H}^\dagger i D_\mu H\right) \left(\Bar{U}^m \gamma^\mu D^n\right) \notag \\[0.7em]
 & + c_{S_d}^{mnij} \left(\Bar{L}^i E^j\right) \left(\Bar{D}^m Q^n\right) + c_{S_u}^{mnij} \left(\Bar{L}^{a,i} E^j\right) \epsilon_{ab} \left(\Bar{Q}^{b,m} U^n\right) \notag \\[0.7em]
 & + c_{T}^{mnij} \left(\Bar{L}^{a,i} \sigma_{\mu\nu} E^j\right) \epsilon_{ab} \left(\Bar{Q}^{b,m} \sigma^{\mu\nu} U^n\right) + \text{h.c.} \bigg\}\,, \label{eq:SMEFT_dim_6}
\end{align}
where $m,n$ and $i,j$ are the quark and lepton flavor indices respectively, $\epsilon_{ab}$ is the anti-symmetric tensor for the $SU(2)_L$ doublet indices $a,b$, and $H$ is the $SU(2)_L$ Higgs doublet with 
\begin{align}
 H^\dagger i\overleftrightarrow{D}_\mu^I H \equiv iH^\dagger \left(\tau^I D_\mu - \overleftarrow{D}_\mu \tau^I\right) H \,, \quad\quad \tilde{H} \equiv i\tau^2 H^\ast \,. 
\end{align}
We do not consider the operators with dimension $n>6$, because their effects will be further suppressed by the NP scale $\Lambda$. 
It should be noted that the operators associated with $c_{H_\ell}$, $c_{H_q}$, and $c_{H_{\tilde q}}$ contribute to the $b\to ql\nu$ transitions only by modifying the left-handed $W$ couplings with fermions and the right-handed $W$ coupling with quarks, respectively. The first two operators can also modify the $Z$ couplings with the left-handed fermions. However, these modified couplings undergo severe constraints from other processes~\cite{Greljo:2023bab}. For example, global and/or individual fits of the SMEFT WCs to the available experimental data, especially to the electroweak precision observables, have demonstrated that a rough bound of $|{c}_{H_{\ell,q, \tilde q}}^{ij,mn}|/\Lambda^2 < \mathcal{O}\left(10^{-2}\right) \,\text{TeV}^{-2}$ is valid for any lepton and quark flavors~\cite{Ellis:2018gqa,Dawson:2019clf,Efrati:2015eaa}. This implies that these terms do not give rise to a sizable LFU violating contribution to the $b\to ql\nu$ transitions of our interest, and we will therefore neglect them in the following discussion. 

To establish possible correlations between the $b \to c l \nu$ and $b \to u l \nu$ decays within the SMEFT, we now resort to the MFV hypothesis~\cite{Chivukula:1987py,DAmbrosio:2002vsn}. According to the description presented in App.~\ref{app:U3andU2}, the MFV hypothesis in the quark sector allows us to write the flavor structure of the WC $c_V$, for instance, as 
\begin{align}
 c_V^{mnij} = \delta_{mn} c_{V\ell}^{ij} \,,
 \label{eq:cVlepton}
\end{align}
where the lepton sector is separately represented as $c_{V\ell}^{ij}$ for later convenience. Applying the MFV hypothesis to the lepton sector and requiring the generation of the desired LFU violation between the first two lepton generations and the third one, we have the form of flavor structure
\begin{align}
 c_{V\ell}^{ij} = (Y_EY_E^\dag)_{ij} c_V^0 \,, 
\end{align}
where $Y_E$ is the SM Yukawa matrix for lepton, and $c_V^0$ denotes a flavor universal coupling. Such a structure can naturally generate a $\tau$-philic interaction. A detailed argument to get the above form, together with its comparison with that of a flavor-changing neutral current (FCNC), can be found in App.~\ref{app:U3andU2}. 
By adopting the same argument to the other terms in Eq.~\eqref{eq:SMEFT_dim_6}, we find the following MFV-based flavor structures of the WCs:
\begin{align}
 c_{S_d}^{mnij} &= (Y_D^{\dagger})_{mn} (Y_E)_{ij} c_{S_d}^0 \,,& &c_{S_u}^{mnij} = (Y_U)_{mn} (Y_E)_{ij} c_{S_u}^0 \,, \\[0.5em]
 c_{T}^{mnij} &= (Y_U)_{mn} (Y_E)_{ij} c_{T}^0 \,,
\end{align}
where $Y_U$ and $Y_D$ are the SM Yukawa matrices for up- and down-type quarks, whereas $c_{S_d}^0$, $c_{S_u}^0$ and $c_{T}^0$ are flavor universal as well as $c_V^0$.
Then, matching the SMEFT Lagrangian of Eq.~\eqref{eq:SMEFT_dim_6} to the LEFT one of Eq.~\eqref{eq:operator} at the tree level, we obtain the following relations between the WCs of these two effective theories:  
\begin{align}
 &C_{V_L}^{q\tau} = -\frac{2m_\tau^2}{\Lambda^2} c_{V}^{0}\,,&  
 &C_{V_R}^{q\tau} = 0\,, \\[0.5em]
 &C_{S_L}^{q\tau} = - \frac{m_\tau m_q}{\Lambda^2} c_{S_u}^{0\ast} \,,&
 &C_{S_R}^{q\tau} = - \frac{m_\tau m_b}{\Lambda^2} c_{S_d}^{0\ast} \,, \\[0.5em]
 &C_T^{q\tau} = - \frac{m_\tau m_q}{\Lambda^2} c_{T}^{0{\ast}} \,,
\end{align}
for $q=u,c$ at the electroweak scale. 
Let us remind the normalization factor of $2\sqrt2G_FV_{qb}$ in Eq.~\eqref{eq:operator} when getting the above matching relations. 
It should also be noted that the dimension-six terms in SMEFT have no LFU-violating contribution to the right-handed WC $C_{V_R}^{ql}$ in LEFT, and the minimum dimension of the operators to generate such kinds of contributions is found to be $n=8$, which implies that their effects are further suppressed by the NP scale $\Lambda$. 
See also footnote~\ref{foonoteonCVR}.
Finally, we have the following exact relations between the $b\to c$ and $b\to u$ sectors:
\begin{align}
 C_{V_L}^{u\tau} = C_{V_L}^{c\tau}\,, \quad
 C_{S_R}^{u\tau} = C_{S_R}^{c\tau}\,, \quad 
 C_{S_L}^{u\tau} = \frac{m_u}{m_c}C_{S_L}^{c\tau}\approx 0 \,, \quad
 C_{T}^{u\tau} = \frac{m_u}{m_c}C_{T}^{c\tau}\approx 0 \,. 
 \label{eq:CXuc}
\end{align}
Thus, only the $V_L$ and $S_R$ types of NP are of importance for the $b \to u\tau\nu$ processes in the framework of SMEFT with MFV. 
To get sizable contributions of our interest, however, the parameters $c_{V}^{0}$ and $c_{S_d}^0$ are required to be large. 
This is due to the fact that the insertion of $Y_EY_E^\dag$ introduces the large suppression of $m_\tau^2/v^2 \approx 5 \times 10^{-5}$. 
For instance, the best-fit point from the $R_{D^{(*)}}$ measurements for the $V_L$-type NP is $C_{V_L}^{c\tau} \approx 0.079$~\cite{Iguro:2024hyk}, which leads to $c_{V}^{0}/\Lambda^2 \approx -1.25 \times 10^4\,\text{TeV}^{-2}$. 
Since the parameter $c_V^0$ itself is not physical, such a large number is not an issue by the way. 

\subsubsection[Third generation-philic interaction and \texorpdfstring{$U(2)^5$}{U(2)5} symmetry]{\boldmath Third generation-philic interaction and \texorpdfstring{$U(2)^5$}{U(2)5} symmetry}

To be more general, Eq.~\eqref{eq:CXuc} can be realized by taking the MFV hypothesis only in the quark sector and simply assuming a $\tau$-philic interaction in the lepton sector. 
For instance, it is sufficient to just consider $c_{V\ell}^{\tau\tau}$ defined in Eq.~\eqref{eq:cVlepton} as a free parameter while assuming that the other components are all zero, 
with which we obtain $c_{V\ell}^{\tau\tau}/\Lambda^2 \approx -1.31\,\text{TeV}^{-2}$ from the combined fit analysis to the $R_{D^{(*)}}$ measurements~\cite{Iguro:2024hyk}. 
For later purpose, we also present its $2\sigma$ allowed range
\begin{align}
 -1.83\,\text{TeV}^{-2} < c_{V\ell}^{\tau\tau}/\Lambda^2 < -0.78\,\text{TeV}^{-2} \,. \label{eq:RDsolCVL}
\end{align} 
In any case, the MFV-based parameterization for the lepton sector is not indispensable for our analysis, and hence we may use $c_{V\ell}^{\tau\tau}$ in the following discussion. 

As explained around Eq.~\eqref{eq:FCNCcase}, the leading term generating the quark FCNC interaction is different from the one generating the flavor-changing charged current (FCCC) $b \to q\tau\nu$ transition. 
Hence, all the FCNC constraints are actually irrelevant to our analysis. 
The serious concern is indeed the flavor diagonal part of the $q\bar{q} \to \tau^+ \tau^-$ processes, which is affected by the SMEFT operators of Eq.~\eqref{eq:SMEFT_dim_6}. 
Under the MFV hypothesis in the quark sector, all the quark flavors $q$ have the same WCs for this diagonal part, \textit{e.g.},  $c_V^{1133}=c_V^{2233}=c_V^{3333}=c_{V\ell}^{\tau\tau}$.  
This immediately leads to a severe constraint by the $\tau^+\tau^-$ search at the LHC~\cite{ATLAS:2020zms,CMS:2022goy}. 
From a recent study of Ref.~\cite{Allwicher:2022gkm}, we see that
\begin{align}
 -0.0099\,\text{TeV}^{-2} & < c_V^{1133}/\Lambda^2 < 0.0040\,\text{TeV}^{-2} \,, \label{eq:Clq3311} \\[0.3cm]
 -0.13\,\text{TeV}^{-2} & < c_V^{2233}/\Lambda^2 < 0.040\,\text{TeV}^{-2} \,, \label{eq:Clq3322} \\[0.3cm]
 -0.80\,\text{TeV}^{-2} & < c_V^{3333}/\Lambda^2 < 0.53\,\text{TeV}^{-2} \,, \label{eq:Clq3333} 
\end{align} 
for the individual WCs in the SMEFT with the most general flavor structure at $95\%$ CL.\footnote{
Two points should be noted here: (i) our notation of the SMEFT WCs corresponds to the well-known convention~\cite{Grzadkowski:2010es}, such as $c_V^{1133} = [\mathcal C_{lq}^{(3)}]_{3311}$; (ii) to be precise, the MFV case needs to take into account all the quark generations for the $\tau^+\tau^-$ production in the analysis, which is different from the individual constraints and should obtain a much stronger bound~\cite{Allwicher:2022gkm}. 
}
Thus, the LHC bound on $c_{V\ell}^{\tau\tau}$ has (almost) no overlap with the range given by Eq.~\eqref{eq:RDsolCVL} that is obtained from the $R_{D^{(*)}}$ measurements. This might indicate that the MFV hypothesis of Eq.~\eqref{eq:cVlepton} does not work at all. 

The problematic LHC bounds of Eqs.~\eqref{eq:Clq3311} and \eqref{eq:Clq3322} can be avoided if we start with the third generation-philic WC of the form 
\begin{align}
 c_V^{mnij} \equiv \delta_{m3} \delta_{n3} \delta_{i3} \delta_{j3} c_{V\ell}^{\tau\tau} \,, 
 \label{eq:3philic}
\end{align} 
with which we can also produce $C_{V_L}^{u\tau}=C_{V_L}^{c\tau}$, as obtained in Eq.~\eqref{eq:CXuc}. 
In this case, we only need to concern the bound for the third generation WC of Eq.~\eqref{eq:Clq3333} for $c_V^{3333} = c_{V\ell}^{\tau\tau}$. 
Indeed, such a specific structure can be realized by imposing the $U(2)^5$ flavor symmetry~\cite{Barbieri:2011ci,Barbieri:2012uh,Blankenburg:2012nx,Faroughy:2020ina,Greljo:2022cah,Allwicher:2023shc} on the SMEFT Lagrangian, 
which distinguishes operators acting on the third generation from that on the first two ones, and constitutes an excellent approximate symmetry of the SM that is broken only at the level of $\mathcal{O}(10^{-2})$. A detailed description can be found in App.~\ref{app:U3andU2}. 
It is still concerned that the $R_{D^{(*)}}$ bound of Eq.~\eqref{eq:RDsolCVL} is already on the edge of the LHC bound of Eq.~\eqref{eq:Clq3333}. However, let us point out that there is a loophole in the analysis of Ref.~\cite{Allwicher:2022gkm}: only the ATLAS search~\cite{ATLAS:2020zms} has been taken into account to obtain Eq.~\eqref{eq:Clq3333}, even though the CMS search is also available~\cite{CMS:2022goy}. Especially, we would like to mention that the CMS collaboration observed an excess, which could result in a milder bound on $c_V^{3333}$, and hence the ATLAS and CMS measurements are not yet conclusive. It would be, therefore, reasonable to take Eq.~\eqref{eq:Clq3333} only as a reference bound for now, and a further improved search for the $\tau^+\tau^-$ final state is expected.  
In addition, the ratio $R_{\Upsilon(3S)}= \Gamma(\Upsilon(3S)\to\tau^+\tau^-)/\Gamma(\Upsilon(3S)\to\ell^+\ell^-)$ can also provide a viable constraint
\begin{align}
 -37.1\,\text{TeV}^{-2} < c_{V\ell}^{\tau\tau}/\Lambda^2 < 3.2\,\text{TeV}^{-2}\,, 
\end{align}
at the $2\sigma$ level, by taking the analytic formula and the experimental data taken from Refs.~\cite{Iguro:2024hyk,Aloni:2017eny,BaBar:2020nlq}. 
We can clearly see that the $R_{D^{(*)}}$ bound of Eq.~\eqref{eq:RDsolCVL} lies within the range obtained from the $R_{\Upsilon(3S)}$ measurement. 

The LHC bounds are also crucial in the case of $c_{S_d}^{mnij}$ with MFV. 
Again, the third generation-philic scenario of
\begin{align}
 c_{S_d}^{mnij} \equiv \delta_{m3} \delta_{n3} \delta_{i3} \delta_{j3} c_{S\ell}^{\tau\tau} \,, 
 \label{eq:CS3philic}
\end{align} 
works for the semi-leptonic $b \to q l \nu$ decays, and hence it is viable as a good approximation of the $U(2)^5$ flavor symmetry that generates $C_{S_R}^{u\tau}=C_{S_R}^{c\tau}$, 
as illustrated in App.~\ref{app:U3andU2}. 
In this case, the LHC bound on $c_{S_d}^{3333}=c_{S\ell}^{\tau\tau}$, read from Ref.~\cite{Allwicher:2022gkm}, can be obtained as  
\begin{align}
 -0.92\,\text{TeV}^{-2} < c_{S\ell}^{\tau\tau}/\Lambda^2 < 0.92\,\text{TeV}^{-2} \,, \label{eq:Cledq} 
\end{align} 
at $95\%$ CL, after running from the scale $\Lambda=1~\text{TeV}$ down to the $m_b$ scale~\cite{Gonzalez-Alonso:2017iyc,Iguro:2024hyk}. 
This is the only relevant constraint, while the $R_{\Upsilon(3S)}$ measurement is not affected by the operator associated with $c_{S_d}^{mnij}$. 
For the cases of $c_{S_u}^{mnij}$ and $c_{T}^{mnij}$, on the other hand, although the $U(2)^5$ flavor symmetry also supports the third generation-philic setup, 
it induces negligible and hierarchical contributions to the semi-leptonic $b \to c$ and $b \to u$ decays; see App.~\ref{app:U3andU2} for further details. 

To summarize, the relation of Eq.~\eqref{eq:CXuc} is still valid in this third generation-philic setup and also consistent with the relevant experimental constraints. 
Finally, one should note that renormalization group running from the electroweak to the $B$-meson scale does not change the relations between $C_X^{u\tau}$ and $C_X^{c\tau}$ specified in Eq.~\eqref{eq:CXuc}. Thus, we can adopt these relations in the following numerical analysis.

\subsubsection{Numerical analysis}

\begin{figure}[t]
\begin{center}
 \includegraphics[viewport=0 0 1250 543, width=0.99\textwidth]{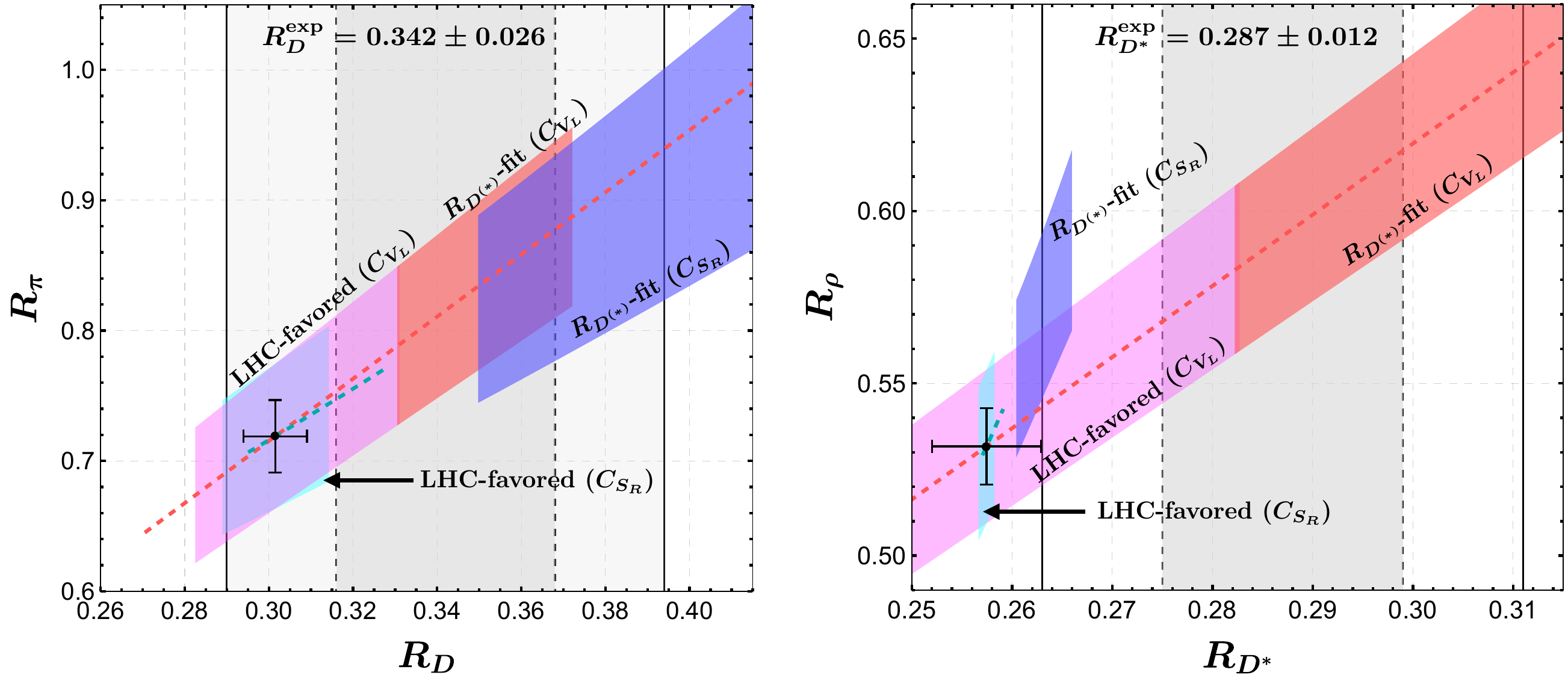}
 \caption{
 Projections on the $R_D$--$R_\pi$ and $R_{D^*}$--$R_\rho$ planes evaluated from the measured $\mathcal{B}(B \to \tau \nu)^\text{exp}$, the latest $R_D$-$R_{D^*}$ fit, and the LHC bounds. The individual measurements of $R_{D^{(*)}}^\text{exp}$ are also shown at the $1\sigma$ ($2\sigma$) level, as indicated by the dashed (solid) vertical lines. 
 The red and dark-cyan dashed lines are obtained by the $\mathcal{B}(B \to \tau \nu)$ measurement for the $V_L$ and $S_R$ types of NP scenarios, respectively. 
The allowed/favored regions from the $R_D$-$R_{D^*}$ fit and the LHC bounds are shown in different colors as described in the figure. 
 \label{fig:uc_correlation}
 }
\end{center}
\end{figure}

With the above setup, the measured branching ratio of $\mathcal{B}(B \to \tau \nu)^\text{exp}$~\cite{ParticleDataGroup:2024cfk}, the latest $R_D$--$R_{D^*}$ fit~\cite{Iguro:2024hyk}, and the LHC bounds~\cite{Allwicher:2022gkm} are most relevant to our combined analysis. 
In Fig.~\ref{fig:uc_correlation}, we put projections on the $R_D$--$R_\pi$ and $R_{D^*}$--$R_\rho$ planes, allowed/favored by these three considerations for the $V_L$ and $S_R$ types of NP scenarios. 
Here, the red and dark-cyan dashed lines, indicating the resulting central values of the ratios $R_X$, are obtained under the constraint of $\mathcal{B}(B \to \tau \nu)^\text{exp}$ for $V_L$ and $S_R$, respectively. 
The red and blue regions are favored by the $R_D$-$R_{D^*}$ fit results within $2\sigma$ for $V_L$ and $S_R$, respectively. The measured $R_{D^{(*)}}^\text{exp}$ are also shown by the gray bands, with the dashed (solid) vertical lines at the $1\sigma$ ($2\sigma$) level. It should be noted that the current best-fit solution of the $S_R$-type NP scenario cannot reproduce the central values $R_{D^{(*)}}^\text{exp}$, although the scenario improves the fit to the $R_{D^{(*)}}$ data~\cite{Iguro:2024hyk}. 
We also show the $2\sigma$ favored regions resulting from the LHC bounds in light-magenta and light-cyan for $V_L$ and $S_R$, respectively. 
However, let us repeat that the LHC bounds are not conclusive yet and taken only as illustrations. 
From Fig.~\ref{fig:uc_correlation}, we can see that the $R_D$-$R_{D^*}$ fit is (not) consistent with the $\mathcal{B}(B \to \tau \nu)^\text{exp}$ constraint in the $V_L$($S_R$)-type NP scenario. Hence, the $S_R$-type NP scenario is excluded as long as the relation of $C_{S_R}^{u\tau}=C_{S_R}^{c\tau}$ is assumed. 
This is a significant outcome of our combined analysis. 

We can also \textit{predict} the ratio $R_p$ in our combined analysis with the specific flavor symmetries. The relation of $C_{V_L}^{u\tau}=C_{V_L}^{c\tau}$ leads to 
\begin{align}
 \text{$V_L$-type NP:} \quad R_p = \frac{R_p^\text{SM}}{R_{\Lambda_c}^\text{SM}} R_{\Lambda_c} \,. 
\end{align} 
Hence, the sum rule fit result of $R_{\Lambda_c}^\text{SR}$ given in Eq.~\eqref{eq:RlamcSR} and the LHCb measurement of $R_{\Lambda_c}^\text{LHCb}$ in Eq.~\eqref{eq:RlamcLHCb}, respectively, imply
\begin{align}
 R_{\Lambda_c}^\text{SR}: & \quad 0.69 \lesssim R_p \lesssim 0.85 \,, \\[0.5em]
 R_{\Lambda_c}^\text{LHCb}: & \quad 0.34 \lesssim R_p \lesssim 0.67 \,. 
\end{align} 
Although these predictions are still quite uncertain, the former is more consistent with the SM value than the latter. Therefore, we see that the discrepancy between $R_{\Lambda_c}^\text{SR}$ and $R_{\Lambda_c}^\text{LHCb}$ can be tested once the ratio $R_p$ is observed. 
To conclude, we expect that measuring the ratios $R_\pi$, $R_\rho$ and $R_p$ for the $b \to u$ sector can provide another check of the $R$ ratio sum rule for the $b \to c$ sector implied from the $R_{D^{(*)}}$ anomalies. 
It can also test the third generation-philic NP scenario discussed here.

\section{Summary}
\label{summary}

As a powerful tool to test the LFU of the SM, the ratios $R_{D^{(*)}}$ and $R_{\Lambda_c}$ have recently received a lot of attention, because any significant deviations between the experimental measurements and the theoretical predictions of them would directly indicate the presence of NP beyond the SM. Interestingly enough, there exists a strong correlation among $R_{D}$, $R_{D^{*}}$ and $R_{\Lambda_c}$, which is referred to as the $R$ ratio sum rule that is almost independent of NP contributions to the processes. Therefore, the $R$ ratio sum rule gives a crucial feedback to the experimental measurements. However, to be definitely conclusive, we need precise calculations of both the SM predictions and the NP contributions. On the other hand, although the same arguments can be applied to the ratios $R_\pi$, $R_\rho$, and $R_p$ as above, we still have fewer measurements and/or theoretical studies of the semi-leptonic $b \to u l \nu$ decays. As such a situation will be significantly improved at the ongoing LHCb run-3 and Belle~II experiments, it is also important to perform a precise study of both the SM and NP effects in the $b\to u$ sector.

Motivated by the above observations, we have in this paper performed a detailed study of all the aforementioned $R_X$ ratios in the presence of general NP contributions 
by using the available lattice QCD and/or LCSR fits to the $B\to P$ ($P=D, \pi$), $B\to V$ ($V=D^\ast, \rho$) and $\Lambda_b\to H$ ($H=\Lambda_c, p$) form factors. 
In particular, we have evaluated the uncertainties of the NP contributions to every $R_X$ ratio, the $R$ ratio sum rules, and the shift factor $\delta_{H}$, inherited from the form factor fits for both the SM and NP currents. 
It is found that our result of $\delta_{\Lambda_c}$ (for the central value) is consistent with that obtained in the previous studies while our result gives the uncertainty for the first time. 
We have also obtained the $R$ ratio sum rule among $R_\pi$, $R_\rho$, and $R_p$ for the first time. 
To establish possible correlations between the $b \to c l \nu$ and $b \to u l \nu$ processes, we have also performed a combined study in the framework of SMEFT with specific flavor symmetries. Our main results can be summarized as below. 
\begin{itemize}
 \item 
 In Sec.~\ref{sec:b2cstudy}, we have obtained the explicit $R_{\Lambda_c}$ prediction from the $R$ ratio sum rule as in Eq.~\eqref{eq:SRfitprecise}, namely, 
 \begin{align}
  R_{\Lambda_c}^\text{SR} = 0.372 \pm 0.017 \big|_{R_X^\text{SM, exp}} \pm (< 0.001) \big|_{\text{SR}} \,, 
 \end{align}
 which indicates that the uncertainty stemming from the sum rule is negligible. 
 This concludes that the current experimental measurement $R_{\Lambda_c}^\text{LHCb} = 0.242 \pm 0.076$ still shows a significant deviation from the sum rule prediction, implying $\delta_{\Lambda_c}^\text{SR/LHCb} = - 0.39 \pm 0.23$ as in Eq.~\eqref{eq:deviation_SRLHCb}. 
 On the other hand, we found that the shift factor $\delta_{\Lambda_c}$ involves non-negligible uncertainties from the form factor inputs. 
 For instance, it turned out that the tensor NP solution to the $R_D$-$R_{D^\ast}$ fit generates $\delta_{\Lambda_c} (C_{T, \text{sol}}^{c\tau} \approx 0.02 \pm i \,0.13) = -0.035 \pm 0.096$ as in Eq.~\eqref{eq:deltaLc}, which may be compared with $\delta_{\Lambda_c}^\text{SR/LHCb}$. 
Thus, we found that this discrepancy may be compensated with the tensor NP contribution, consistent with the NP solution to $R_{D^{(*)}}^\text{exp}$, within the current uncertainty. 
This is one of the novel points of our study, which is available only by evaluating the uncertainties from the form factor inputs. 
Our result also offers an extensive motivation for reducing the uncertainties of the form factor inputs in the $b \to c$ sector. 
\item 
In Sec.~\ref{sec:b2ustudy}, we have obtained, from the $R$ ratio sum rule, the relation between $R_\rho$ and $R_p$ that have not been measured yet. 
To see a conclusive correlation, however, the precise measurements of $R_\pi$ and $R_\rho$ are necessary. 
Apart from the sum rule, we have additionally investigated the NP effects on the $R_\pi$-$R_\rho$ and $R_\pi$-$R_p$ relations, in which the uncertainties are also considered. 
It turned out that the NP type can be distinguished by the $R_\pi$-$R_\rho$ plane even with the current uncertainty. 
On the other hand, it is difficult to see the specific relation on $R_\pi$-$R_p$ distinct for each NP type since $R_p$ includes the large uncertainties. 
It also shows a good motivation for evaluating the form factor inputs from the theoretical side and measuring $R_X$ from the experimental side in the $b \to u$ sector. 
\item
In addition, we have configured two novel $R$ ratio sum rules by employing the ratios $R_{J/\psi}$ from $B_c\to J/\psi l\nu$ and $R_{X_c}$ from $B\to X_c l\nu$ decays, respectively. 
The sum rule given by Eq.~\eqref{eq:RJpsi} would become significant once the $R_{J/\psi}$ measurement is improved, while the sum rule among $R_{X_c}$, $R_D$ and $R_{D^\ast}$ could provide another complementary test of the dynamics behind these decays.
\item 
We have performed a combined study of the $b \to c l \nu$ and $b \to u l \nu$ decays based on the SMEFT with both $U(3)^5$ and $U(2)^5$ flavor symmetries.
Under the assumption, the corresponding WCs can be correlated such as $C_{V_L}^{u\tau}=C_{V_L}^{c\tau}$ and $C_{S_R}^{u\tau}=C_{S_R}^{c\tau}$, resulting in the interesting correlations between the $b \to c$ and $b \to u$ decays.
It is found that the $R_D$-$R_{D^*}$ fit is (not) consistent with the $\mathcal{B}(B \to \tau \nu)^\text{exp}$ constraint in the $V_L$($S_R$)-type NP scenario. 
Hence, the $S_R$-type NP scenario is excluded as long as the relation of $C_{S_R}^{u\tau}=C_{S_R}^{c\tau}$ is assumed.  
On the other hand, the $V_L$-type NP gives the explicit relation as $R_p/R_p^\text{SM} = R_{\Lambda_c}/R_{\Lambda_c}^\text{SM}$. 
This could provide a further consistency check of $R_{\Lambda_c}^\text{SR}$ and $R_{\Lambda_c}^\text{LHCb}$ in such a flavor specific NP scenario, if $R_p$ is observed by the future experiments. 
\end{itemize}

With the successful running of Belle~II and LHCb run-3, together with the progress of lattice QCD and LCSR calculations of the form factors, we will have more precise predictions of the $R_X$ ratios and subsequently the $R$ ratio sum rules, which can be utilized to identify the potential NP scenarios responsible for the $b \to c \tau \nu$ anomaly and possible deviations in the $b\to u\tau\nu$ processes in the future.

\section*{Acknowledgement}

We appreciate Takashi Kaneko for the fruitful discussion about the lattice results of the form factor inputs. 
We also thank Stefan Meinel for providing his preliminary evaluation of the $\Lambda_b \to p$ tensor form factors. 
This work is supported by the National Natural Science Foundation of China under Grant Nos.~12475094, 12135006 and 12075097, as well as by the Fundamental Research Funds for the Central Universities under Grant Nos.~CCNU22LJ004 and CCNU24AI003. S.I. enjoys support from JSPS KAKENHI Grant Number 24K22879 and JPJSCCA20200002.

\appendix
\section{Status of experimental measurements}
\label{app:status}

Regarding the semi-leptonic $B$-meson decays mediated by the quark-level $b \to c l \nu$ transition with $l= e, \mu, \tau$, we have the following observations: 
\begin{itemize}
 \item 
 The exclusive processes $B \to D^{(*)} \ell \nu$ for $\ell = e, \mu$ have been measured to extract $|V_{cb}|$ by the Belle~\cite{Belle:2015pkj,Belle:2018ezy,Belle:2023bwv}, Belle~II~\cite{Belle-II:2023okj} and BaBar~\cite{BaBar:2019vpl,BaBar:2023kug} collaborations. The experimental average is presently given as $|V_{cb}|^\text{ex} = (39.8 \pm 0.6) \times 10^{-3}$~\cite{ParticleDataGroup:2024cfk}, although it relies on the modeling of the $B \to D^{(*)}$ transition form factors. This value is then compared with $|V_{cb}|^\text{in} = (42.2 \pm 0.5) \times 10^{-3}$~\cite{ParticleDataGroup:2024cfk} extracted from the inclusive $B \to X_c \ell \nu$ decays, which is analyzed based on the heavy quark expansion. Thus, we see a $3\sigma$ discrepancy between the values extracted with these two methods; see Refs.~\cite{Bernlochner:2024sfg,Jay:2024ygl} for a recent review. 
  
 \item 
 The semi-tauonic counterparts, $B \to D^{(*)} \tau\nu$, have been analyzed by using the ratios defined by $R_{D^{(*)}} = \Gamma(B \to D^{(*)} \tau\nu)/\Gamma(B \to D^{(*)} \ell \nu)$, so that their dependence on the CKM matrix element $V_{cb}$ is cancelled out and the uncertainties due to the $B \to D^{(*)}$ transition form factors are reduced to a large extent. Thus, these ratios can be used to test the LFU of the SM. They have been measured by Belle~\cite{Belle:2015qfa,Belle:2016ure,Belle:2016dyj,Belle:2017ilt,Belle:2019rba}, BaBar~\cite{BaBar:2012obs,BaBar:2013mob}, LHCb (\textit{i.e.}, run-1~\cite{Aaij:2015yra,Aaij:2017uff,Aaij:2017deq,LHCb:2023zxo,LHCb:2023uiv} and run-2~\cite{LHCb:2024jll}), and Belle~II~\cite{Belle-II:2024ami} collaborations. The latest world average is reported in Ref.~\cite{HFLAV2024winter} as $R_{D}^\text{exp} = 0.342 \pm 0.026$ and $R_{D^*}^\text{exp} = 0.287 \pm 0.012$. 
 A complementary probe of the LFU to $R_{D^{(*)}}$ can be seen in the inclusive ratio of $R_{X_c} = \Gamma(B \to X_c \tau\nu)/\Gamma(B \to X_c \ell \nu)$~\cite{Freytsis:2015qca}, which has been recently measured by the Belle~II experiment with $R_{X_c}^\text{exp} =0.228 \pm 0.039$~\cite{Belle-II:2023aih}.
\end{itemize}
On the other hand, we have the following situation for the $b \to u l\nu$ semi-leptonic $B$-meson decays:
\begin{itemize}
 \item 
 The exclusive $B \to \pi \ell\nu$ and inclusive $B \to X_u \ell\nu$ processes have been measured to obtain $|V_{ub}|$. The present official status is summarized as $|V_{ub}|^\text{ex} = (3.70 \pm 0.10 \pm 0.12) \times 10^{-3}$ and $|V_{ub}|^\text{in} = (4.13 \pm 0.12 \pm 0.13 \pm 0.18) \times 10^{-3}$~\cite{ParticleDataGroup:2024cfk}, consistent with each other within $2\sigma$. A more recent measurement of $B^0\to \pi^-\ell^+ \nu$ from Belle~II~\cite{Belle-II:2024xwh} reduces the deviation in $|V_{ub}|$ to be within $1\sigma$ level. Although the other exclusive processes of $B\to\rho \ell\nu$ and $B\to\omega \ell\nu$ have been observed~\cite{Belle-II:2024xwh,HFLAV:2022esi} and theoretical studies for the $|V_{ub}|^\text{ex}$ extraction are available in Refs.~\cite{Bernlochner:2021rel,Leljak:2023gna}, it has not yet been included in the official world average~\cite{HFLAV:2022esi,ParticleDataGroup:2024cfk}. Notice that we have to rely on the light-cone sum rule (LCSR) calculations for the $B \to \rho, \omega$ form factors and the narrow-width approximation for the $|V_{ub}|^\text{ex}$ extraction, which results in lower values of $|V_{ub}|$~\cite{Bernlochner:2021rel,Leljak:2023gna}.
 
 \item 
 The semi-tauonic process $B \to \pi \tau\nu$ was first measured by the Belle collaboration~\cite{Belle:2015qal}, with a branching ratio of
 $\mathcal{B}(\bar B^0 \to \pi^+ \tau^- \nu) < 2.5 \times 10^{-4}$ at $90\%$ confidence level (CL), or equivalently $\mathcal{B}(\bar B^0 \to \pi^+ \tau^- \nu) = (1.52 \pm 0.72 \pm 0.13) \times 10^{-4}$. Taking $\mathcal{B}(\bar B^0 \to \pi^+ \ell^- \nu) = (1.50 \pm 0.06) \times 10^{-4}$~\cite{HFLAV:2022esi}, we find a value of $R_\pi^\text{exp} = \Gamma (B \to \pi \tau\nu)/\Gamma (B \to \pi \ell\nu) \approx 1.01 \pm 0.49$. On the other hand, the $B \to \rho \tau\nu$ and $B \to \omega \tau\nu$ decays have not been measured yet, but are expected to be measurable at the LHCb run-3~\cite{LHCb:2018roe,LHCb:2022ine} and Belle~II~\cite{Belle-II:2018jsg,Belle-II:2022cgf} experiments. 
\end{itemize}
We also have other interesting processes relevant for $b \to (c,u) l\nu$, such as the semi-leptonic $\Lambda_b$ and $B_c$ decays. Their status can be summarized as below.
\begin{itemize}
 \item 
 The baryonic processes $\Lambda_b \to \Lambda_c \mu\nu$ and $\Lambda_b \to \Lambda_c \tau\nu$ have been measured by the DELPHI~\cite{DELPHI:2003qft} and LHCb~\cite{LHCb:2022piu} collaborations, respectively. The latter also reports $R_{\Lambda_c}^\text{exp} = \Gamma (\Lambda_b \to \Lambda_c \tau\nu)/\Gamma (\Lambda_b \to \Lambda_c \mu\nu) = 0.242 \pm 0.026 \pm 0.040 \pm 0.059$. Normalizing the LHCb measurement to the SM prediction for $\Gamma(\Lambda_b \to \Lambda_c \mu\nu)$ instead of the DELPHI measurement, a modestly increased value of $R_{\Lambda_c}^\text{exp} = (0.04/|V_{cb}|)^2 (0.285 \pm 0.073)$ would be obtained~\cite{Bernlochner:2022hyz}. 
 Regarding $b\to u \mu \nu$, Ref.~\cite{LHCb:2015eia} observed $\Lambda_b \to p \mu\nu$ in terms of the ratio to $\Lambda_b \to \Lambda_c \mu\nu$ for limited $q^2$ ranges, with $q^2$ being the invariant mass squared of the lepton-neutrino system. It also gives a measurement of $|V_{ub}/V_{cb}| = 0.083 \pm 0.004 \pm 0.004$ and $\mathcal{B}(\Lambda_b \to p \mu\nu) = (4.1 \pm 1.0)\times10^{-4}$ by extrapolating to the full $q^2$ range. However, the semi-tauonic mode $\Lambda_b \to p \tau\nu$ has not been measured yet. 
 
 \item 
 The semi-leptonic $B_c\to J/\psi l\nu$ decays arise from the same quark-level $b \to c l\nu$ transition as in the $B \to D^{(*)} l\nu$ and $\Lambda_b \to \Lambda_c l\nu$ processes, and thus they are on the same playground as $R_{D}$, $R_{D^*}$ and $R_{\Lambda_c}$. Recently, the $\tau/\mu$ ratio has been observed by the LHCb~\cite{Aaij:2017tyk} and CMS~\cite{CMS:2024seh,CMSRJpsi2} collaborations, although these measurements are still plagued by large uncertainties. A naive average is given as $R_{J/\psi}^\text{exp} = \Gamma (B_c \to J/\psi \tau\nu)/\Gamma (B_c \to J/\psi \mu\nu) = 0.61 \pm 0.18$~\cite{Iguro:2024hyk}. 
\end{itemize}
As a matter of fact, both the electronic and muonic processes are used to extract the CKM matrix elements, in which the LFU between the first two generations is always assumed. On the other hand, as the experimental measurements have shown some discrepancies from the SM predictions in the ratios $R_{D^{(*)}}$ since 2012, many theoretical studies have paid attention to NP possibilities in the tauonic modes, which will also be supposed in this paper. In this case, the ratios $R_X/R_X^\mathrm{SM}$ can be considered as optimal observables for testing the LFU, where $R_X^\mathrm{SM}$ stands for the SM prediction and $R_X$ indicates both the SM and NP contributions in the tauonic processes.

The LFU test has also been studied in the flavor changing neutral sector of $R_{K^{(*)}} = \Gamma (B \to K^{(*)} \mu^+\mu^-)/\Gamma (B \to K^{(*)} e^+e^-)$. 
At present, recent measurements of $R_{K^{(*)}}$~\cite{LHCb:2022qnv,LHCb:2022vje} agree with the SM. 
Nevertheless, these processes are still interesting because the branching ratios and angular observables of some $b \to s \mu^+\mu^-$ processes are still inconsistent with the SM. 
Hence, a comprehensive NP study including $b \to s \mu^+\mu^-$ is still important. 
See Refs.~\cite{Bhattacharya:2016mcc,Kumar:2018kmr} for this direction of the study.

\section{Form Factor descriptions}
\label{app:FFinputs}

\subsection*{\boldmath{$B \to D^{(*)}$}: SM current}

The outer functions $\phi_F(z)$ for individual form factors $F$ are written as~\cite{Boyd:1997kz}
\begin{align}
 \phi_{f_+^D}(z) & = \chi_{f_+^D} (1+z)^2(1-z)^{1/2} \left[(1+r)(1-z)+2 \sqrt{r}(1+z)\right]^{-5} \,,  \\[1em]
 \phi_{f_0^D}(z) & =\chi_{f_0^D} (1+z)(1-z)^{3/2} \left[(1+r)(1-z)+2 \sqrt{r}(1+z)\right]^{-4} \,, \\[1em]
 \phi_g(z) & =\chi_{g} (1+z)^2(1-z)^{-1/2} \left[(1+r)(1-z)+2 \sqrt{r}(1+z)\right]^{-4} \,, \\[1em]
 \phi_f(z) & =\chi_{f} (1+z)(1-z)^{{3/2}} \left[(1+r)(1-z)+2 \sqrt{r}(1+z)\right]^{-4} \,, \\[1em]
 \phi_{F_1}(z) & =\chi_{F_1} (1+z)(1-z)^{{5/2}} \left[(1+r)(1-z)+2 \sqrt{r}(1+z)\right]^{-5} \,, \\[1em]
 \phi_{F_2}(z) & =\chi_{F_2} (1+z)^2 (1-z)^{-1/2} \left[(1+r)(1-z)+2 \sqrt{r}(1+z)\right]^{-4} \,, 
\end{align}
where $r \equiv m_{D^{(\ast)}}/m_B$ is the mass ratio, and the prefactors $\chi_F$ can be numerically given as~\cite{Bigi:2016mdz,Bigi:2017njr,Bigi:2017jbd}
\begin{align}
 & \chi_{f_+^D} = 12.43 \,,& &\chi_{f_0^D} = 10.11 \,,&  &\chi_g = 53.79 \,, \\[0.5em]
 & \chi_f = 1.454  \,,&  &\chi_{F_1} = 0.195 \,,&  &\chi_{F_2} = 10.71 \,.
\end{align}
The Blaschke factors are given in the following form: 
\begin{align}
 P_F(z) = \prod_n \frac{z - z_n^F}{1 -z\,z_n^F} \,, 
\end{align} 
where each form factor $F$ has its own $z^F_n$ that account for the resonance poles. 
We do not exhibit their explicit definitions, but only quote their numerical values as~\cite{Bigi:2016mdz,Bigi:2017njr,Bigi:2017jbd}
\begin{align}
 &z_1^{f_0^D} = -0.433 \,,& &z_2^{f_0^D} = -0.819 \,,  \\[1em]
 & z_1^{f_+^D} = -0.308 \,,& &z_2^{f_+^D} = -0.555 \,,& &z_3^{f_+^D} = -0.646 \,,  \\[1em]
 & z_1^{F_2} = -0.274 \,,& &z_2^{F_2} = -0.443 \,,&  &z_3^{F_2} = -0.791 \,,   \\[1em]
 & z_1^{g} = -0.286 \,,&  &z_2^{g} = -0.479 \,,&  &z_3^{g} = -0.537 \,,&  &z_4^{g} = -0.890 \,, \\[1em]
 & z_1^{f,F_1} = -0.402 \,,&  &z_2^{f,F_1} = -0.406 \,,& &z_3^{f,F_1} = -0.637 \,,&  &z_4^{f,F_1} = -0.642 \,.  
\end{align} 
Based on the above setup, the BGL series coefficients $a_n^F$ have been fitted to the lattice QCD and LCSR evaluations with $N_F =2$ for the SM currents in Ref.~\cite{Cui:2023jiw}. 
They are given explicitly as 
\begin{align}
 \label{eq:BDinput}
 &\hspace{-0.8cm} 
 \big\{
 a_0^{f_{+}^D} ,\,
 a_1^{f_{+}^D} ,\, 
 a_2^{f_{+}^D} ,\,
 a_1^{f_0^D} ,\, 
 a_2^{f_0^D} ,\, 
 a_0^g ,\,
 a_1^g ,\,
 a_2^g ,\,
 a_0^f ,\,
 a_1^f ,\,
 a_2^f ,\,
 a_1^{F_1} ,\,
 a_2^{F_1} ,\,
 a_1^{F_2} ,\,
 a_2^{F_2}
\big\}  \notag \\[0.5em]
=
 \big\{
  & 0.0137(1),
 -0.0417(33),
  0.0415(1124),
 -0.2072(147),
  0.1880(5330), \notag \\[0.3em]
 & 0.0256(9),
 -0.1005(456),
  0.2587(6564),
  0.0109(2),
  0.0081(101), \notag \\[0.3em]
 & 0.0693(2140),
 -0.0024(20),
 -0.0155(388),
 -0.2097(581),
  0.5667(8789)
 \big\} \,, 
\end{align} 
while the remaining coefficients $a_0^{f_0^D}$, $a_0^{F_1}$, and $a_0^{F_2}$ are determined by the kinematic conditions of 
$f_0^D(q^2=0) = f_+^D(q^2=0)$, $f(z=0) = F_1(z=0)/\left(m_B-m_{D^\ast}\right)$, and $F_1(q^2=0) = \left(m_B^2 - m_{D^\ast}^2\right) F_2(q^2 = 0)/2$. 
That is 
\begin{align}
 a_0^{f_0^D} &= 4.945 \,a_0^{f_+^D} + 0.319 \,a_1^{f_+^D} +0.021  \,a_2^{f_+^D} - 0.064 \,a_1^{f_0^D} -0.004 \,a_2^{f_0^D}   \,, \\[0.5em]
 a_0^{F_1} &= 0.167 \,a_0^f \,, \\[0.5em]
 a_0^{F_2} &= 3.563 \,a_0^{f} + 1.193 \,a_1^{F_1} + 0.067 \,a_2^{F_1} - 0.056 \,a_1^{F_2} -0.003 \,a_2^{F_2} \,. 
\end{align} 
We will also take into account the correlation matrices for the BGL expansion coefficients in Eq.~\eqref{eq:BDinput}, as provided in Ref.~\cite{Cui:2023jiw}, when evaluating the error propagations in the observables.

\subsection*{\boldmath{$B \to D^{(*)}$}: Tensor current}

The resonance masses $M_{F}$ are given numerically as
\begin{align}
 M_{f_T^D} = M_{T_1^{D^\ast}} = 6.330\,\text{GeV} \,, \quad\quad M_{T_2^{D^\ast}} = M_{T_{23}^{D^\ast}} = 6.767\,\text{GeV} \,,  \label{b2c_tensor}
\end{align}
for $F = f_T^D$, $T_1^{D^\ast}$, $T_2^{D^\ast}$, and $T_{23}^{D^\ast}$, where $T_{23}^{D^\ast}$ is fitted instead of $T_3^{D^\ast}$ itself and related to $T_3^{D^\ast}$ via
\begin{align}
 T_{23}^{D^\ast} (q^2) = \frac{T_2^{D^\ast}(q^2) \left(m_B^2 + 3m_{D^\ast}^2 - q^2\right) \left(m_B^2 - m_{D^\ast}^2\right) - T_3^{D^\ast}(q^2) Q_+^{D^\ast} Q_-^{D^\ast}}{8m_B m_{D^\ast}^2 \left(m_B - m_{D^\ast}\right)} \,. \label{FF_T23}
\end{align}
Then, the fit results for the tensor form factors are given as~\cite{Gubernari:2018wyi}
\begin{align}
  &\big\{a_0^{f_T^D} ,\, a_1^{f_T^D} ,\, a_2^{f_T^D} \big\}  = \big\{ 0.565 (36), -2.5 (2.2), 10.6 (31.8)  \big\} \,, \\[1em] 
  &\big\{a_0^{T_1^{D^\ast},\,T_2^{D^\ast}} ,\, a_1^{T_1^{D^\ast}} ,\, a_2^{T_1^{D^\ast}} ,\, a_1^{T_2^{D^\ast}} ,\, a_2^{T_2^{D^\ast}} ,\, a_0^{T_{23}^{D^\ast}} ,\, a_1^{T_{23}^{D^\ast}} ,\, a_2^{T_{23}^{D^\ast}}  \big\} \notag \\[0.5em]
  &= \big\{  0.630 (78) \,, -1.418 (2.630) \,, -0.688 (39.922) \,, 1.523 (3.097) \,, 0.007 (47.537) \,, \notag \\[0.3em]
  &\hspace{2em} 0.806 (93) \,, 0.586 (3.644) \,, 4.667 (57.422) \big\} \,, 
\end{align}
where $a_0^{T_2^{D^\ast}} = a_0^{T_1^{D^\ast}}$ is determined from the kinematic relation of $T_2^{D^\ast}(q^2=0) = T_1^{D^\ast}(q^2=0)$. 
In addition, the correlations among these expansion coefficients, provided in the same reference, are also taken into account in our analysis.

\subsection*{\boldmath{$\Lambda_b \to \Lambda_c$}: SM and Tensor currents}

The resonance mass for each form factor is given as 
\begin{align}
 &M_{F_+^{\Lambda_c}} = M_{F_\perp^{\Lambda_c}} = M_{h_+^{\Lambda_c}} = M_{h_\perp^{\Lambda_c}} = 6.332 \,\text{GeV} \,,& &M_{F_0^{\Lambda_c}} = 6.725 \,\text{GeV} \,, \\[0.5em] 
 &M_{G_+^{\Lambda_c}} = M_{G_\perp^{\Lambda_c}} = M_{\tilde h_+^{\Lambda_c}} = M_{\tilde h_\perp^{\Lambda_c}} = 6.768 \,\text{GeV} \,,& &M_{G_0^{\Lambda_c}} = 6.276 \,\text{GeV} \,.
\end{align} 
Equipped with the above setup of the parameterization as well as the endpoint constraints on the form factors,\footnote{
To be more precise, the endpoint constraints are imposed on the lattice spacing, while the final fit results in the physical limit are obtained by extrapolations from the lattice fit. 
Due to this procedure, the endpoint constraints $F_0^{\Lambda_c}(q^2=0)=F_+^{\Lambda_c}(q^2=0)$ and $G_0^{\Lambda_c}(q^2=0)=G_+^{\Lambda_c}(q^2=0)$ are not exact in their fit results but consistent within the errors. 
} 
$F_0^{\Lambda_c}(q^2=0)=F_+^{\Lambda_c}(q^2=0)$\,, $G_0^{\Lambda_c}(q^2=0)=G_+^{\Lambda_c}(q^2=0)$\,, $G_{\perp}^{\Lambda_c}(q^2=t_0)=G_+^{\Lambda_c}(q^2=t_0)$\,, and $\tilde h_\perp^{\Lambda_c} \left(q^2 = t_0\right) = \tilde h_+^{\Lambda_c} \left(q^2 = t_0\right)$\,,
we can obtain the series coefficients $a_n^F$ with $N_F=1$ by fitting to the lattice data points, which read~\cite{Detmold:2015aaa,Datta:2017aue}
\begin{align}
 & \big\{ a_0^{F_+^{\Lambda_c}} ,\, a_1^{F_+^{\Lambda_c}} ,\, a_0^{F_0^{\Lambda_c}} ,\, a_1^{F_0^{\Lambda_c}} ,\, a_0^{F_\perp^{\Lambda_c}} ,\, a_1^{F_\perp^{\Lambda_c}} ,\, a_0^{G_+^{\Lambda_c},G_\perp^{\Lambda_c}} ,\, a_1^{G_+^{\Lambda_c}} ,\, a_0^{G_0^{\Lambda_c}} ,\, a_1^{G_0^{\Lambda_c}} ,\, a_1^{G_\perp^{\Lambda_c}} \big\}  \notag \\[0.5em] 
 & = \big\{ 
 0.8146 (167), -4.8988 (5425), 0.7439 (125), -4.6477 (6083), 1.0780 (256), \,\notag \\[0.3em]
 &\hspace{0.7em} -6.4171 (8480), 0.6847 (86), -4.4312 (3572), 0.7396 (143), -4.3665 (3314), \,\notag \\[0.3em]
 &\hspace{0.7em} -4.4634 (3613) \big\} \,, \\[0.5em] 
 &\big\{ a_0^{h_+^{\Lambda_c}} ,\, a_1^{h_+^{\Lambda_c}} ,\, a_0^{h_\perp^{\Lambda_c}} ,\, a_1^{h_\perp^{\Lambda_c}} ,\, a_0^{\tilde h_+^{\Lambda_c}, \tilde h_\perp^{\Lambda_c}} ,\, a_1^{\tilde h_+^{\Lambda_c}} ,\, a_1^{\tilde h_\perp^{\Lambda_c}} \big\} = \big\{ 
 0.9752 (303), -5.5000 (1.2361), \,\notag \\[0.3em]
 &\hspace{1em} 0.7054 (137), -4.3578 (5114), 0.6728 (88), -4.4322 (3882), -4.4928 (3584) \big\} \,,
\end{align}
where the covariance matrix of the fit results is explicitly given in the same references and will be taken into account in our study. 
Although this nominal fit is supposed to be the fundamental result of Refs.~\cite{Detmold:2015aaa,Datta:2017aue}, the authors have also provided a procedure to calculate the ``systematic'' as stated in the main text. 
We explicitly follow Eqs.~(82)--(84) in Ref.~\cite{Detmold:2015aaa}.

\subsection*{\boldmath{$B \to \pi$}: SM and Tensor currents}

The BCL series coefficients $b_F^n$ are taken from Ref.~\cite{Cui:2022zwm} such as  
\begin{align}
 \big\{b_0^{f_+^\pi},\, &b_1^{f_+^\pi},\, b_2^{f_+^\pi},\, b_0^{f_0^\pi},\, b_1^{f_0^\pi}\big\} \notag\\[0.3em]
 = \big\{&0.404 (13), -0.618 (63), -0.473 (215), 0.496 (19), -1.537 (56)\big\}, \\[0.5em]
 \big\{ b_0^{f_T^\pi},\, &b_1^{f_T^\pi},\, b_2^{f_T^\pi}\big\} 
 = \big\{0.396 (15), -0.553 (73), -0.248 (235)\big\},
\end{align}
for the truncation order $N_F = 3$, where the remaining coefficient $b_2^{f_0^\pi}$ is related to the others as
\begin{align}
  b_2^{f_0^\pi} = 12.78(b_0^{f_+^\pi}-b_0^{f_0^\pi})+3.48 b_1^{f_+^\pi}+1.19 b_2^{f_+^\pi}-3.58 b_1^{f_0^\pi} \,, 
\end{align}
from the kinematic condition of $f_+^\pi (q^2 = 0) = f_0^\pi (q^2 = 0)$. In addition, the correlation matrix is provided by Table~2 in the same reference.

\subsection*{\boldmath{$B \to \rho$}: SM and Tensor currents}

The resonance masses now read  
\begin{align}
 &M_{A_0^\rho} = 5.279~{\rm GeV} \,, \qquad M_{V^\rho}  = M_{T_1^\rho} = 5.325~{\rm GeV} \,,  \\[0.5em]
 &M_{A_1^\rho}  = M_{A_{12}^\rho} = M_{T_{2}^\rho} = M_{T_{23}^\rho} = 5.724~{\rm GeV} \,, 
\end{align}
for $F = A_0^\rho$, $V^\rho$, $A_1^\rho$, $A_{12}^\rho$, $T_1^\rho$, $T_2^\rho$, and $T_{23}^\rho$, 
where $A_{12}^\rho$ and $T_{23}^\rho$ are fitted, instead of $A_2^\rho$ and $T_3^\rho$ themselves, with the following relations~\cite{Bharucha:2015bzk}:
\begin{align}
 A_{12}^\rho (q^2) &= \frac{A_1^\rho (q^2) \left(m_B^2 - m_\rho^2 - q^2\right) \left(m_B+m_\rho\right)^2 - A_2^\rho (q^2) Q_+^\rho Q_-^\rho}{16m_B m_\rho^2 \left(m_B+m_\rho\right)} \,,\\[0.5em]
 T_{23}^\rho (q^2) &= \frac{T_2^\rho(q^2) \left(m_B^2 + 3m_{\rho}^2 - q^2\right) \left(m_B^2 - m_{\rho}^2\right) - T_3^\rho(q^2) Q_+^\rho Q_-^\rho}{8m_B m_{\rho}^2 \left(m_B - m_{\rho}\right)} \,. 
\end{align} 
Then, the fit results are imported from Ref.~\cite{Bharucha:2015bzk} such that 
\begin{align}
 &\big\{ a_1^{A_0^\rho},\,  a_2^{A_0^\rho},\,  a_0^{A_1^\rho},\,  a_1^{A_1^\rho},\,  a_2^{A_1^\rho},\,  a_0^{A_{12}^\rho},\,  a_1^{A_{12}^\rho},\,  a_2^{A_{12}^\rho},\,  a_0^{V^\rho},\,  a_1^{V^\rho},\,  a_2^{V^\rho} \big\}  \notag \\[0.5em]
 & = \big\{-0.833 (204), 1.331 (1.050), 0.262 (26), 0.393 (139), 0.163 (408), 0.297 (35), \notag \\[0.2em]
 & \hspace{2em} \, 0.759 (197), 0.465 (756), 0.327 (31), -0.860 (183), 1.802 (965) \big\} \,, \\[1em]
 &\big\{a_0^{T_1^\rho},\, a_1^{T_1^\rho},\,  a_2^{T_1^\rho},\,  a_1^{T_2^\rho},\,  a_2^{T_2^\rho}, a_0^{T_{23}^\rho},\,  a_1^{T_{23}^\rho},\,  a_2^{T_{23}^\rho} \big\} \notag \\[0.5em]
 &= \big\{ 0.272 (26), -0.742 (143), 1.453 (773), 0.471 (134), 0.576 (465), 0.747 (76), \notag \\[0.2em]
 &\hspace{2em} 1.896 (428), 2.930 (1.807) \big\} \,, 
\end{align}
for $N_F = 2$, and the full correlation matrices are also provided as ancillary files in the same reference. 
The rest of the expansion coefficients, $a_0^{A_0^\rho}$ and $a_0^{T_2^\rho}$, are given as
\begin{align}
 a_0^{T_2^\rho} = a_0^{T_1^\rho} \,, \quad\quad a_0^{A_0^\rho} = 1.201\, a_0^{A_{12}^\rho} \,, 
\end{align}
from the kinematic relations of $T_1^\rho(q^2=0) = T_2^\rho (q^2=0)$ and $A_{12}^\rho (q^2=0) = \left(m_B^2-m_\rho^2\right)/\left(8m_B m_\rho\right) A_0^\rho(q^2=0)$, respectively.

For the semi-leptonic $B\to \rho \ell \nu$ decay, it is important to note that the actual signal final state is $ \pi\pi \ell \nu$, since the $\rho$ meson is not stable and has a relatively large decay width. 
This necessitates careful consideration of how non-resonant background, finite-width effects, and $S$- and $P$-wave contributions are treated or distinguished in both theoretical computations and experimental measurements. 
As detailed in Ref.~\cite{Bharucha:2015bzk}, these issues can be addressed in a reasonable manner, ensuring that the LCSR computation of $B\to \rho$ form factors does not suffer from significant additional uncertainties due to these effects.

\subsection*{\boldmath{$\Lambda_b \to p$}: SM current} 

The resonance masses are taken as 
\begin{align}
	M_{F_+^p} &= M_{F_\perp^p} 
    = 5.325~{\rm GeV} ,   \qquad M_{F_0^p} = 5.656~{\rm GeV} , \\[0.5em]
	M_{G_+^p} &= M_{G_\perp^p} 
    = 5.706~{\rm GeV} ,   \qquad M_{G_0^p} = 5.279~{\rm GeV} .  
\end{align}
Then, the lattice fit results are given from Ref.~\cite{Detmold:2015aaa} as 
\begin{align}
 &\big\{a_0^{F_+^p} ,\, a_1^{F_+^p} ,\, a_0^{F_0^p} ,\, a_1^{F_0^p} ,\, a_0^{F_\perp^p} ,\, a_1^{F_\perp^p} ,\, a_0^{G_+^p,G_\perp^p} ,\, a_1^{G_+^p} ,\, a_0^{G_0^p} ,\, a_1^{G_0^p} ,\, a_1^{G_\perp^p} 
 \big\}   \notag \\[0.5em] 
 &= \big\{ 
   0.438 (31), -0.645 (209), 0.419 (26), -0.786 (204), 0.539 (44), -0.807 (304),  \notag \\[0.3em] 
 &\hspace{2em} 0.391 (20), -0.817 (175), 0.453 (29), -0.782 (189), -0.906 (196)
 \big\} \,, 
\end{align}
for the nominal fit with $N_F = 1$ and their covariances are also provided in the same reference. 
Similar to the case of the $\Lambda_b\to \Lambda_c$ form factors, we will also consider in our analysis the systematic uncertainty caused by their assumptions~\cite{Detmold:2015aaa}.

A description for the $\Lambda_b \to p$ tensor form factor is given in the main text.

\section{Flavor symmetries for the semi-leptonic decays}
\label{app:U3andU2}

\subsection{MFV hypothesis}

Within the SM, the Yukawa sector is the only breaking source of the symmetry group $\mathcal{G}_F$ defined by Eq.~\eqref{eq:U(3)_symmetry}. 
However, once we promote the SM Yukawa couplings $Y_{U,D,E}$ to become some \textit{spurions}, \textit{i.e.}, non-dynamical fields with the well-defined transformation properties under the non-Abelian part of $\mathcal{G}_F$,
\begin{equation}
 Y_U \sim \left(\bf 3,\bf \bar 3, \bf 1, \bf 1, \bf 1\right), \quad Y_D \sim \left(\bf 3, \bf 1, \bf \bar 3, \bf 1, \bf 1\right), \quad Y_E \sim \left(\bf 1, \bf 1, \bf 1, \bf 3, \bf \bar 3\right) \,, 
\end{equation}
for each $SU(3)$ factor in the same order as in Eq.~\eqref{eq:U(3)_symmetry}, the whole SM Lagrangian will be formally invariant under the global symmetry of $\mathcal{G}_F$. The above configuration can be adapted to the SMEFT Lagrangian as well. This means that all the higher-dimensional operators built out of the SM field contents and the Yukawa spurions $Y_{U,D,E}$ in the SMEFT can be made formally invariant under $\mathcal{G}_F$ to all orders, and the breaking of the symmetry occurs only via the appropriate insertions of the spurions $Y_{U,D,E}$, which induces the flavor violating interactions below the electroweak scale~\cite{Chivukula:1987py,DAmbrosio:2002vsn}. 
Notice that, with the MFV hypothesis, the dynamics of flavor violation is completely determined by the structure of the ordinary Yukawa couplings, both within the SM and beyond~\cite{Isidori:2023pyp,Faroughy:2020ina}.

As an illustration, let us discuss the term associated with $c_V$ introduced in Eq.~\eqref{eq:SMEFT_dim_6}. 
The MFV criterion implies that the quark and lepton sectors are factorized such that $c_V^{mnij}=c_{Vq}^{mn}\, c_{V\ell}^{ij}$, where $c_{Vq}^{mn}$ and $c_{V\ell}^{ij}$ transform as $\left(\bf 1 \oplus 8, \bf 1, \bf 1\right)$ and $\left(\bf 1 \oplus 8, \bf 1\right)$ under the $SU(3)_Q \otimes SU(3)_U \otimes SU(3)_D$ and $SU(3)_L \otimes SU(3)_E$ subgroups of $\mathcal{G}_F$, respectively. 
The most general form of $c_{Vq}$ can be represented in terms of the building blocks $Y_UY_U^\dag$ and $Y_DY_D^\dag$ as~\cite{Grinstein:2023njq,Hou:2024vyw} 
\begin{equation}
 c_{Vq}^{mn} = \left( b_0^{(q)} {\bf 1} + b_1^{(q)} (Y_UY_U^\dag) + b_2^{(q)} (Y_DY_D^\dag) + \cdots \right)_{mn} \,, 
 \label{eq:cvMFV}
\end{equation}
where $b_k^{(q)}$ are undetermined but flavor-blind coefficients, and the ellipsis denotes other terms with any possible products of $Y_UY_U^\dag$ and $Y_DY_D^\dag$ with some finite powers. The complete expression can be found in Ref.~\cite{Hou:2024vyw}, which is however not necessary for our case, as will be described below. 
Explicitly, the quark sector has the following contributions to the left-handed charged current of $\bar u_L^m \gamma^\mu d_L^n$:  
\begin{align}
 \delta_{mn} \left(\bar u_L^m \gamma^\mu d_L^n \right) & \to V_{u^m d^n} \left(\bar u_L^m \gamma^\mu d_L^n \right) \,, \\[0.2cm]
 (Y_UY_U^\dag)_{mn} \left(\bar u_L^m \gamma^\mu d_L^n \right) & \to  \frac{2m_{u^m}^2}{v^2} V_{u^m d^n} \left(\bar u_L^m \gamma^\mu d_L^n \right) \,, \label{eq:chargedYuYu} \\[0.15cm]
 (Y_DY_D^\dag)_{mn} \left(\bar u_L^m \gamma^\mu d_L^n \right) & \to  \frac{2m_{d^n}^2}{v^2} V_{u^m d^n} \left(\bar u_L^m \gamma^\mu d_L^n \right) \,, 
\end{align}
where the quark fields on the right-hand side are already transformed from the weak to the mass eigenstate basis, $f_{L} \to f_{L}^\prime= W_{f}^\dagger f_{L}$, with $W_{f}$ being the $3\times 3$ unitary matrices in flavor space and the prime being omitted in the above equations. 
We can see that in the fermion mass eigenbasis, the FCCC transition induced by the $b_{k\neq0}^{(q)}$ terms occurs via the elements of the CKM matrix $V=W_{u}^\dagger W_{d}$, together with the suppression factor $(m_q^2/v^2)^k$. 
Therefore, all the terms with $b_{k\neq0}^{(q)}$ are negligible for $b \to (u,c)$ transitions as long as all the coefficients $b_{k}^{(q)}$ are assumed to be of the same order.
As a consequence, we need only keep the term with $b_{0}^{(q)}$ for the quark sector and, up to here, the WC $c_V$ can be expressed as
\begin{align}
 c_V^{mnij} = \delta_{mn} c_{V\ell}^{ij} \,. 
\end{align}
In the same way, the MFV hypothesis in the lepton sector allows us to represent the flavor structure of $c_{V\ell}^{ij}$ as~\cite{Cirigliano:2005ck,Hou:2024vyw}  
\begin{equation}
 c_{V\ell}^{ij} = \left( \sum_{k=0} b_k^{(\ell)} (Y_EY_E^\dag)^k \right)_{ij} \,, 
\end{equation}
and, after performing the necessary mass basis rotation, we have 
\begin{align}
 (Y_EY_E^\dag)_{ij} \left(\bar\ell_L^i\gamma^\mu\nu_L^j\right) \to \frac{2 m_{\ell^i}^2}{v^2} \,\delta_{ij} \left( \bar\ell_L^i\gamma^\mu\nu_L^j\right) \,.
\end{align}
It is observed that the terms with $b_{k\neq0}^{(\ell)}$ naturally induce a $\tau$-philic interaction and $b_{1}^{(\ell)}$ gives the largest contribution, while $b_0^{(\ell)}=0$ is necessary to obtain the desired LFU violation. 
To conclude, the natural setup of the WC $c_V$ under the MFV hypothesis can finally be represented as 
\begin{align}
 c_V^{mnij} = \delta_{mn} (Y_EY_E^\dag)_{ij} c_V^0 \,, 
\end{align}
where $c_V^0 \equiv b_0^{(q)} b_1^{(\ell)}$, and we have taken $b_0^{(\ell)}=0$ in the lepton sector.

From the form of Eq.~\eqref{eq:cvMFV}, we can also see that the quark flavor structures differ between the charged and neutral currents.
For instance, the quark sector $\bar{Q}^m \gamma^\mu \tau^I Q^n$ of our interest also induces the neutral current $\bar d^m_L \gamma^\mu d_L^n$. 
In this case, the $b_{k\neq0}^{(q)}$ terms provide the following dominant contribution from the mass basis rotation:
\begin{equation}
 (Y_UY_U^\dag)_{mn} \left(\bar d^m_L \gamma^\mu d_L^n\right) \to \sum_l \frac{m_{u^l}^2}{v^2} V_{u^l d^m}^* V_{u^l d^n}\left(\bar d^m_L \gamma^\mu d_L^n\right) \approx \frac{m_{t}^2}{v^2} V_{t d^m}^* V_{t d^n} \left(\bar d^m_L \gamma^\mu d_L^n\right) \,, 
 \label{eq:FCNCcase}
\end{equation}
which is different from Eq.~\eqref{eq:chargedYuYu} and generates a considerable effect on the FCNC process like $b \to s\ell\ell$. 
This is the original idea to introduce the MFV hypothesis to constrain the coefficients $b_{k\neq0}^{(q)}$. 
Note also that the term with $b_0^{(q)}$ of Eq.~\eqref{eq:cvMFV}, while producing the leading contribution to the semi-leptonic $b \to ql\nu$ transitions, does not induce the tree-level FCNC interactions.

\subsection{Third generation-philic flavor structure}

The $U(2)^5$ approach is somewhat similar to what has been discussed for the $U(3)^5$ flavor symmetry, with the only difference that we are now requiring the SMEFT Lagrangian to be invariant under the smaller group~\cite{Barbieri:2011ci,Barbieri:2012uh,Blankenburg:2012nx}  
\begin{equation} \label{eq:u(5)symmetry}
U(2)^5 = U(2)_Q \otimes U(2)_U \otimes U(2)_D \otimes U(2)_L \otimes U(2)_E.
\end{equation}
By construction, the symmetry acts on each of the five SM fermion species with different gauge quantum numbers, as indicated by the subscripts in Eq.~\eqref{eq:u(5)symmetry}. Within each of these species, the first two generations transform as a doublet, while the third one as a singlet, under the corresponding $U(2)$ subgroup. Explicitly, the $U(2)^5$ representations of the five SM fermion species can be written, respectively, as
\begin{equation}
Q = \left[\begin{array}{l} Q^a \sim \left(\bf 2, \bf 1, \bf 1, \bf 1, \bf 1\right) \\[0.2cm] 
                           Q^3 \sim \left(\bf 1, \bf 1, \bf 1, \bf 1, \bf 1\right) 
    \end{array}\right],\qquad 
L = \left[\begin{array}{l} L^a \sim \left(\bf 1, \bf 1, \bf 1, \bf 2, \bf 1\right) \\[0.2cm] 
                           L^3 \sim \left(\bf 1, \bf 1, \bf 1, \bf 1, \bf 1\right) 
    \end{array}\right], 
\end{equation}
and
\begin{equation}
U = \left[\begin{array}{l} U^a \sim \left(\bf 1, \bf 2, \bf 1, \bf 1, \bf 1\right) \\[0.2cm]
                    U^3 \sim \left(\bf 1, \bf 1, \bf 1, \bf 1, \bf 1\right) 
    \end{array}\right],
D = \left[\begin{array}{l} D^a \sim \left (\bf 1, \bf 1, \bf 2, \bf 1, \bf 1\right) \\[0.2 cm]
                           D^3 \sim \left(\bf 1, \bf 1, \bf 1, \bf 1, \bf 1\right)
    \end{array}\right],
E = \left[\begin{array}{l} E^a \sim \left(\bf 1, \bf 1, \bf 1, \bf 1, \bf 2\right) \\[0.2cm]
                           E^3 \sim \left(\bf 1, \bf 1, \bf 1, \bf 1, \bf 1\right)
    \end{array}\right],
\end{equation}
with the flavor index $a = 1,2$. To obtain the $U(2)^5$ invariant subset from the full SMEFT basis of Eq.~\eqref{eq:SMEFT_dim_6}, let us introduce the minimal set of spurions needed to reproduce the observed SM fermion masses and mixings~\cite{Faroughy:2020ina,Greljo:2022cah}
\begin{align}
V_Q &\sim \left(\bf 2, \bf 1, \bf 1, \bf 1, \bf 1\right), \qquad & V_L &\sim \left(\bf 1, \bf 1, \bf 1, \bf 2, \bf 1\right), & \notag \\[0.2cm]
\Delta_U &\sim \left(\bf 2, \bf \bar 2, \bf 1, \bf 1, \bf 1\right), \qquad & \Delta_D &\sim \left(\bf 2, \bf 1, \bf \bar 2, \bf 1, \bf 1\right), \qquad & \Delta_E \sim \left(\bf 1, \bf 1, \bf 1, \bf 2, \bf \bar 2\right),
\end{align}
where $V_{Q}$ and $V_{L}$ are two-dimensional complex vectors, while $\Delta_U$, $\Delta_D$ and $\Delta_E$ are $2\times 2$ complex matrices. In terms of these spurions, the $3\times 3$ Yukawa couplings $Y_{U,D,E}$ can be decomposed, respectively, as~\cite{Faroughy:2020ina,Greljo:2022cah}
\begin{equation}
Y_{U}=y_{t}\left(\begin{array}{cc} \Delta_{U} & x_{t} V_Q \\[0.15cm]
                                            0 & 1
                   \end{array}\right), \quad 
Y_{D}=y_{b}\left(\begin{array}{cc} \Delta_{D} & x_{b} V_Q \\[0.15cm]
                                            0 & 1
                   \end{array}\right), \quad 
Y_E=y_\tau\left(\begin{array}{cc} \Delta_E & x_\tau V_{L} \\[0.15cm]
                                         0 & 1
                   \end{array}\right),
\end{equation}
where $x_{t,b,\tau}$ and $y_{t,b,\tau}$ are free complex parameters that are all expected to be of $\mathcal{O}(1)$. Unlike the case of $U(3)^5$ symmetry, the spurions introduced here cannot be completely determined in terms of the fermion masses and mixings, and we can only estimate their sizes by requiring no fine-tuning in these $\mathcal{O}(1)$ parameters. This implies that $|V_{Q,L}|=\mathcal{O}(10^{-1})$, while the entries in $\Delta_{U,D,E}$ are significantly smaller than $|V_{Q,L}|$, with a maximal size of $\mathcal{O}(10^{-2})$ in the quark sector~\cite{Fuentes-Martin:2019mun}.\footnote{It should be noted that, in the limit of vanishing neutrino masses, the size of $|V_{L}|$ cannot be unambiguously determined, and the SM lepton Yukawa coupling $Y_E$ can even be reproduced by setting $V_{L}=0$. On the other hand, assuming a common structure for the three Yukawa couplings, as suggested by the similar hierarchies observed in their eigenvalues, it is natural to assume that $|V_{Q}|\sim |V_{L}|$. This in turn ensures the hypothesis of a common origin for the two leading $U(2)^5$ breaking terms in the quark and lepton sectors~\cite{Isidori:2023pyp,Fuentes-Martin:2019mun}.} 

If we consider the insertion of at most one power of the leading spurion $V_Q$ and one power of the leading spurion $V_L$ while neglecting all the sub-leading spurions $\Delta_{U,D,E}$, the $U(2)^5$ invariant form of the third term on the right-hand side of Eq.~\eqref{eq:SMEFT_dim_6} can then be written as~\cite{Fuentes-Martin:2019mun}
\begin{equation}
c_{V}^{\prime 0} \big(\Gamma_L^\dagger\big)^{in} \big(\Gamma_L\big)^{mj} \left(\Bar{Q}^m \gamma^\mu \tau^I Q^n\right) \left(\Bar{L}^i \gamma_\mu \tau^I L^j\right),
\end{equation}
where $c_{V}^{\prime 0}$ determines the overall strength of the NP effect, and the flavor structure is parametrized by the product of two factorized tensors, $\big(\Gamma_L^\dagger\big)^{in} \big(\Gamma_L\big)^{mj}$. They are normalized by $\big(\Gamma_L^\dagger\big)^{33} \big(\Gamma_L\big)^{33}=1$, which is the only term surviving in the exact $U(2)^5$ symmetric limit. The explicit expression of the tensor $\Gamma_L^{mj}$ is given in the interaction basis by~\cite{Fuentes-Martin:2019mun}
\begin{equation} \label{eq:Gamma_L}
    \Gamma_L^{mj} = \left(\begin{array}{cc} x_{QL} V_Q^m \big(V_L^j\big)^\ast & x_{Q} V_Q^m \\[0.2cm]
                                            x_L \big(V_L^j\big)^\ast & 1
                            \end{array}\right),
\end{equation}
where $x_{Q}$, $x_{L}$ and $x_{QL}$ are $\mathcal{O}(1)$ coefficients and all the higher-order terms in $V_{Q,L}$ have been neglected. Transforming to the mass eigenstate bases of down-type quarks and charged leptons, with 
\begin{equation}
    Q_L^m = \left(\begin{array}{c} V_{nm}^\ast u_L^n \\[0.15cm]
                                   d_L^m
                            \end{array}\right), \qquad
    L_L^j = \left(\begin{array}{c} \nu_L^j \\[0.15cm]
                                     e_L^j
                            \end{array}\right),                            
\end{equation} 
we have $\Gamma_L \to \hat{\Gamma}_L = W_d^\dagger \Gamma_L W_e$, where $W_d$ and $W_e$ are the unitary transformations introduced to diagonalize the Yukawa couplings $Y_D$ and $Y_E$, respectively. Making use of the general forms of $W_d$ and $W_e$ given in Ref.~\cite{Fuentes-Martin:2019mun} and neglecting the tiny terms suppressed by more than two powers of $V_{Q,L}$, we have~\cite{Fuentes-Martin:2019mun}  
\begin{equation}
    \hat{\Gamma}_L = \text{e}^{i\phi_Q}
    \left(\begin{array}{ccc} \Delta_{QL}^{de} & \Delta_{QL}^{d\mu} & \lambda_Q^d \\[0.15cm]
                             \Delta_{QL}^{se} & \Delta_{QL}^{s\mu} & \lambda_Q^s \\[0.15cm]
                             \lambda_L^e & \lambda_L^\mu & x_{QL}^{b\tau}
            \end{array}\right)
                  \approx \text{e}^{i\phi_Q}
    \left(\begin{array}{ccc} 0 & 0 & \lambda_Q^d \\[0.15cm]
                             0 & \Delta_{QL}^{s\mu} & \lambda_Q^s \\[0.15cm]
                             \lambda_L^e & \lambda_L^\mu & 1
            \end{array}\right),
\end{equation}
with $x_{QL}^{b\tau} = \mathcal{O}(1)$, $\lambda_{Q}^s = \mathcal{O}(|V_Q|)$, $\lambda_{L}^\mu = \mathcal{O}(|V_{L}|)$, $\Delta_{QL}^{s\mu} = \mathcal{O}(\lambda_Q^s \lambda_{L}^\mu)$, and 
\begin{equation}
\frac{\lambda_Q^d}{\lambda_Q^s} = \frac{\Delta_{QL}^{de}}{\Delta_{QL}^{se}} = \frac{\Delta_{QL}^{d\mu}}{\Delta_{QL}^{s\mu}} = \frac{V_{t d}^\ast}{V_{t s}^\ast}, \qquad \frac{\lambda_{L}^e}{\lambda_{L}^\mu} = \frac{\Delta_{QL}^{s e}}{\Delta_{QL}^{s \mu}} = \frac{\Delta_{QL}^{d e}}{\Delta_{QL}^{d \mu}} = s_e,
\end{equation}
where $s_e=\sin\theta_e$ denotes the sine of the mixing angle between the first two generations in lepton sector.

It is now ready to obtain the strength governing the $b \to q \tau \nu$ transition due to the third term on the right-hand side of Eq.~\eqref{eq:SMEFT_dim_6},
\begin{align}
C_{V_L}^{q\tau} & = - \frac{v^2}{\Lambda^2} c_{V}^{\prime0} \left[1 + \lambda_Q^s \left(\frac{V_{q s}}{V_{q b}} + \frac{V_{q d}}{V_{q b}} \frac{V_{t d}^*}{V_{t s}^*}\right)\right] \\[0.2cm]
& =- \frac{v^2}{\Lambda^2} c_{V}^{\prime0} \left(1-\lambda_Q^s \frac{V_{t b}^*}{V_{t s}^*}\right), \label{eq:U2(5)_CXuc}
\end{align}
where the second line results from the unitarity of the CKM matrix. It is of particular interest to note that, relative to the SM, the $b \to u$ and $b \to c$ transitions share the same size of NP effects, which is a distinctive feature of the minimally broken $U(2)^5$ hypothesis~\cite{Fuentes-Martin:2019mun}. On the other hand, the flavor diagonal strengths $C_{V_L}^{mm33}$ governing the $q\bar{q} \to \tau^+ \tau^-$ processes display the following hierarchical structures:
\begin{equation}
    c_{V}^{1133} = |\lambda_Q^d|^2 c_V^{\prime 0} \quad\leq\quad c_{V}^{2233} = |\lambda_Q^s|^2 c_V^{\prime 0} \quad\ll\quad c_{V}^{3333} = c_V^{\prime 0}.
\end{equation}
Therefore, the LHC bounds given by Eqs.~\eqref{eq:Clq3311} and \eqref{eq:Clq3322}, once transformed into the constraint on $c_V^{\prime 0}$, can be safely avoided in the SMEFT with a minimally broken $U(2)^5$ hypothesis. 
Hence, the flavor structure of Eq.~\eqref{eq:3philic} can be regarded as a good approximation of the exact $U(2)^5$ hypothesis.

Similarly, we can apply the same setup to the remaining three terms in the third and fourth lines of Eq.~\eqref{eq:SMEFT_dim_6}. Guided by the
$U(2)^5$ flavor symmetry and considering at most one power of $V_{Q}$ and one power of $V_{L}$, we can decompose the flavor structures of these three terms as
\begin{align}
& c_{S_d}^{\prime 0} \big(\Gamma_L^\dagger\big)^{in} \big(\Gamma_R\big)^{mj} \left(\bar{L}^i E^j\right)\left(\bar{D}^m Q^n\right), \\[0.5em]
& c_{S_u}^{\prime 0} \big(\Gamma_L^\dagger\big)^{im} \big(\Gamma_R\big)^{nj} \left(\bar{L}^{a,i} E^j\right) \epsilon_{ab} \left(\bar{Q}^{b,m} U^n\right), \\[0.5em]
& c_T^{\prime 0} \big(\Gamma_L^\dagger\big)^{im} \big(\Gamma_R\big)^{nj} \left(\bar{L}^{a,i} \sigma_{\mu\nu} E^j\right) \epsilon_{ab} \left(\bar{Q}^{b,m} \sigma^{\mu\nu} U^n\right).
\end{align}
Here, $\Gamma_L$ is already defined in Eq.~\eqref{eq:Gamma_L}, with $V_{Q,L}$ given by~\cite{Fuentes-Martin:2019mun} 
\begin{equation}
V_{Q,L} = V_{Q,L}^\ast = \begin{pmatrix} 0 \\ \left|V_{Q,L}\right|
\end{pmatrix},
\end{equation}
while $\Gamma_R$ is defined by 
\begin{equation}
\Gamma_R \equiv \begin{pmatrix} 0 & 0 \\ 0 & 1
\end{pmatrix}.
\end{equation}
The matrices $\Gamma_L$ and $\Gamma_R$ are both given in the weak eigenbasis, and obtained by neglecting the $U(2)_{U,D,E}$ breaking spurions and making the only surviving term in the exact $U(2)^5$ limit normalized to one (\textit{i.e.} $\big(\Gamma_L^\dagger\big)^{33} \big(\Gamma_R\big)^{33}=1$). In addition, the overall strengths of the NP effects for the three terms are controlled by the prefactors $c_{S_d}^{\prime 0}$, $c_{S_u}^{\prime 0}$ and $c_T^{\prime 0}$, respectively. After transforming to the fermion mass eigenbasis, the LEFT WCs governing the charged $b\to q\tau\nu$ transitions are derived, respectively, as
\begin{equation}
C_{S_R}^{q\tau} = - \frac{v^2}{2\Lambda^2} c_{S_u}^{\prime 0} \left(1-\lambda_{Q}^s\frac{V_{tb}^\ast}{V_{ts}^\ast}\right)\,,  \label{eq:CSR_U(2)5}
\end{equation}
and 
\begin{equation}
C_{S_L}^{u\tau} = C_T^{u\tau} \simeq 0, \qquad C_{S_L}^{c\tau} = C_T^{c\tau} \propto m_c/m_t\,, \label{eq:CSL_CT_U(2)5}
\end{equation}
at the electroweak scale. From Eqs.~\eqref{eq:U2(5)_CXuc} and \eqref{eq:CSR_U(2)5}, we can see that the left-handed vector (right-handed scalar) NP contribution has the same size in both $b\to u\tau\nu$ and $b\to c\tau\nu$ transitions, even though the original flavor structures of the corresponding operators may be independent from each other. For the left-handed scalar and tensor types of NP effects, on the other hand, we can find from Eq.~\eqref{eq:CSL_CT_U(2)5} that their contributions to the $b\to u\tau\nu$ and $b\to c\tau\nu$ transitions show an obviously hierarchical structure, being proportional to the light up-type quark masses, although their effects are practically negligible. These relations among the LEFT WCs coincide with what we have derived in the SMEFT with MFV hypothesis, and they hold not only at the electroweak but also at the $m_b$ scale when we consider the renormalization group running effects.

\section{Covariance tables for our results}
\label{app:covariancetable}

\begin{table}[t]
\renewcommand{\arraystretch}{1.5}
\centering
\resizebox{0.7\textwidth}{!}{
\begin{tabular}{c|ccccc} 
 \hline
 Covariance  & $R_D^\text{SM}$ & $a_D^{SS}$ & $a_D^{TT}$ & $a_D^{VS}$ & $a_D^{VT}$ \\ \hline
 $R_D^\text{SM}$ & $0.000057$ & $0.000033$ & $0.000081$ & $0.000028$ & $0.000058$ \\
 $a_D^{SS}$ & $0.000033$ & $0.000030$ & $0.000061$ & $0.000031$ & $0.000037$ \\
 $a_D^{TT}$ & $0.000081$ & $0.000061$ & $0.116347$ & $0.000054$ & $0.079300$ \\
 $a_D^{VS}$ & $0.000028$ & $0.000031$ & $0.000054$ & $0.000033$ & $0.000030$ \\
 $a_D^{VT}$ & $0.000058$ & $0.000037$ & $0.079300$ & $0.000030$ & $0.054057$ \\ \hline
 Central value & $0.3015$ & $1.0698$ & $0.7215$ & $1.5276$ & $1.0146$ \\ \hline
\end{tabular}
}
\caption{
Covariance and central values for the ratio $R_D$ corresponding to Eq.~\eqref{eq:RDform}. 
\label{tab:covaRD}
}
\end{table}

\begin{table}[t]
\renewcommand{\arraystretch}{1.5}
\centering
\begin{adjustbox}{width=0.99\columnwidth,center}
\begin{tabular}{c|ccccccc}
 \hline
 Covariance & $R_{D^*}^\text{SM}$ & $a_{D^*}^{SS}$ & $a_{D^*}^{TT}$ & $a_{D^*}^{V_LV_R}$ & $a_{D^*}^{VS}$ & $a_{D^*}^{V_LT}$ & $a_{D^*}^{V_RT}$ \\ \hline
 $R_{D^*}^\text{SM}$ & $\phantom{-}0.000030$ & $-0.000003$ & $\phantom{-}0.001526$ & $-0.000012$ & $\phantom{-}0.000009$ & $-0.000292$ & $\phantom{-}0.000302$ \\
 $a_{D^*}^{SS}$ & $-0.000003$ & $\phantom{-}0.000002$ & $-0.000443$ & $\phantom{-}0.000001$ & $-0.000007$ & $\phantom{-}0.000094$ & $-0.000110$ \\
 $a_{D^*}^{TT}$ & $\phantom{-}0.001526$ & $-0.000443$ & $\phantom{-}87.784597$ & $\phantom{-}0.001661$ & $\phantom{-}0.001310$ & $-15.344725$ & $\phantom{-}18.095736$ \\
 $a_{D^*}^{V_LV_R}$ & $-0.000012$ & $\phantom{-}0.000001$ & $\phantom{-}0.001661$ & $\phantom{-}0.000230$ & $-0.000003$ & $\phantom{-}0.000403$ & $\phantom{-}0.000612$ \\
 $a_{D^*}^{VS}$ & $\phantom{-}0.000009$ & $-0.000007$ & $\phantom{-}0.001310$ & $-0.000003$ & $\phantom{-}0.000019$ & $-0.000273$ & $\phantom{-}0.000320$ \\
 $a_{D^*}^{V_LT}$ & $-0.000292$ & $\phantom{-}0.000094$ & $-15.344725$ & $\phantom{-}0.000403$ & $-0.000273$ & $\phantom{-}2.949721$ & $-3.269753$ \\
 $a_{D^*}^{V_RT}$ & $\phantom{-}0.000302$ & $-0.000110$ & $\phantom{-}18.095736$ & $\phantom{-}0.000612$ & $\phantom{-}0.000320$ & $-3.269753$ & $\phantom{-}3.778571$ \\ \hline
 Central value & $\phantom{-}0.2575$ & $\phantom{-}0.0427$ & $\phantom{-}17.7607$ & $-1.7971$ & $-0.1130$ & $-5.4698$ & $\phantom{-}7.0979$ \\ \hline
\end{tabular}
\end{adjustbox}
\caption{
Covariance and central values for the ratio $R_{D^\ast}$ corresponding to Eq.~\eqref{eq:RDstform}. 
\label{tab:covaRDst}
}
\end{table}

In Tables~\ref{tab:covaRD} and \ref{tab:covaRDst}, we provide covariance for all the coefficients of our numerical formulae for $R_D$ and $R_{D^*}$ along with their central values.

Here, we also give instructions on how to estimate the uncertainties of the ratio $R_{\Lambda_c}$ and the shift factor $\delta_{\Lambda_c}$. 
Besides the statistical uncertainty, we have also followed the recommended procedure of Ref.~\cite{Detmold:2015aaa} to calculate the systematic uncertainty from the $\Lambda_b \to \Lambda_c$ form factor inputs. 
The total uncertainty is then obtained by adding the statistical and systematic ones in quadrature. 
The central values and statistical error covariance are obtained by their fit analysis for the $\Lambda_b \to \Lambda_c$ form factors, called the normal-order (NO) fit, 
which corresponds to the regular central values $c_a$ and errors $\sigma_a$ for each observable $a$. 
In addition, a similar fit analysis but with a different setup for the form factor parameters, called the higher-order (HO) fit, is also performed, which gives $c_a^\text{HO}$ and $\sigma_a^\text{HO}$ for the same observable $a$. 
Then, what they called the ``systematic'' error is defined as 
\begin{align}
 \sigma_a^\text{sys} = \textrm{max}\left[|c_a- c_a^\text{HO}|, \sqrt{ |\sigma_a^{2} - (\sigma_a^\text{HO})^2 | } \right] \,, 
 \label{eq:syserror}
\end{align}
and the total error is taken as $\sqrt{\sigma_a^2+(\sigma_a^\text{sys})^2}$. 
This is how we evaluate the uncertainties of $R_{\Lambda_c}$ and $\delta_{\Lambda_c}$ and also all the numerical results relevant to the $\Lambda_b \to \Lambda_c$ form factors.
In Tables~\ref{tab:covaRLc} and \ref{tab:covaSr}, we show the covariance tables of $R_{\Lambda_c}$ and $\delta_{\Lambda_c}$ for the form factor inputs from the NO and HO fit results. 
Note that $a_\text{SR}^{SS}$, $a_\text{SR}^{S_LS_R}$, $a_\text{SR}^{TT}$, $a_\text{SR}^{V_RT}$, $a_\text{SR}^{V_LV_R}$, $a_\text{SR}^{V_LT}$, and $a_\text{SR}^{VS}$ correspond to the coefficients of 
$|C_{S_L}^{c\tau}|^2 + |C_{S_R}^{c\tau}|^2$, 
$\textrm{Re}\left( C_{S_L}^{c\tau} C_{S_R}^{c\tau\ast} \right)$,
$|C_T^{c\tau}|^2$, 
$ \textrm{Re}\left( C_{V_R}^{c\tau} C_{T}^{c\tau\ast} \right)$, 
$\textrm{Re} \left[ \left(1 +C_{V_L}^{c\tau} \right)C_{V_R}^{c\tau\ast} \right]$, 
$\textrm{Re} \left[ \left(1 +C_{V_L}^{c\tau} \right)C_T^{c\tau\ast} \right]$, and
$\textrm{Re} \left[ \left(1 +C_{V_L}^{c\tau} \right) C_{S_R}^{c\tau\ast} + C_{S_L}^{c\tau} C_{V_R}^{c\tau\ast} \right]$ in Eq.~\eqref{eq:delta_Lc}, respectively. 
One can utilize these tables to take into account the correlation among the different terms of our numerical formulae for $R_{\Lambda_c}$ and $\delta_{\Lambda_c}$, with which the total uncertainties are finally obtained by the above prescription.

\begin{table}[t]
\renewcommand{\arraystretch}{1.5}
\centering
\resizebox{\textwidth}{!}{
\begin{tabular}{c|ccccccccc} 
 \hline
 Covariance (NO) & $R_{\Lambda_c}^\text{SM}$ & $a_{\Lambda_c}^{SS}$ & $a_{\Lambda_c}^{TT}$ & $a_{\Lambda_c}^{V_LV_R}$ & $a_{\Lambda_c}^{S_LS_R}$ & $a_{\Lambda_c}^{VS_1}$ & $a_{\Lambda_c}^{VS_2}$ & $a_{\Lambda_c}^{V_LT}$ & $a_{\Lambda_c}^{V_RT}$ \\ \hline
 $R_{\Lambda_c}^\text{SM}$ & $\phantom{-}0.000054$ & $\phantom{-}0.000005$ & $\phantom{-}0.000372$ & $-0.000089$ & $\phantom{-}0.000010$ & $-0.000004$ & $-0.000008$ & $-0.000220$ & $\phantom{-}0.000135$ \\
 $a_{\Lambda_c}^{SS}$ & $\phantom{-}0.000005$ & $\phantom{-}0.000025$ & $-0.000014$ & $\phantom{-}0.000086$ & $\phantom{-}0.000053$ & $\phantom{-}0.000040$ & $\phantom{-}0.000035$ & $\phantom{-}0.000091$ & $-0.000064$ \\
 $a_{\Lambda_c}^{TT}$ & $\phantom{-}0.000372$ & $-0.000014$ & $\phantom{-}0.042716$ & $-0.000818$ & $\phantom{-}0.000012$ & $-0.000075$ & $-0.000144$ & $-0.005770$ & $\phantom{-}0.010311$ \\
 $a_{\Lambda_c}^{V_LV_R}$ & $-0.000089$ & $\phantom{-}0.000086$ & $-0.000818$ & $\phantom{-}0.000745$ & $\phantom{-}0.000186$ & $\phantom{-}0.000175$ & $\phantom{-}0.000158$ & $\phantom{-}0.001324$ & $-0.000572$ \\
 $a_{\Lambda_c}^{S_LS_R}$ & $\phantom{-}0.000010$ & $\phantom{-}0.000053$ & $\phantom{-}0.000012$ & $\phantom{-}0.000186$ & $\phantom{-}0.000123$ & $\phantom{-}0.000096$ & $\phantom{-}0.000074$ & $\phantom{-}0.000153$ & $-0.000117$ \\
 $a_{\Lambda_c}^{VS_1}$ & $-0.000004$ & $\phantom{-}0.000040$ & $-0.000075$ & $\phantom{-}0.000175$ & $\phantom{-}0.000096$ & $\phantom{-}0.000080$ & $\phantom{-}0.000060$ & $\phantom{-}0.000174$ & $-0.000125$ \\
 $a_{\Lambda_c}^{VS_2}$ & $-0.000008$ & $\phantom{-}0.000035$ & $-0.000144$ & $\phantom{-}0.000158$ & $\phantom{-}0.000074$ & $\phantom{-}0.000060$ & $\phantom{-}0.000056$ & $\phantom{-}0.000217$ & $-0.000145$ \\
 $a_{\Lambda_c}^{V_LT}$ & $-0.000220$ & $\phantom{-}0.000091$ & $-0.005770$ & $\phantom{-}0.001324$ & $\phantom{-}0.000153$ & $\phantom{-}0.000174$ & $\phantom{-}0.000217$ & $\phantom{-}0.004173$ & $-0.001990$ \\
 $a_{\Lambda_c}^{V_RT}$ & $\phantom{-}0.000135$ & $-0.000064$ & $\phantom{-}0.010311$ & $-0.000572$ & $-0.000117$ & $-0.000125$ & $-0.000145$ & $-0.001990$ & $\phantom{-}0.002859$ \\ \hline
 Central value (NO) & $\phantom{-}0.3318$ & $\phantom{-}0.3214$ & $\phantom{-}10.4377$ & $-0.7218$ & $\phantom{-}0.5134$ & $\phantom{-}0.3339$ & $\phantom{-}0.4968$ & $-3.1096$ & $\phantom{-}4.8800$ \\ \hline\hline
 Covariance (HO) & $R_{\Lambda_c}^\text{SM}$ & $a_{\Lambda_c}^{SS}$ & $a_{\Lambda_c}^{TT}$ & $a_{\Lambda_c}^{V_LV_R}$ & $a_{\Lambda_c}^{S_LS_R}$ & $a_{\Lambda_c}^{VS_1}$ & $a_{\Lambda_c}^{VS_2}$ & $a_{\Lambda_c}^{V_LT}$ & $a_{\Lambda_c}^{V_RT}$ \\ \hline
 $R_{\Lambda_c}^\text{SM}$ & $\phantom{-}0.000104$ & $\phantom{-}0.000019$ & $\phantom{-}0.001093$ & $-0.000106$ & $\phantom{-}0.000040$ & $\phantom{-}0.000011$ & $\phantom{-}0.000003$ & $-0.000368$ & $\phantom{-}0.000321$ \\
 $a_{\Lambda_c}^{SS}$ & $\phantom{-}0.000019$ & $\phantom{-}0.000200$ & $\phantom{-}0.001571$ & $\phantom{-}0.000810$ & $\phantom{-}0.000427$ & $\phantom{-}0.000322$ & $\phantom{-}0.000286$ & $\phantom{-}0.000640$ & $-0.000066$ \\
 $a_{\Lambda_c}^{TT}$ & $\phantom{-}0.001093$ & $\phantom{-}0.001571$ & $\phantom{-}0.868537$ & $\phantom{-}0.005756$ & $\phantom{-}0.003359$ & $\phantom{-}0.002285$ & $\phantom{-}0.001933$ & $-0.156204$ & $\phantom{-}0.202509$ \\
 $a_{\Lambda_c}^{V_LV_R}$ & $-0.000106$ & $\phantom{-}0.000810$ & $\phantom{-}0.005756$ & $\phantom{-}0.004394$ & $\phantom{-}0.001778$ & $\phantom{-}0.001398$ & $\phantom{-}0.001195$ & $\phantom{-}0.004539$ & $-0.000638$ \\
 $a_{\Lambda_c}^{S_LS_R}$ & $\phantom{-}0.000040$ & $\phantom{-}0.000427$ & $\phantom{-}0.003359$ & $\phantom{-}0.001778$ & $\phantom{-}0.000950$ & $\phantom{-}0.000726$ & $\phantom{-}0.000601$ & $\phantom{-}0.001314$ & $-0.000084$ \\
 $a_{\Lambda_c}^{VS_1}$ & $\phantom{-}0.000011$ & $\phantom{-}0.000322$ & $\phantom{-}0.002285$ & $\phantom{-}0.001398$ & $\phantom{-}0.000726$ & $\phantom{-}0.000565$ & $\phantom{-}0.000457$ & $\phantom{-}0.001077$ & $-0.000137$ \\
 $a_{\Lambda_c}^{VS_2}$ & $\phantom{-}0.000003$ & $\phantom{-}0.000286$ & $\phantom{-}0.001933$ & $\phantom{-}0.001195$ & $\phantom{-}0.000601$ & $\phantom{-}0.000457$ & $\phantom{-}0.000421$ & $\phantom{-}0.001047$ & $-0.000221$ \\
 $a_{\Lambda_c}^{V_LT}$ & $-0.000368$ & $\phantom{-}0.000640$ & $-0.156204$ & $\phantom{-}0.004539$ & $\phantom{-}0.001314$ & $\phantom{-}0.001077$ & $\phantom{-}0.001047$ & $\phantom{-}0.048578$ & $-0.040611$ \\
 $a_{\Lambda_c}^{V_RT}$ & $\phantom{-}0.000321$ & $-0.000066$ & $\phantom{-}0.202509$ & $-0.000638$ & $-0.000084$ & $-0.000137$ & $-0.000221$ & $-0.040611$ & $\phantom{-}0.049201$ \\ \hline 
 Central value (HO) & $\phantom{-}0.3284$ & $\phantom{-}0.3193$ & $\phantom{-}10.6645$ & $-0.7157$ & $\phantom{-}0.5085$ & $\phantom{-}0.3309$ & $\phantom{-}0.4946$ & $-3.0972$ & $\phantom{-}4.9345$ \\ \hline
\end{tabular}
}
\caption{
 Covariance and central values for the ratio $R_{\Lambda_c}$ corresponding to Eq.~\eqref{eq:RLcform} for the NO (upper panel) and HO (lower panel) parameter fits. 
 The systematic errors are evaluated by means of Eq.~\eqref{eq:syserror} based on Ref.~\cite{Detmold:2015aaa}.
\label{tab:covaRLc}
}
\end{table}

\begin{table}[t]
\renewcommand{\arraystretch}{1.5}
\resizebox{\textwidth}{!}{
\begin{tabular}{c|ccccccc}
 \hline
 Covariance (NO)
 & $a_\text{SR}^{SS}$ 
 & $a_\text{SR}^{S_LS_R}$ 
 & $a_\text{SR}^{TT}$
 & $a_\text{SR}^{V_RT}$ 
 & $a_\text{SR}^{V_LV_R}$ 
 & $a_\text{SR}^{V_LT}$
 & $a_\text{SR}^{VS}$ \\ \hline
 $a_\text{SR}^{SS}$ & $\phantom{-}0.000015$ & $\phantom{-}0.000012$ & $-0.000775$ & $-0.000209$ & $\phantom{-}0.000021$ & $\phantom{-}0.000186$ & $\phantom{-}0.000024$ \\
 $a_\text{SR}^{S_LS_R}$ & $\phantom{-}0.000012$ & $\phantom{-}0.000014$ & $-0.000430$ & $-0.000103$ & $-0.000012$ & $\phantom{-}0.000062$ & $\phantom{-}0.000014$ \\
 $a_\text{SR}^{TT}$ & $-0.000775$ & $-0.000430$ & $\phantom{-}46.545257$ & $\phantom{-}9.602331$ & $-0.002067$ & $-8.128861$ & $-0.001652$ \\
 $a_\text{SR}^{V_RT}$ & $-0.000209$ & $-0.000103$ & $\phantom{-}9.602331$ & $\phantom{-}2.008910$ & $-0.000567$ & $-1.730379$ & $-0.000456$ \\
 $a_\text{SR}^{V_LV_R}$ & $\phantom{-}0.000021$ & $-0.000012$ & $-0.002067$ & $-0.000567$ & $\phantom{-}0.000568$ & $\phantom{-}0.001674$ & $\phantom{-}0.000075$ \\
 $a_\text{SR}^{V_LT}$ & $\phantom{-}0.000186$ & $\phantom{-}0.000062$ & $-8.128861$ & $-1.730379$ & $\phantom{-}0.001674$ & $\phantom{-}1.570848$ & $\phantom{-}0.000442$ \\
 $a_\text{SR}^{VS}$ & $\phantom{-}0.000024$ & $\phantom{-}0.000014$ & $-0.001652$ & $-0.000456$ & $\phantom{-}0.000075$ & $\phantom{-}0.000442$ & $\phantom{-}0.000047$ \\ \hline
 Central value (NO) & $-0.0011$ & $-0.0075$ & $-2.6808$ & $-0.5606$ & $\phantom{-}0.0408$ & $\phantom{-}0.5936$ & $-0.0017$ \\ \hline\hline
 Covariance (HO)
 & $a_\text{SR}^{SS}$ 
 & $a_\text{SR}^{S_LS_R}$ 
 & $a_\text{SR}^{TT}$
 & $a_\text{SR}^{V_RT}$ 
 & $a_\text{SR}^{V_LV_R}$ 
 & $a_\text{SR}^{V_LT}$
 & $a_\text{SR}^{VS}$ \\ \hline
 $a_\text{SR}^{SS}$ & $\phantom{-}0.000028$ & $\phantom{-}0.000022$ & $-0.000902$ & $-0.000287$ & $\phantom{-}0.000032$ & $\phantom{-}0.000259$ & $\phantom{-}0.000045$ \\
 $a_\text{SR}^{S_LS_R}$ & $\phantom{-}0.000022$ & $\phantom{-}0.000024$ & $-0.000573$ & $-0.000157$ & $-0.000013$ & $\phantom{-}0.000110$ & $\phantom{-}0.000026$ \\
 $a_\text{SR}^{TT}$ & $-0.000902$ & $-0.000573$ & $\phantom{-}47.712307$ & $\phantom{-}9.871231$ & $\phantom{-}0.000090$ & $-8.341200$ & $-0.001835$ \\
 $a_\text{SR}^{V_RT}$ & $-0.000287$ & $-0.000157$ & $\phantom{-}9.871231$ & $\phantom{-}2.072108$ & $-0.000232$ & $-1.781711$ & $-0.000601$ \\
 $a_\text{SR}^{V_LV_R}$ & $\phantom{-}0.000032$ & $-0.000013$ & $\phantom{-}0.000090$ & $-0.000232$ & $\phantom{-}0.001155$ & $\phantom{-}0.002406$ & $\phantom{-}0.000106$ \\
 $a_\text{SR}^{V_LT}$ & $\phantom{-}0.000259$ & $\phantom{-}0.000110$ & $-8.341200$ & $-1.781711$ & $\phantom{-}0.002406$ & $\phantom{-}1.623709$ & $\phantom{-}0.000579$ \\
 $a_\text{SR}^{VS}$ & $\phantom{-}0.000045$ & $\phantom{-}0.000026$ & $-0.001835$ & $-0.000601$ & $\phantom{-}0.000106$ & $\phantom{-}0.000579$ & $\phantom{-}0.000087$ \\ \hline
 Central value (HO) & $-0.0013$ & $-0.0082$ & $-2.4860$ & $-0.5175$ & $\phantom{-}0.0541$ & $\phantom{-}0.6181$ & $-0.0011$ \\ \hline
\end{tabular}
}
\caption{
 Covariance and central values for the shift factor $\delta_{\Lambda_c}$ corresponding to Eq.~\eqref{eq:delta_Lc} for the NO (upper panel) and HO (lower panel) parameter fits. 
 The systematic errors are evaluated by means of Eq.~\eqref{eq:syserror} based on Ref.~\cite{Detmold:2015aaa}. 
\label{tab:covaSr}
}
\end{table}

In principle, we can also provide similar tables for the $b \to u$ case. However, it is not much useful for now because we are still using the preliminary results for the $\Lambda_b \to p$ tensor form factors. In addition, we need more precise fit studies of the $B \to \rho$ form factors as well as the experimental measurements of the ratios $R_\pi$, $R_\rho$, and $R_p$. Otherwise, the correlation among the different NP contributions is less significant. Thus, we do not show these results here.

\FloatBarrier

\bibliographystyle{JHEP}
\bibliography{bulnu}

\providecommand{\href}[2]{#2}\begingroup\raggedright\begin{thebibliography}{100}

\bibitem{Iguro:2024hyk}
S.~Iguro, T.~Kitahara, and R.~Watanabe, {\it {Global fit to
  b\textrightarrow{}c\ensuremath{\tau}\ensuremath{\nu} anomalies as of Spring
  2024}},  {\em Phys. Rev. D} {\bf 110} (2024), no.~7 075005,
  [\href{http://arxiv.org/abs/2405.06062}{{\tt arXiv:2405.06062}}].

\bibitem{Blanke:2018yud}
M.~Blanke, A.~Crivellin, S.~de~Boer, T.~Kitahara, M.~Moscati, U.~Nierste, and
  I.~Ni\v{s}and\v{z}i\'c, {\it {Impact of polarization observables and $ B_c\to
  \tau \nu$ on new physics explanations of the $b\to c \tau \nu$ anomaly}},
  {\em Phys. Rev. D} {\bf 99} (2019), no.~7 075006,
  [\href{http://arxiv.org/abs/1811.09603}{{\tt arXiv:1811.09603}}].

\bibitem{Blanke:2019qrx}
M.~Blanke, A.~Crivellin, T.~Kitahara, M.~Moscati, U.~Nierste, and
  I.~Ni\v{s}and\v{z}i\'c, {\it {Addendum to \textquotedblleft{}Impact of
  polarization observables and $B_c\to \tau \nu$ on new physics explanations of
  the $b\to c \tau \nu$ anomaly''}},
  \href{http://arxiv.org/abs/1905.08253}{{\tt arXiv:1905.08253}}. [Addendum:
  Phys.Rev.D 100, 035035 (2019)].

\bibitem{Fedele:2022iib}
M.~Fedele, M.~Blanke, A.~Crivellin, S.~Iguro, T.~Kitahara, U.~Nierste, and
  R.~Watanabe, {\it {Impact of $\Lambda_b \to \Lambda_c \tau \nu$ measurement
  on new physics in $b\to c l \nu$ transitions}},  {\em Phys. Rev. D} {\bf 107}
  (2023), no.~5 055005, [\href{http://arxiv.org/abs/2211.14172}{{\tt
  arXiv:2211.14172}}].

\bibitem{Endo:2025fke}
M.~Endo, S.~Iguro, S.~Mishima, and R.~Watanabe, {\it {Heavy quark symmetry
  behind $b \to c$ semileptonic sum rule}},
  \href{http://arxiv.org/abs/2501.09382}{{\tt arXiv:2501.09382}}.

\bibitem{ParticleDataGroup:2024cfk}
{\bf Particle Data Group} Collaboration, S.~Navas et~al., {\it {Review of
  particle physics}},  {\em Phys. Rev. D} {\bf 110} (2024), no.~3 030001.

\bibitem{Belle:2015qal}
{\bf Belle} Collaboration, P.~Hamer et~al., {\it {Search for $B^0 \to \pi^-
  \tau^+ \nu_\tau$ with hadronic tagging at Belle}},  {\em Phys. Rev. D} {\bf
  93} (2016), no.~3 032007, [\href{http://arxiv.org/abs/1509.06521}{{\tt
  arXiv:1509.06521}}].

\bibitem{Belle-II:2024xwh}
{\bf Belle-II} Collaboration, I.~Adachi et~al., {\it {Determination of
  $|V_{ub}|$ from simultaneous measurements of untagged $B^0\to\pi^- \ell^+
  \nu_{\ell}$ and $B^+\to\rho^0 \ell^+\nu_{\ell}$ decays}},
  \href{http://arxiv.org/abs/2407.17403}{{\tt arXiv:2407.17403}}.

\bibitem{LHCb:2015eia}
{\bf LHCb} Collaboration, R.~Aaij et~al., {\it {Determination of the quark
  coupling strength $|V_{ub}|$ using baryonic decays}},  {\em Nature Phys.}
  {\bf 11} (2015) 743--747, [\href{http://arxiv.org/abs/1504.01568}{{\tt
  arXiv:1504.01568}}].

\bibitem{LHCb:2018roe}
{\bf LHCb} Collaboration, R.~Aaij et~al., {\it {Physics case for an LHCb
  Upgrade II - Opportunities in flavour physics, and beyond, in the HL-LHC
  era}},  \href{http://arxiv.org/abs/1808.08865}{{\tt arXiv:1808.08865}}.

\bibitem{LHCb:2022ine}
{\bf LHCb} Collaboration, {\it {Future physics potential of LHCb}}, .
  {\url{https://cds.cern.ch/record/2806113/files/LHCbPUB-2022-012.pdf}}.

\bibitem{Belle-II:2018jsg}
{\bf Belle-II} Collaboration, W.~Altmannshofer et~al., {\it {The Belle II
  Physics Book}},  {\em PTEP} {\bf 2019} (2019), no.~12 123C01,
  [\href{http://arxiv.org/abs/1808.10567}{{\tt arXiv:1808.10567}}]. [Erratum:
  PTEP 2020, 029201 (2020)].

\bibitem{Belle-II:2022cgf}
{\bf Belle-II} Collaboration, L.~Aggarwal et~al., {\it {Snowmass White Paper:
  Belle II physics reach and plans for the next decade and beyond}},
  \href{http://arxiv.org/abs/2207.06307}{{\tt arXiv:2207.06307}}.

\bibitem{Buchmuller:1985jz}
W.~Buchmuller and D.~Wyler, {\it {Effective Lagrangian Analysis of New
  Interactions and Flavor Conservation}},  {\em Nucl. Phys. B} {\bf 268} (1986)
  621--653.

\bibitem{Grzadkowski:2010es}
B.~Grzadkowski, M.~Iskrzynski, M.~Misiak, and J.~Rosiek, {\it {Dimension-Six
  Terms in the Standard Model Lagrangian}},  {\em JHEP} {\bf 10} (2010) 085,
  [\href{http://arxiv.org/abs/1008.4884}{{\tt arXiv:1008.4884}}].

\bibitem{Brivio:2017vri}
I.~Brivio and M.~Trott, {\it {The Standard Model as an Effective Field
  Theory}},  {\em Phys. Rept.} {\bf 793} (2019) 1--98,
  [\href{http://arxiv.org/abs/1706.08945}{{\tt arXiv:1706.08945}}].

\bibitem{Isidori:2023pyp}
G.~Isidori, F.~Wilsch, and D.~Wyler, {\it {The standard model effective field
  theory at work}},  {\em Rev. Mod. Phys.} {\bf 96} (2024), no.~1 015006,
  [\href{http://arxiv.org/abs/2303.16922}{{\tt arXiv:2303.16922}}].

\bibitem{Faroughy:2020ina}
D.~A. Faroughy, G.~Isidori, F.~Wilsch, and K.~Yamamoto, {\it {Flavour
  symmetries in the SMEFT}},  {\em JHEP} {\bf 08} (2020) 166,
  [\href{http://arxiv.org/abs/2005.05366}{{\tt arXiv:2005.05366}}].

\bibitem{Greljo:2022cah}
A.~Greljo, A.~Palavri\'c, and A.~E. Thomsen, {\it {Adding Flavor to the
  SMEFT}},  {\em JHEP} {\bf 10} (2022) 010,
  [\href{http://arxiv.org/abs/2203.09561}{{\tt arXiv:2203.09561}}].

\bibitem{Iguro:2018qzf}
S.~Iguro and Y.~Omura, {\it {Status of the semileptonic $B$ decays and muon g-2
  in general 2HDMs with right-handed neutrinos}},  {\em JHEP} {\bf 05} (2018)
  173, [\href{http://arxiv.org/abs/1802.01732}{{\tt arXiv:1802.01732}}].

\bibitem{Robinson:2018gza}
D.~J. Robinson, B.~Shakya, and J.~Zupan, {\it {Right-handed neutrinos and
  $R(D^{(\ast)})$}},  {\em JHEP} {\bf 02} (2019) 119,
  [\href{http://arxiv.org/abs/1807.04753}{{\tt arXiv:1807.04753}}].

\bibitem{Babu:2018vrl}
K.~S. Babu, B.~Dutta, and R.~N. Mohapatra, {\it {A theory of $R(D^{*}, D)$
  anomaly with right-handed currents}},  {\em JHEP} {\bf 01} (2019) 168,
  [\href{http://arxiv.org/abs/1811.04496}{{\tt arXiv:1811.04496}}].

\bibitem{Mandal:2020htr}
R.~Mandal, C.~Murgui, A.~Pe\~nuelas, and A.~Pich, {\it {The role of
  right-handed neutrinos in $b \to c \tau \bar{\nu}$ anomalies}},  {\em JHEP}
  {\bf 08} (2020), no.~08 022, [\href{http://arxiv.org/abs/2004.06726}{{\tt
  arXiv:2004.06726}}].

\bibitem{Penalva:2021wye}
N.~Penalva, E.~Hern\'andez, and J.~Nieves, {\it {The role of right-handed
  neutrinos in $b \rightarrow c \tau(\pi \nu_\tau, \rho \nu_\tau, \mu\bar{\nu}
  _{\mu} \nu\tau) \bar{\nu}_\tau$ from visible final-state kinematics}},  {\em
  JHEP} {\bf 10} (2021) 122, [\href{http://arxiv.org/abs/2107.13406}{{\tt
  arXiv:2107.13406}}].

\bibitem{Datta:2022czw}
A.~Datta, H.~Liu, and D.~Marfatia, {\it {$\bar{B}\rightarrow D^{(*)} \ell
  \bar{X}$ decays in effective field theory with massive right-handed
  neutrinos}},  {\em Phys. Rev. D} {\bf 106} (2022), no.~1 L011702,
  [\href{http://arxiv.org/abs/2204.01818}{{\tt arXiv:2204.01818}}].

\bibitem{Jenkins:2017jig}
E.~E. Jenkins, A.~V. Manohar, and P.~Stoffer, {\it {Low-Energy Effective Field
  Theory below the Electroweak Scale: Operators and Matching}},  {\em JHEP}
  {\bf 03} (2018) 016, [\href{http://arxiv.org/abs/1709.04486}{{\tt
  arXiv:1709.04486}}]. [Erratum: JHEP 12, 043 (2023)].

\bibitem{Aebischer:2015fzz}
J.~Aebischer, A.~Crivellin, M.~Fael, and C.~Greub, {\it {Matching of gauge
  invariant dimension-six operators for $b\to s$ and $b\to c$ transitions}},
  {\em JHEP} {\bf 05} (2016) 037, [\href{http://arxiv.org/abs/1512.02830}{{\tt
  arXiv:1512.02830}}].

\bibitem{Iguro:2020keo}
S.~Iguro, M.~Takeuchi, and R.~Watanabe, {\it {Testing leptoquark/EFT in
  $\bar{B} \to D^{(*)} l \bar{\nu}$ at the LHC}},  {\em Eur. Phys. J. C} {\bf
  81} (2021), no.~5 406, [\href{http://arxiv.org/abs/2011.02486}{{\tt
  arXiv:2011.02486}}].

\bibitem{Greljo:2018tzh}
A.~Greljo, J.~Martin~Camalich, and J.~D. Ruiz-\'Alvarez, {\it {Mono-$\tau$
  Signatures at the LHC Constrain Explanations of $B$-decay Anomalies}},  {\em
  Phys. Rev. Lett.} {\bf 122} (2019), no.~13 131803,
  [\href{http://arxiv.org/abs/1811.07920}{{\tt arXiv:1811.07920}}].

\bibitem{LHCb:2022piu}
{\bf LHCb} Collaboration, R.~Aaij et~al., {\it {Observation of the decay
  $\Lambda_b^0\rightarrow \Lambda_c^+\tau^-\overline{\nu}_{\tau}$}},  {\em
  Phys. Rev. Lett.} {\bf 128} (2022), no.~19 191803,
  [\href{http://arxiv.org/abs/2201.03497}{{\tt arXiv:2201.03497}}].

\bibitem{Bernlochner:2022hyz}
F.~U. Bernlochner, Z.~Ligeti, M.~Papucci, and D.~J. Robinson, {\it
  {Interpreting LHCb's $\Lambda_b\to \Lambda_c\tau\bar\nu$ measurement and
  puzzles in semileptonic $\Lambda_b$ decays}},  {\em Phys. Rev. D} {\bf 107}
  (2023), no.~1 L011502, [\href{http://arxiv.org/abs/2206.11282}{{\tt
  arXiv:2206.11282}}].

\bibitem{HFLAV2024winter}
{\bf HFLAV} Collaboration. {``Preliminary average of $R(D)$ and $R(D^\ast)$ for
  Moriond 2024'' at
  \url{https://hflav-eos.web.cern.ch/hflav-eos/semi/moriond24/html/RDsDsstar/RDRDs.html}}.

\bibitem{Chen:2006nua}
C.-H. Chen and C.-Q. Geng, {\it {Charged Higgs on $B^- \to \tau
  \bar{\nu}_{\tau}$ and $\bar{B} \to P(V) \ell \bar{\nu}_\ell$}},  {\em JHEP}
  {\bf 10} (2006) 053, [\href{http://arxiv.org/abs/hep-ph/0608166}{{\tt
  hep-ph/0608166}}].

\bibitem{Celis:2012dk}
A.~Celis, M.~Jung, X.-Q. Li, and A.~Pich, {\it {Sensitivity to charged scalars
  in $B\to D^{(*)}\tau\nu_\tau$ and $B\to\tau\nu_\tau$ decays}},  {\em JHEP}
  {\bf 01} (2013) 054, [\href{http://arxiv.org/abs/1210.8443}{{\tt
  arXiv:1210.8443}}].

\bibitem{Tanaka:2012nw}
M.~Tanaka and R.~Watanabe, {\it {New physics in the weak interaction of $\bar
  B\to D^{(*)}\tau\bar\nu$}},  {\em Phys. Rev. D} {\bf 87} (2013), no.~3
  034028, [\href{http://arxiv.org/abs/1212.1878}{{\tt arXiv:1212.1878}}].

\bibitem{Fajfer:2012vx}
S.~Fajfer, J.~F. Kamenik, and I.~Nisandzic, {\it {On the $B \to D^* \tau \bar
  \nu_{\tau}$ Sensitivity to New Physics}},  {\em Phys. Rev. D} {\bf 85} (2012)
  094025, [\href{http://arxiv.org/abs/1203.2654}{{\tt arXiv:1203.2654}}].

\bibitem{Sakaki:2013bfa}
Y.~Sakaki, M.~Tanaka, A.~Tayduganov, and R.~Watanabe, {\it {Testing leptoquark
  models in $\bar B \to D^{(*)} \tau \bar\nu$}},  {\em Phys. Rev. D} {\bf 88}
  (2013), no.~9 094012, [\href{http://arxiv.org/abs/1309.0301}{{\tt
  arXiv:1309.0301}}].

\bibitem{Bernlochner:2015mya}
F.~U. Bernlochner, {\it {$B \to \pi \tau \overline \nu_\tau$ decay in the
  context of type II 2HDM}},  {\em Phys. Rev. D} {\bf 92} (2015), no.~11
  115019, [\href{http://arxiv.org/abs/1509.06938}{{\tt arXiv:1509.06938}}].

\bibitem{Tanaka:2016ijq}
M.~Tanaka and R.~Watanabe, {\it {New physics contributions in
  $B\to\pi\tau\bar\nu$ and $B\to\tau\bar\nu$}},  {\em PTEP} {\bf 2017} (2017),
  no.~1 013B05, [\href{http://arxiv.org/abs/1608.05207}{{\tt
  arXiv:1608.05207}}].

\bibitem{Li:2016pdv}
X.-Q. Li, Y.-D. Yang, and X.~Zhang, {\it {$\Lambda_b\to \Lambda_c\tau
  \bar{\nu}_{\tau}$ decay in scalar and vector leptoquark scenarios}},  {\em
  JHEP} {\bf 02} (2017) 068, [\href{http://arxiv.org/abs/1611.01635}{{\tt
  arXiv:1611.01635}}].

\bibitem{Hu:2018veh}
Q.-Y. Hu, X.-Q. Li, and Y.-D. Yang, {\it {$b\to c\tau\nu$ transitions in the
  standard model effective field theory}},  {\em Eur. Phys. J. C} {\bf 79}
  (2019), no.~3 264, [\href{http://arxiv.org/abs/1810.04939}{{\tt
  arXiv:1810.04939}}].

\bibitem{Hu:2020axt}
Q.-Y. Hu, X.-Q. Li, Y.-D. Yang, and D.-H. Zheng, {\it {The measurable angular
  distribution of $ {\Lambda}_b^0\to {\Lambda}_c^{+}\left(\to
  {\Lambda}^0{\pi}^{+}\right){\tau}^{-}\left(\to
  {\pi}^{-}{\nu}_{\tau}\right){\bar{\nu}}_{\tau } $ decay}},  {\em JHEP} {\bf
  02} (2021) 183, [\href{http://arxiv.org/abs/2011.05912}{{\tt
  arXiv:2011.05912}}].

\bibitem{Bernlochner:2021rel}
F.~U. Bernlochner, M.~T. Prim, and D.~J. Robinson, {\it {$B\rightarrow \rho l
  \nu$ and $\omega l \nu$ in and beyond the Standard Model: Improved
  predictions and $|V_{ub}|$}},  {\em Phys. Rev. D} {\bf 104} (2021), no.~3
  034032, [\href{http://arxiv.org/abs/2104.05739}{{\tt arXiv:2104.05739}}].

\bibitem{Biswas:2021cyd}
A.~Biswas and S.~Nandi, {\it {A closer look at observables from exclusive
  semileptonic $B\to(\pi,\rho)\ell\nu_{\ell}$ decays}},  {\em JHEP} {\bf 09}
  (2021) 127, [\href{http://arxiv.org/abs/2105.01732}{{\tt arXiv:2105.01732}}].

\bibitem{Du:2015tda}
D.~Du, A.~X. El-Khadra, S.~Gottlieb, A.~S. Kronfeld, J.~Laiho, E.~Lunghi, R.~S.
  Van~de Water, and R.~Zhou, {\it {Phenomenology of semileptonic B-meson decays
  with form factors from lattice QCD}},  {\em Phys. Rev. D} {\bf 93} (2016),
  no.~3 034005, [\href{http://arxiv.org/abs/1510.02349}{{\tt
  arXiv:1510.02349}}].

\bibitem{Ball:2004rg}
P.~Ball and R.~Zwicky, {\it {$B_{d,s} \to \rho, \omega, K^*, \phi$ decay
  form-factors from light-cone sum rules revisited}},  {\em Phys. Rev. D} {\bf
  71} (2005) 014029, [\href{http://arxiv.org/abs/hep-ph/0412079}{{\tt
  hep-ph/0412079}}].

\bibitem{Bharucha:2015bzk}
A.~Bharucha, D.~M. Straub, and R.~Zwicky, {\it {$B\to V\ell^+\ell^-$ in the
  Standard Model from light-cone sum rules}},  {\em JHEP} {\bf 08} (2016) 098,
  [\href{http://arxiv.org/abs/1503.05534}{{\tt arXiv:1503.05534}}].

\bibitem{Chen:2001zc}
C.-H. Chen and C.~Q. Geng, {\it {Baryonic rare decays of $\Lambda_b \to \Lambda
  \ell^+ \ell^-$}},  {\em Phys. Rev. D} {\bf 64} (2001) 074001,
  [\href{http://arxiv.org/abs/hep-ph/0106193}{{\tt hep-ph/0106193}}].

\bibitem{Feldmann:2011xf}
T.~Feldmann and M.~W.~Y. Yip, {\it {Form factors for $\Lambda_b \to \Lambda$
  transitions in the soft-collinear effective theory}},  {\em Phys. Rev. D}
  {\bf 85} (2012) 014035, [\href{http://arxiv.org/abs/1111.1844}{{\tt
  arXiv:1111.1844}}]. [Erratum: Phys.Rev.D 86, 079901 (2012)].

\bibitem{Detmold:2015aaa}
W.~Detmold, C.~Lehner, and S.~Meinel, {\it {$\Lambda_b \to p \ell^-
  \bar{\nu}_\ell$ and $\Lambda_b \to \Lambda_c \ell^- \bar{\nu}_\ell$ form
  factors from lattice QCD with relativistic heavy quarks}},  {\em Phys. Rev.
  D} {\bf 92} (2015), no.~3 034503,
  [\href{http://arxiv.org/abs/1503.01421}{{\tt arXiv:1503.01421}}].

\bibitem{Detmold:2016pkz}
W.~Detmold and S.~Meinel, {\it {$\Lambda_b \to \Lambda \ell^+ \ell^-$ form
  factors, differential branching fraction, and angular observables from
  lattice QCD with relativistic $b$ quarks}},  {\em Phys. Rev. D} {\bf 93}
  (2016), no.~7 074501, [\href{http://arxiv.org/abs/1602.01399}{{\tt
  arXiv:1602.01399}}].

\bibitem{Caprini:1997mu}
I.~Caprini, L.~Lellouch, and M.~Neubert, {\it {Dispersive bounds on the shape
  of $\bar{B} \to D^{(*)} \ell \bar{\nu}$ form-factors}},  {\em Nucl. Phys. B}
  {\bf 530} (1998) 153--181, [\href{http://arxiv.org/abs/hep-ph/9712417}{{\tt
  hep-ph/9712417}}].

\bibitem{Boyd:1995cf}
C.~G. Boyd, B.~Grinstein, and R.~F. Lebed, {\it {Model independent extraction
  of $|V_{cb}|$ using dispersion relations}},  {\em Phys. Lett. B} {\bf 353}
  (1995) 306--312, [\href{http://arxiv.org/abs/hep-ph/9504235}{{\tt
  hep-ph/9504235}}].

\bibitem{Bourrely:2008za}
C.~Bourrely, I.~Caprini, and L.~Lellouch, {\it {Model-independent description
  of $B \to \pi l \nu$ decays and a determination of $|V_{ub}|$}},  {\em Phys.
  Rev. D} {\bf 79} (2009) 013008, [\href{http://arxiv.org/abs/0807.2722}{{\tt
  arXiv:0807.2722}}]. [Erratum: Phys.Rev.D 82, 099902 (2010)].

\bibitem{Golonka:2005pn}
P.~Golonka and Z.~Was, {\it {PHOTOS Monte Carlo: A Precision tool for QED
  corrections in $Z$ and $W$ decays}},  {\em Eur. Phys. J. C} {\bf 45} (2006)
  97--107, [\href{http://arxiv.org/abs/hep-ph/0506026}{{\tt hep-ph/0506026}}].

\bibitem{deBoer:2018ipi}
S.~de~Boer, T.~Kitahara, and I.~Nisandzic, {\it {Soft-Photon Corrections to
  $\bar{B} \to D \tau^{-} \bar{\nu}_{\tau}$ Relative to $\bar{B} \to D \mu^{-}
  \bar{\nu}_{\mu}$}},  {\em Phys. Rev. Lett.} {\bf 120} (2018), no.~26 261804,
  [\href{http://arxiv.org/abs/1803.05881}{{\tt arXiv:1803.05881}}].

\bibitem{Beneke:2017vpq}
M.~Beneke, C.~Bobeth, and R.~Szafron, {\it {Enhanced electromagnetic correction
  to the rare $B$-meson decay $B_{s,d} \to \mu^+ \mu^-$}},  {\em Phys. Rev.
  Lett.} {\bf 120} (2018), no.~1 011801,
  [\href{http://arxiv.org/abs/1708.09152}{{\tt arXiv:1708.09152}}].

\bibitem{Beneke:2019slt}
M.~Beneke, C.~Bobeth, and R.~Szafron, {\it {Power-enhanced leading-logarithmic
  QED corrections to $B_q \to \mu^+\mu^-$}},  {\em JHEP} {\bf 10} (2019) 232,
  [\href{http://arxiv.org/abs/1908.07011}{{\tt arXiv:1908.07011}}]. [Erratum:
  JHEP 11, 099 (2022)].

\bibitem{Cornella:2022ubo}
C.~Cornella, M.~K\"onig, and M.~Neubert, {\it {Structure-dependent QED effects
  in exclusive B decays at subleading power}},  {\em Phys. Rev. D} {\bf 108}
  (2023), no.~3 L031502, [\href{http://arxiv.org/abs/2212.14430}{{\tt
  arXiv:2212.14430}}].

\bibitem{Boer:2023vsg}
P.~B\"oer and T.~Feldmann, {\it {Structure-dependent QED effects in exclusive
  $B$-meson decays}},  {\em Eur. Phys. J. ST} {\bf 233} (2024), no.~2 299--323,
  [\href{http://arxiv.org/abs/2312.12885}{{\tt arXiv:2312.12885}}].

\bibitem{Rowe:2024jml}
M.~Rowe and R.~Zwicky, {\it {Structure-dependent QED in $ {B}^{-}\to
  {\ell}^{-}\overline{\nu}\left(\gamma \right) $}},  {\em JHEP} {\bf 07} (2024)
  249, [\href{http://arxiv.org/abs/2404.07648}{{\tt arXiv:2404.07648}}].

\bibitem{Rowe:2024pfs}
M.~Rowe, {\em {Structure-dependent quantum electrodynamics in heavy meson
  physics}}.
\newblock PhD thesis, Edinburgh U., 2024.

\bibitem{Isidori:2020acz}
G.~Isidori, S.~Nabeebaccus, and R.~Zwicky, {\it {QED corrections in $
  \overline{B}\to \overline{K}{\mathrm{\ell}}^{+}{\mathrm{\ell}}^{-} $ at the
  double-differential level}},  {\em JHEP} {\bf 12} (2020) 104,
  [\href{http://arxiv.org/abs/2009.00929}{{\tt arXiv:2009.00929}}].

\bibitem{Zwicky:2021olr}
R.~Zwicky, {\it {Notes on QED Corrections in Weak Decays}},  {\em Symmetry}
  {\bf 13} (2021), no.~11 2036, [\href{http://arxiv.org/abs/2205.06194}{{\tt
  arXiv:2205.06194}}].

\bibitem{MILC:2015uhg}
{\bf MILC} Collaboration, J.~A. Bailey et~al., {\it {$B\rightarrow D \ell \nu$
  form factors at nonzero recoil and $|V_{cb}|$ from 2+1-flavor lattice QCD}},
  {\em Phys. Rev. D} {\bf 92} (2015), no.~3 034506,
  [\href{http://arxiv.org/abs/1503.07237}{{\tt arXiv:1503.07237}}].

\bibitem{Na:2015kha}
{\bf HPQCD} Collaboration, H.~Na, C.~M. Bouchard, G.~P. Lepage, C.~Monahan, and
  J.~Shigemitsu, {\it {$B \rightarrow D l \nu$ form factors at nonzero recoil
  and extraction of $|V_{cb}|$}},  {\em Phys. Rev. D} {\bf 92} (2015), no.~5
  054510, [\href{http://arxiv.org/abs/1505.03925}{{\tt arXiv:1505.03925}}].
  [Erratum: Phys.Rev.D 93, 119906 (2016)].

\bibitem{FermilabLattice:2021cdg}
{\bf Fermilab Lattice, MILC, Fermilab Lattice, MILC} Collaboration, A.~Bazavov
  et~al., {\it {Semileptonic form factors for $B\rightarrow D^*\ell \nu $ at
  nonzero recoil from $2+1$-flavor lattice QCD: Fermilab
  Lattice~and~MILC~Collaborations}},  {\em Eur. Phys. J. C} {\bf 82} (2022),
  no.~12 1141, [\href{http://arxiv.org/abs/2105.14019}{{\tt
  arXiv:2105.14019}}]. [Erratum: Eur.Phys.J.C 83, 21 (2023)].

\bibitem{Aoki:2023qpa}
{\bf JLQCD} Collaboration, Y.~Aoki, B.~Colquhoun, H.~Fukaya, S.~Hashimoto,
  T.~Kaneko, R.~Kellermann, J.~Koponen, and E.~Kou, {\it
  {B\textrightarrow{}D*\ensuremath{\ell}\ensuremath{\nu}\ensuremath{\ell}
  semileptonic form factors from lattice QCD with M\"obius domain-wall
  quarks}},  {\em Phys. Rev. D} {\bf 109} (2024), no.~7 074503,
  [\href{http://arxiv.org/abs/2306.05657}{{\tt arXiv:2306.05657}}].

\bibitem{Harrison:2023dzh}
{\bf HPQCD, (HPQCD Collaboration)\textdaggerdbl{}} Collaboration, J.~Harrison
  and C.~T.~H. Davies, {\it {$B\to D^\ast$ and $B_s\to D_s^\ast$ vector,
  axial-vector and tensor form factors for the full $q^2$ range from lattice
  QCD}},  {\em Phys. Rev. D} {\bf 109} (2024), no.~9 094515,
  [\href{http://arxiv.org/abs/2304.03137}{{\tt arXiv:2304.03137}}].

\bibitem{Gubernari:2018wyi}
N.~Gubernari, A.~Kokulu, and D.~van Dyk, {\it {$B\to P$ and $B\to V$ Form
  Factors from $B$-Meson Light-Cone Sum Rules beyond Leading Twist}},  {\em
  JHEP} {\bf 01} (2019) 150, [\href{http://arxiv.org/abs/1811.00983}{{\tt
  arXiv:1811.00983}}].

\bibitem{Cui:2023jiw}
B.-Y. Cui, Y.-K. Huang, Y.-M. Wang, and X.-C. Zhao, {\it {Shedding new light on
  $R(D_{(s)}^{(*)})$ and $|V_{cb}|$ from semileptonic $\bar{B}_{(s)}\rightarrow
  D_{(s)}^{(*)} \ell \bar{\nu}_{\ell}$ decays}},  {\em Phys. Rev. D} {\bf 108}
  (2023), no.~7 L071504, [\href{http://arxiv.org/abs/2301.12391}{{\tt
  arXiv:2301.12391}}].

\bibitem{Datta:2017aue}
A.~Datta, S.~Kamali, S.~Meinel, and A.~Rashed, {\it {Phenomenology of
  ${\Lambda}_b\to {\Lambda}_c\tau {\overline{\nu}}_{\tau } $ using lattice QCD
  calculations}},  {\em JHEP} {\bf 08} (2017) 131,
  [\href{http://arxiv.org/abs/1702.02243}{{\tt arXiv:1702.02243}}].

\bibitem{Kapoor:2024ufg}
T.~Kapoor, Z.-R. Huang, and E.~Kou, {\it {New physics search via angular
  distribution of $B \rightarrow D^* \ell {\nu}_{\ell}$ decay in the light of
  the new lattice data}},  \href{http://arxiv.org/abs/2401.11636}{{\tt
  arXiv:2401.11636}}.

\bibitem{Boyd:1997kz}
C.~G. Boyd, B.~Grinstein, and R.~F. Lebed, {\it {Precision corrections to
  dispersive bounds on form-factors}},  {\em Phys. Rev. D} {\bf 56} (1997)
  6895--6911, [\href{http://arxiv.org/abs/hep-ph/9705252}{{\tt
  hep-ph/9705252}}].

\bibitem{Flynn:2015mha}
J.~M. Flynn, T.~Izubuchi, T.~Kawanai, C.~Lehner, A.~Soni, R.~S. Van~de Water,
  and O.~Witzel, {\it {$B \to \pi \ell \nu$ and $B_s \to K \ell \nu$ form
  factors and $|V_{ub}|$ from 2+1-flavor lattice QCD with domain-wall light
  quarks and relativistic heavy quarks}},  {\em Phys. Rev. D} {\bf 91} (2015),
  no.~7 074510, [\href{http://arxiv.org/abs/1501.05373}{{\tt
  arXiv:1501.05373}}].

\bibitem{FermilabLattice:2015mwy}
{\bf Fermilab Lattice, MILC} Collaboration, J.~A. Bailey et~al., {\it
  {$|V_{ub}|$ from $B\to\pi\ell\nu$ decays and (2+1)-flavor lattice QCD}},
  {\em Phys. Rev. D} {\bf 92} (2015), no.~1 014024,
  [\href{http://arxiv.org/abs/1503.07839}{{\tt arXiv:1503.07839}}].

\bibitem{Colquhoun:2022atw}
{\bf JLQCD} Collaboration, B.~Colquhoun, S.~Hashimoto, T.~Kaneko, and
  J.~Koponen, {\it {Form factors of $B\rightarrow \pi \ell \nu$ and a
  determination of $|V_{ub}|$ with M\"obius domain-wall fermions}},  {\em Phys.
  Rev. D} {\bf 106} (2022), no.~5 054502,
  [\href{http://arxiv.org/abs/2203.04938}{{\tt arXiv:2203.04938}}].

\bibitem{FermilabLattice:2015cdh}
{\bf Fermilab Lattice, MILC} Collaboration, J.~A. Bailey et~al., {\it
  {$B\to\pi\ell\ell$ form factors for new-physics searches from lattice QCD}},
  {\em Phys. Rev. Lett.} {\bf 115} (2015), no.~15 152002,
  [\href{http://arxiv.org/abs/1507.01618}{{\tt arXiv:1507.01618}}].

\bibitem{Ball:2004ye}
P.~Ball and R.~Zwicky, {\it {New results on $B \to \pi, K, \eta$ decay
  formfactors from light-cone sum rules}},  {\em Phys. Rev. D} {\bf 71} (2005)
  014015, [\href{http://arxiv.org/abs/hep-ph/0406232}{{\tt hep-ph/0406232}}].

\bibitem{Duplancic:2008ix}
G.~Duplancic, A.~Khodjamirian, T.~Mannel, B.~Melic, and N.~Offen, {\it
  {Light-cone sum rules for $B \to \pi$ form factors revisited}},  {\em JHEP}
  {\bf 04} (2008) 014, [\href{http://arxiv.org/abs/0801.1796}{{\tt
  arXiv:0801.1796}}].

\bibitem{Wang:2015vgv}
Y.-M. Wang and Y.-L. Shen, {\it {QCD corrections to $B\rightarrow \pi$ form
  factors from light-cone sum rules}},  {\em Nucl. Phys. B} {\bf 898} (2015)
  563--604, [\href{http://arxiv.org/abs/1506.00667}{{\tt arXiv:1506.00667}}].

\bibitem{Khodjamirian:2017fxg}
A.~Khodjamirian and A.~V. Rusov, {\it {$B_{s}\to K \ell \nu_\ell$ and $B_{(s)}
  \to \pi (K) \ell^+\ell^-$ decays at large recoil and CKM matrix elements}},
  {\em JHEP} {\bf 08} (2017) 112, [\href{http://arxiv.org/abs/1703.04765}{{\tt
  arXiv:1703.04765}}].

\bibitem{Lu:2018cfc}
C.-D. L\"u, Y.-L. Shen, Y.-M. Wang, and Y.-B. Wei, {\it {QCD calculations of $B
  \to \pi, K$ form factors with higher-twist corrections}},  {\em JHEP} {\bf
  01} (2019) 024, [\href{http://arxiv.org/abs/1810.00819}{{\tt
  arXiv:1810.00819}}].

\bibitem{Leljak:2021vte}
D.~Leljak, B.~Meli\'c, and D.~van Dyk, {\it {The $\bar{B} \rightarrow \pi$ form
  factors from QCD and their impact on $|V_{ub}|$}},  {\em JHEP} {\bf 07}
  (2021) 036, [\href{http://arxiv.org/abs/2102.07233}{{\tt arXiv:2102.07233}}].

\bibitem{Cui:2022zwm}
B.-Y. Cui, Y.-K. Huang, Y.-L. Shen, C.~Wang, and Y.-M. Wang, {\it {Precision
  calculations of $B_{d,s} \rightarrow \pi, K$ decay form factors in
  soft-collinear effective theory}},  {\em JHEP} {\bf 03} (2023) 140,
  [\href{http://arxiv.org/abs/2212.11624}{{\tt arXiv:2212.11624}}].

\bibitem{Gao:2019lta}
J.~Gao, C.-D. L\"u, Y.-L. Shen, Y.-M. Wang, and Y.-B. Wei, {\it {Precision
  calculations of $B \to V$ form factors from soft-collinear effective theory
  sum rules on the light-cone}},  {\em Phys. Rev. D} {\bf 101} (2020), no.~7
  074035, [\href{http://arxiv.org/abs/1907.11092}{{\tt arXiv:1907.11092}}].

\bibitem{Wang:2009hra}
Y.-M. Wang, Y.-L. Shen, and C.-D. Lu, {\it {$\Lambda_b \to p, \Lambda$
  transition form factors from QCD light-cone sum rules}},  {\em Phys. Rev. D}
  {\bf 80} (2009) 074012, [\href{http://arxiv.org/abs/0907.4008}{{\tt
  arXiv:0907.4008}}].

\bibitem{Faller:2008tr}
S.~Faller, A.~Khodjamirian, C.~Klein, and T.~Mannel, {\it {$B\to D^{(*)}$ Form
  Factors from QCD Light-Cone Sum Rules}},  {\em Eur. Phys. J. C} {\bf 60}
  (2009) 603--615, [\href{http://arxiv.org/abs/0809.0222}{{\tt
  arXiv:0809.0222}}].

\bibitem{Huang:2022lfr}
K.-S. Huang, W.~Liu, Y.-L. Shen, and F.-S. Yu, {\it {$\Lambda _b \rightarrow p,
  N^*(1535)$ form factors from QCD light-cone sum rules}},  {\em Eur. Phys. J.
  C} {\bf 83} (2023), no.~4 272, [\href{http://arxiv.org/abs/2205.06095}{{\tt
  arXiv:2205.06095}}].

\bibitem{StefanComment}
S.~Meinel. Private conversation. To appear on arXiv.

\bibitem{Meinel:2023wyg}
S.~Meinel, {\it {Status of next-generation $\Lambda_b \to p, \Lambda,
  \Lambda_c$ form-factor calculations}},  {\em PoS} {\bf LATTICE2023} (2024)
  275, [\href{http://arxiv.org/abs/2309.01821}{{\tt arXiv:2309.01821}}].

\bibitem{Bernlochner:2018bfn}
F.~U. Bernlochner, Z.~Ligeti, D.~J. Robinson, and W.~L. Sutcliffe, {\it
  {Precise predictions for $\Lambda_b \to \Lambda_c$ semileptonic decays}},
  {\em Phys. Rev. D} {\bf 99} (2019), no.~5 055008,
  [\href{http://arxiv.org/abs/1812.07593}{{\tt arXiv:1812.07593}}].

\bibitem{Dutta:2015ueb}
R.~Dutta, {\it {$\Lambda_b \to (\Lambda_c,\,p)\,\tau\,\nu$ decays within
  standard model and beyond}},  {\em Phys. Rev. D} {\bf 93} (2016), no.~5
  054003, [\href{http://arxiv.org/abs/1512.04034}{{\tt arXiv:1512.04034}}].

\bibitem{Tanaka:1994ay}
M.~Tanaka, {\it {Charged Higgs effects on exclusive semitauonic $B$ decays}},
  {\em Z. Phys. C} {\bf 67} (1995) 321--326,
  [\href{http://arxiv.org/abs/hep-ph/9411405}{{\tt hep-ph/9411405}}].

\bibitem{Freytsis:2015qca}
M.~Freytsis, Z.~Ligeti, and J.~T. Ruderman, {\it {Flavor models for $\bar{B}
  \to D^{(*)} \tau \bar{\nu}$}},  {\em Phys. Rev. D} {\bf 92} (2015), no.~5
  054018, [\href{http://arxiv.org/abs/1506.08896}{{\tt arXiv:1506.08896}}].

\bibitem{Capdevila:2023yhq}
B.~Capdevila, A.~Crivellin, and J.~Matias, {\it {Review of semileptonic B
  anomalies}},  {\em Eur. Phys. J. ST} {\bf 1} (2023) 20,
  [\href{http://arxiv.org/abs/2309.01311}{{\tt arXiv:2309.01311}}].

\bibitem{London:2021lfn}
D.~London and J.~Matias, {\it {$B$ Flavour Anomalies: 2021 Theoretical Status
  Report}},  {\em Ann. Rev. Nucl. Part. Sci.} {\bf 72} (2022) 37--68,
  [\href{http://arxiv.org/abs/2110.13270}{{\tt arXiv:2110.13270}}].

\bibitem{Bernlochner:2021vlv}
F.~U. Bernlochner, M.~F. Sevilla, D.~J. Robinson, and G.~Wormser, {\it
  {Semitauonic b-hadron decays: A lepton flavor universality laboratory}},
  {\em Rev. Mod. Phys.} {\bf 94} (2022), no.~1 015003,
  [\href{http://arxiv.org/abs/2101.08326}{{\tt arXiv:2101.08326}}].

\bibitem{Bifani:2018zmi}
S.~Bifani, S.~Descotes-Genon, A.~Romero~Vidal, and M.-H. Schune, {\it {Review
  of Lepton Universality tests in $B$ decays}},  {\em J. Phys. G} {\bf 46}
  (2019), no.~2 023001, [\href{http://arxiv.org/abs/1809.06229}{{\tt
  arXiv:1809.06229}}].

\bibitem{Aaij:2017tyk}
{\bf LHCb} Collaboration, R.~Aaij et~al., {\it {Measurement of the ratio of
  branching fractions
  $\mathcal{B}(B_c^+\,\to\,J/\psi\tau^+\nu_\tau)$/$\mathcal{B}(B_c^+\,\to\,J/\psi\mu^+\nu_\mu)$}},
  {\em Phys. Rev. Lett.} {\bf 120} (2018), no.~12 121801,
  [\href{http://arxiv.org/abs/1711.05623}{{\tt arXiv:1711.05623}}].

\bibitem{CMS:2024seh}
{\bf CMS} Collaboration, A.~Hayrapetyan et~al., {\it {Test of lepton flavor
  universality in semileptonic B$^+_\text{c}$ meson decays in proton-proton
  collisions at $\sqrt{s}$ = 13 TeV}},
  \href{http://arxiv.org/abs/2408.00678}{{\tt arXiv:2408.00678}}.

\bibitem{CMSRJpsi2}
{\bf CMS} Collaboration, C.~Rovelli, {\it {Lepton flavour (universality)
  violation studies at CMS}}, .
  {\url{https://indico.cern.ch/event/1291157/contributions/5878345/}}.

\bibitem{Yasmeen:2024cki}
T.~Yasmeen, I.~Ahmed, S.~Shafaq, M.~Arslan, and M.~J. Aslam, {\it {Probing New
  Physics in Light of Recent Developments in $b \rightarrow c \ell \nu$
  Transitions}},  {\em PTEP} {\bf 2024} (2024), no.~7 073B07,
  [\href{http://arxiv.org/abs/2401.02334}{{\tt arXiv:2401.02334}}].

\bibitem{Harrison:2020nrv}
{\bf LATTICE-HPQCD} Collaboration, J.~Harrison, C.~T.~H. Davies, and A.~Lytle,
  {\it {$R(J/\psi)$ and $B_c^- \rightarrow J/\psi \ell^-\bar{\nu}_\ell$ Lepton
  Flavor Universality Violating Observables from Lattice QCD}},  {\em Phys.
  Rev. Lett.} {\bf 125} (2020), no.~22 222003,
  [\href{http://arxiv.org/abs/2007.06956}{{\tt arXiv:2007.06956}}].

\bibitem{Grossman:1994ax}
Y.~Grossman and Z.~Ligeti, {\it {The Inclusive $\bar{B} \to \tau \bar\nu X$
  decay in two Higgs doublet models}},  {\em Phys. Lett. B} {\bf 332} (1994)
  373--380, [\href{http://arxiv.org/abs/hep-ph/9403376}{{\tt hep-ph/9403376}}].

\bibitem{Goldberger:1999yh}
W.~D. Goldberger, {\it {Semileptonic B decays as a probe of new physics}},
  \href{http://arxiv.org/abs/hep-ph/9902311}{{\tt hep-ph/9902311}}.

\bibitem{Colangelo:2016ymy}
P.~Colangelo and F.~De~Fazio, {\it {Tension in the inclusive versus exclusive
  determinations of $|V_{cb}|$: a possible role of new physics}},  {\em Phys.
  Rev. D} {\bf 95} (2017), no.~1 011701,
  [\href{http://arxiv.org/abs/1611.07387}{{\tt arXiv:1611.07387}}].

\bibitem{Celis:2016azn}
A.~Celis, M.~Jung, X.-Q. Li, and A.~Pich, {\it {Scalar contributions to $b\to c
  (u) \tau \nu$ transitions}},  {\em Phys. Lett. B} {\bf 771} (2017) 168--179,
  [\href{http://arxiv.org/abs/1612.07757}{{\tt arXiv:1612.07757}}].

\bibitem{Mannel:2017jfk}
T.~Mannel, A.~V. Rusov, and F.~Shahriaran, {\it {Inclusive semitauonic $B$
  decays to order ${\cal O} (\Lambda_{QCD}^3/m_b^3)$}},  {\em Nucl. Phys. B}
  {\bf 921} (2017) 211--224, [\href{http://arxiv.org/abs/1702.01089}{{\tt
  arXiv:1702.01089}}].

\bibitem{Kamali:2018fhr}
S.~Kamali, A.~Rashed, and A.~Datta, {\it {New physics in inclusive $B \to
  X_c\ell \bar{\nu}$ decay in light of $R(D^{(*)})$ measurements}},  {\em Phys.
  Rev. D} {\bf 97} (2018), no.~9 095034,
  [\href{http://arxiv.org/abs/1801.08259}{{\tt arXiv:1801.08259}}].

\bibitem{Bhattacharya:2018kig}
S.~Bhattacharya, S.~Nandi, and S.~Kumar~Patra, {\it {$b \rightarrow c \tau \nu
  _{\tau }$ Decays: a catalogue to compare, constrain, and correlate new
  physics effects}},  {\em Eur. Phys. J. C} {\bf 79} (2019), no.~3 268,
  [\href{http://arxiv.org/abs/1805.08222}{{\tt arXiv:1805.08222}}].

\bibitem{Kamali:2018bdp}
S.~Kamali, {\it {New physics in inclusive semileptonic $B$ decays including
  nonperturbative corrections}},  {\em Int. J. Mod. Phys. A} {\bf 34} (2019),
  no.~06n07 1950036, [\href{http://arxiv.org/abs/1811.07393}{{\tt
  arXiv:1811.07393}}].

\bibitem{Fael:2022wfc}
M.~Fael, M.~Rahimi, and K.~K. Vos, {\it {New physics contributions to moments
  of inclusive $b \rightarrow c$ semileptonic decays}},  {\em JHEP} {\bf 02}
  (2023) 086, [\href{http://arxiv.org/abs/2208.04282}{{\tt arXiv:2208.04282}}].

\bibitem{ToAppearXQL}
L.-F. Lai, X.-Q. Li, Y.-Y. Li, and Y.-D. Yang. To appear on arXiv.

\bibitem{Ligeti:2014kia}
Z.~Ligeti and F.~J. Tackmann, {\it {Precise predictions for $B \to X_c \tau
  \bar \nu$ decay distributions}},  {\em Phys. Rev. D} {\bf 90} (2014), no.~3
  034021, [\href{http://arxiv.org/abs/1406.7013}{{\tt arXiv:1406.7013}}].

\bibitem{Rahimi:2022vlv}
M.~Rahimi and K.~K. Vos, {\it {Standard Model predictions for lepton flavour
  universality ratios of inclusive semileptonic B decays}},  {\em JHEP} {\bf
  11} (2022) 007, [\href{http://arxiv.org/abs/2207.03432}{{\tt
  arXiv:2207.03432}}].

\bibitem{Belle-II:2023aih}
{\bf Belle-II} Collaboration, I.~Adachi et~al., {\it {First Measurement of
  $R(X_{\tau/\ell})$ as an Inclusive Test of the $b\rightarrow c \tau \nu$
  Anomaly}},  {\em Phys. Rev. Lett.} {\bf 132} (2024), no.~21 211804,
  [\href{http://arxiv.org/abs/2311.07248}{{\tt arXiv:2311.07248}}].

\bibitem{Chivukula:1987py}
R.~S. Chivukula and H.~Georgi, {\it {Composite Technicolor Standard Model}},
  {\em Phys. Lett. B} {\bf 188} (1987) 99--104.

\bibitem{DAmbrosio:2002vsn}
G.~D'Ambrosio, G.~F. Giudice, G.~Isidori, and A.~Strumia, {\it {Minimal flavor
  violation: An Effective field theory approach}},  {\em Nucl. Phys. B} {\bf
  645} (2002) 155--187, [\href{http://arxiv.org/abs/hep-ph/0207036}{{\tt
  hep-ph/0207036}}].

\bibitem{Cirigliano:2005ck}
V.~Cirigliano, B.~Grinstein, G.~Isidori, and M.~B. Wise, {\it {Minimal flavor
  violation in the lepton sector}},  {\em Nucl. Phys. B} {\bf 728} (2005)
  121--134, [\href{http://arxiv.org/abs/hep-ph/0507001}{{\tt hep-ph/0507001}}].

\bibitem{Dekens:2019ept}
W.~Dekens and P.~Stoffer, {\it {Low-energy effective field theory below the
  electroweak scale: matching at one loop}},  {\em JHEP} {\bf 10} (2019) 197,
  [\href{http://arxiv.org/abs/1908.05295}{{\tt arXiv:1908.05295}}]. [Erratum:
  JHEP 11, 148 (2022)].

\bibitem{Jenkins:2017dyc}
E.~E. Jenkins, A.~V. Manohar, and P.~Stoffer, {\it {Low-Energy Effective Field
  Theory below the Electroweak Scale: Anomalous Dimensions}},  {\em JHEP} {\bf
  01} (2018) 084, [\href{http://arxiv.org/abs/1711.05270}{{\tt
  arXiv:1711.05270}}]. [Erratum: JHEP 12, 042 (2023)].

\bibitem{Naterop:2023dek}
L.~Naterop and P.~Stoffer, {\it {Low-energy effective field theory below the
  electroweak scale: one-loop renormalization in the 't Hooft-Veltman scheme}},
   {\em JHEP} {\bf 02} (2024) 068, [\href{http://arxiv.org/abs/2310.13051}{{\tt
  arXiv:2310.13051}}].

\bibitem{Gerard:1982mm}
J.~M. Gerard, {\it {FERMION MASS SPECTRUM IN $SU(2)_L \times U(1)$}},  {\em Z.
  Phys. C} {\bf 18} (1983) 145.

\bibitem{Greljo:2023bab}
A.~Greljo, J.~Salko, A.~Smolkovi\v{c}, and P.~Stangl, {\it {SMEFT restrictions
  on exclusive $b \rightarrow u \ell \nu$ decays}},  {\em JHEP} {\bf 11} (2023)
  023, [\href{http://arxiv.org/abs/2306.09401}{{\tt arXiv:2306.09401}}].

\bibitem{Ellis:2018gqa}
J.~Ellis, C.~W. Murphy, V.~Sanz, and T.~You, {\it {Updated Global SMEFT Fit to
  Higgs, Diboson and Electroweak Data}},  {\em JHEP} {\bf 06} (2018) 146,
  [\href{http://arxiv.org/abs/1803.03252}{{\tt arXiv:1803.03252}}].

\bibitem{Dawson:2019clf}
S.~Dawson and P.~P. Giardino, {\it {Electroweak and QCD corrections to $Z$ and
  $W$ pole observables in the standard model EFT}},  {\em Phys. Rev. D} {\bf
  101} (2020), no.~1 013001, [\href{http://arxiv.org/abs/1909.02000}{{\tt
  arXiv:1909.02000}}].

\bibitem{Efrati:2015eaa}
A.~Efrati, A.~Falkowski, and Y.~Soreq, {\it {Electroweak constraints on
  flavorful effective theories}},  {\em JHEP} {\bf 07} (2015) 018,
  [\href{http://arxiv.org/abs/1503.07872}{{\tt arXiv:1503.07872}}].

\bibitem{ATLAS:2020zms}
{\bf ATLAS} Collaboration, G.~Aad et~al., {\it {Search for heavy Higgs bosons
  decaying into two tau leptons with the ATLAS detector using $pp$ collisions
  at $\sqrt{s}=13$ TeV}},  {\em Phys. Rev. Lett.} {\bf 125} (2020), no.~5
  051801, [\href{http://arxiv.org/abs/2002.12223}{{\tt arXiv:2002.12223}}].

\bibitem{CMS:2022goy}
{\bf CMS} Collaboration, A.~Tumasyan et~al., {\it {Searches for additional
  Higgs bosons and for vector leptoquarks in $\tau\tau$ final states in
  proton-proton collisions at $\sqrt{s}$ = 13 TeV}},  {\em JHEP} {\bf 07}
  (2023) 073, [\href{http://arxiv.org/abs/2208.02717}{{\tt arXiv:2208.02717}}].

\bibitem{Allwicher:2022gkm}
L.~Allwicher, D.~A. Faroughy, F.~Jaffredo, O.~Sumensari, and F.~Wilsch, {\it
  {Drell-Yan tails beyond the Standard Model}},  {\em JHEP} {\bf 03} (2023)
  064, [\href{http://arxiv.org/abs/2207.10714}{{\tt arXiv:2207.10714}}].

\bibitem{Barbieri:2011ci}
R.~Barbieri, G.~Isidori, J.~Jones-Perez, P.~Lodone, and D.~M. Straub, {\it
  {$U(2)$ and Minimal Flavour Violation in Supersymmetry}},  {\em Eur. Phys. J.
  C} {\bf 71} (2011) 1725, [\href{http://arxiv.org/abs/1105.2296}{{\tt
  arXiv:1105.2296}}].

\bibitem{Barbieri:2012uh}
R.~Barbieri, D.~Buttazzo, F.~Sala, and D.~M. Straub, {\it {Flavour physics from
  an approximate $U(2)^3$ symmetry}},  {\em JHEP} {\bf 07} (2012) 181,
  [\href{http://arxiv.org/abs/1203.4218}{{\tt arXiv:1203.4218}}].

\bibitem{Blankenburg:2012nx}
G.~Blankenburg, G.~Isidori, and J.~Jones-Perez, {\it {Neutrino Masses and LFV
  from Minimal Breaking of $U(3)^5$ and $U(2)^5$ flavor Symmetries}},  {\em
  Eur. Phys. J. C} {\bf 72} (2012) 2126,
  [\href{http://arxiv.org/abs/1204.0688}{{\tt arXiv:1204.0688}}].

\bibitem{Allwicher:2023shc}
L.~Allwicher, C.~Cornella, G.~Isidori, and B.~A. Stefanek, {\it {New physics in
  the third generation. A comprehensive SMEFT analysis and future prospects}},
  {\em JHEP} {\bf 03} (2024) 049, [\href{http://arxiv.org/abs/2311.00020}{{\tt
  arXiv:2311.00020}}].

\bibitem{Aloni:2017eny}
D.~Aloni, A.~Efrati, Y.~Grossman, and Y.~Nir, {\it {$\Upsilon$ and $\psi$
  leptonic decays as probes of solutions to the $R_D^{(*)}$ puzzle}},  {\em
  JHEP} {\bf 06} (2017) 019, [\href{http://arxiv.org/abs/1702.07356}{{\tt
  arXiv:1702.07356}}].

\bibitem{BaBar:2020nlq}
{\bf BaBar} Collaboration, J.~P. Lees et~al., {\it {Precision measurement of
  the ${\cal B}(\Upsilon(3S)\to\tau^+\tau^-)/{\cal
  B}(\Upsilon(3S)\to\mu^+\mu^-)$ ratio}},  {\em Phys. Rev. Lett.} {\bf 125}
  (2020) 241801, [\href{http://arxiv.org/abs/2005.01230}{{\tt
  arXiv:2005.01230}}].

\bibitem{Gonzalez-Alonso:2017iyc}
M.~Gonz\'alez-Alonso, J.~Martin~Camalich, and K.~Mimouni, {\it
  {Renormalization-group evolution of new physics contributions to
  (semi)leptonic meson decays}},  {\em Phys. Lett. B} {\bf 772} (2017)
  777--785, [\href{http://arxiv.org/abs/1706.00410}{{\tt arXiv:1706.00410}}].

\bibitem{Belle:2015pkj}
{\bf Belle} Collaboration, R.~Glattauer et~al., {\it {Measurement of the decay
  $B\to D\ell\nu_\ell$ in fully reconstructed events and determination of the
  Cabibbo-Kobayashi-Maskawa matrix element $|V_{cb}|$}},  {\em Phys. Rev. D}
  {\bf 93} (2016), no.~3 032006, [\href{http://arxiv.org/abs/1510.03657}{{\tt
  arXiv:1510.03657}}].

\bibitem{Belle:2018ezy}
{\bf Belle} Collaboration, E.~Waheed et~al., {\it {Measurement of the CKM
  matrix element $|V_{cb}|$ from $B^0\to D^{*-}\ell^ {+} \nu_\ell$ at Belle}},
  {\em Phys. Rev. D} {\bf 100} (2019), no.~5 052007,
  [\href{http://arxiv.org/abs/1809.03290}{{\tt arXiv:1809.03290}}]. [Erratum:
  Phys.Rev.D 103, 079901 (2021)].

\bibitem{Belle:2023bwv}
{\bf Belle} Collaboration, M.~T. Prim et~al., {\it {Measurement of differential
  distributions of $B\to D^\ast \ell \nu_{\ell}$ and implications on
  $|V_{cb}|$}},  {\em Phys. Rev. D} {\bf 108} (2023), no.~1 012002,
  [\href{http://arxiv.org/abs/2301.07529}{{\tt arXiv:2301.07529}}].

\bibitem{Belle-II:2023okj}
{\bf Belle-II} Collaboration, I.~Adachi et~al., {\it {Determination of
  $|V_{cb}|$ using $\bar{B}^0\rightarrow D^{\ast+} \ell^-\bar{\nu}_\ell$ decays
  with Belle II}},  {\em Phys. Rev. D} {\bf 108} (2023), no.~9 092013,
  [\href{http://arxiv.org/abs/2310.01170}{{\tt arXiv:2310.01170}}].

\bibitem{BaBar:2019vpl}
{\bf BaBar} Collaboration, J.~P. Lees et~al., {\it {Extraction of form Factors
  from a Four-Dimensional Angular Analysis of $\overline{B} \rightarrow D^\ast
  \ell^- \overline{\nu}_\ell$}},  {\em Phys. Rev. Lett.} {\bf 123} (2019),
  no.~9 091801, [\href{http://arxiv.org/abs/1903.10002}{{\tt
  arXiv:1903.10002}}].

\bibitem{BaBar:2023kug}
{\bf BaBar} Collaboration, J.~P. Lees et~al., {\it {Model-independent
  extraction of form factors and $|V_{cb}|$ in $B^-\rightarrow D \ell^-
  \bar\nu_\ell$ with hadronic tagging at BABAR}},  {\em Phys. Rev. D} {\bf 110}
  (2024), no.~3 032018, [\href{http://arxiv.org/abs/2311.15071}{{\tt
  arXiv:2311.15071}}].

\bibitem{Bernlochner:2024sfg}
F.~U. Bernlochner, M.~T. Prim, and K.~K. Vos, {\it {$|V_{ub}|$ and $|V_{cb}|$
  from exclusive semileptonic decays}},  {\em Eur. Phys. J. ST} {\bf 233}
  (2024), no.~2 347--358.

\bibitem{Jay:2024ygl}
W.~I. Jay, R.~van Tonder, and R.~Watanabe, {\it {Summary of the CKM 2023
  Working Group on $V_{ub}$, $V_{cb}$ and semileptonic/leptonic B decays
  including $\tau$}},  in {\em {12th International Workshop on the CKM
  Unitarity Triangle}}, 3, 2024.
\newblock \href{http://arxiv.org/abs/2403.18175}{{\tt arXiv:2403.18175}}.

\bibitem{Belle:2015qfa}
{\bf Belle} Collaboration, M.~Huschle et~al., {\it {Measurement of the
  branching ratio of $\bar{B} \to D^{(\ast)} \tau^- \bar{\nu}_\tau$ relative to
  $\bar{B} \to D^{(\ast)} \ell^- \bar{\nu}_\ell$ decays with hadronic tagging
  at Belle}},  {\em Phys. Rev. D} {\bf 92} (2015), no.~7 072014,
  [\href{http://arxiv.org/abs/1507.03233}{{\tt arXiv:1507.03233}}].

\bibitem{Belle:2016ure}
{\bf Belle} Collaboration, Y.~Sato et~al., {\it {Measurement of the branching
  ratio of $\bar{B}^0 \rightarrow D^{*+} \tau^- \bar{\nu}_{\tau}$ relative to
  $\bar{B}^0 \rightarrow D^{*+} \ell^- \bar{\nu}_{\ell}$ decays with a
  semileptonic tagging method}},  {\em Phys. Rev. D} {\bf 94} (2016), no.~7
  072007, [\href{http://arxiv.org/abs/1607.07923}{{\tt arXiv:1607.07923}}].

\bibitem{Belle:2016dyj}
{\bf Belle} Collaboration, S.~Hirose et~al., {\it {Measurement of the $\tau$
  lepton polarization and $R(D^*)$ in the decay $\bar{B} \to D^* \tau^-
  \bar{\nu}_\tau$}},  {\em Phys. Rev. Lett.} {\bf 118} (2017), no.~21 211801,
  [\href{http://arxiv.org/abs/1612.00529}{{\tt arXiv:1612.00529}}].

\bibitem{Belle:2017ilt}
{\bf Belle} Collaboration, S.~Hirose et~al., {\it {Measurement of the $\tau$
  lepton polarization and $R(D^*)$ in the decay $\bar{B} \rightarrow D^* \tau^-
  \bar{\nu}_\tau$ with one-prong hadronic $\tau$ decays at Belle}},  {\em Phys.
  Rev. D} {\bf 97} (2018), no.~1 012004,
  [\href{http://arxiv.org/abs/1709.00129}{{\tt arXiv:1709.00129}}].

\bibitem{Belle:2019rba}
{\bf Belle} Collaboration, G.~Caria et~al., {\it {Measurement of
  $\mathcal{R}(D)$ and $\mathcal{R}(D^*)$ with a semileptonic tagging method}},
   {\em Phys. Rev. Lett.} {\bf 124} (2020), no.~16 161803,
  [\href{http://arxiv.org/abs/1910.05864}{{\tt arXiv:1910.05864}}].

\bibitem{BaBar:2012obs}
{\bf BaBar} Collaboration, J.~P. Lees et~al., {\it {Evidence for an excess of
  $\bar{B} \to D^{(*)} \tau^-\bar{\nu}_\tau$ decays}},  {\em Phys. Rev. Lett.}
  {\bf 109} (2012) 101802, [\href{http://arxiv.org/abs/1205.5442}{{\tt
  arXiv:1205.5442}}].

\bibitem{BaBar:2013mob}
{\bf BaBar} Collaboration, J.~P. Lees et~al., {\it {Measurement of an Excess of
  $\bar{B} \to D^{(*)}\tau^- \bar{\nu}_\tau$ Decays and Implications for
  Charged Higgs Bosons}},  {\em Phys. Rev. D} {\bf 88} (2013), no.~7 072012,
  [\href{http://arxiv.org/abs/1303.0571}{{\tt arXiv:1303.0571}}].

\bibitem{Aaij:2015yra}
{\bf LHCb} Collaboration, R.~Aaij et~al., {\it {Measurement of the ratio of
  branching fractions $\mathcal{B}(\bar{B}^0 \to
  D^{*+}\tau^{-}\bar{\nu}_{\tau})/\mathcal{B}(\bar{B}^0 \to
  D^{*+}\mu^{-}\bar{\nu}_{\mu})$}},  {\em Phys. Rev. Lett.} {\bf 115} (2015),
  no.~11 111803, [\href{http://arxiv.org/abs/1506.08614}{{\tt
  arXiv:1506.08614}}]. [Erratum: Phys.Rev.Lett. 115, 159901 (2015)].

\bibitem{Aaij:2017uff}
{\bf LHCb} Collaboration, R.~Aaij et~al., {\it {Measurement of the ratio of the
  $B^0 \to D^{*-} \tau^+ \nu_{\tau}$ and $B^0 \to D^{*-} \mu^+ \nu_{\mu}$
  branching fractions using three-prong $\tau$-lepton decays}},  {\em Phys.
  Rev. Lett.} {\bf 120} (2018), no.~17 171802,
  [\href{http://arxiv.org/abs/1708.08856}{{\tt arXiv:1708.08856}}].

\bibitem{Aaij:2017deq}
{\bf LHCb} Collaboration, R.~Aaij et~al., {\it {Test of Lepton Flavor
  Universality by the measurement of the $B^0 \to D^{*-} \tau^+ \nu_{\tau}$
  branching fraction using three-prong $\tau$ decays}},  {\em Phys. Rev. D}
  {\bf 97} (2018), no.~7 072013, [\href{http://arxiv.org/abs/1711.02505}{{\tt
  arXiv:1711.02505}}].

\bibitem{LHCb:2023zxo}
{\bf LHCb} Collaboration, R.~Aaij et~al., {\it {Measurement of the ratios of
  branching fractions $\mathcal{R}(D^{*})$ and $\mathcal{R}(D^{0})$}},  {\em
  Phys. Rev. Lett.} {\bf 131} (2023) 111802,
  [\href{http://arxiv.org/abs/2302.02886}{{\tt arXiv:2302.02886}}].

\bibitem{LHCb:2023uiv}
{\bf LHCb} Collaboration, R.~Aaij et~al., {\it {Test of lepton flavor
  universality using $B^0\rightarrow D^{*-} \tau^+ \nu_\tau$ decays with
  hadronic $\tau$ channels}},  {\em Phys. Rev. D} {\bf 108} (2023), no.~1
  012018, [\href{http://arxiv.org/abs/2305.01463}{{\tt arXiv:2305.01463}}].

\bibitem{LHCb:2024jll}
{\bf LHCb} Collaboration, R.~Aaij et~al., {\it {Measurement of the branching
  fraction ratios $R(D^{+})$ and $R(D^{*+})$ using muonic $\tau$ decays}},
  \href{http://arxiv.org/abs/2406.03387}{{\tt arXiv:2406.03387}}.

\bibitem{Belle-II:2024ami}
{\bf Belle-II} Collaboration, I.~Adachi et~al., {\it {A test of lepton flavor
  universality with a measurement of $R(D^{*})$ using hadronic $B$ tagging at
  the Belle II experiment}},  \href{http://arxiv.org/abs/2401.02840}{{\tt
  arXiv:2401.02840}}.

\bibitem{HFLAV:2022esi}
{\bf HFLAV} Collaboration, Y.~S. Amhis et~al., {\it {Averages of b-hadron,
  c-hadron, and $\tau$-lepton properties as of 2021}},  {\em Phys. Rev. D} {\bf
  107} (2023), no.~5 052008, [\href{http://arxiv.org/abs/2206.07501}{{\tt
  arXiv:2206.07501}}].

\bibitem{Leljak:2023gna}
D.~Leljak, B.~Meli\'c, F.~Novak, M.~Reboud, and D.~van Dyk, {\it {Toward a
  complete description of b \textrightarrow{} u\ensuremath{\ell}$^{-}
  \overline{\nu} $ decays within the Weak Effective Theory}},  {\em JHEP} {\bf
  08} (2023) 063, [\href{http://arxiv.org/abs/2302.05268}{{\tt
  arXiv:2302.05268}}].

\bibitem{DELPHI:2003qft}
{\bf DELPHI} Collaboration, J.~Abdallah et~al., {\it {Measurement of the
  Lambda0(b) decay form-factor}},  {\em Phys. Lett. B} {\bf 585} (2004) 63--84,
  [\href{http://arxiv.org/abs/hep-ex/0403040}{{\tt hep-ex/0403040}}].

\bibitem{LHCb:2022qnv}
{\bf LHCb} Collaboration, R.~Aaij et~al., {\it {Test of lepton universality in
  $b \rightarrow s \ell^+ \ell^-$ decays}},  {\em Phys. Rev. Lett.} {\bf 131}
  (2023), no.~5 051803, [\href{http://arxiv.org/abs/2212.09152}{{\tt
  arXiv:2212.09152}}].

\bibitem{LHCb:2022vje}
{\bf LHCb} Collaboration, R.~Aaij et~al., {\it {Measurement of lepton
  universality parameters in $B^+\to K^+\ell^+\ell^-$ and $B^0\to
  K^{*0}\ell^+\ell^-$ decays}},  {\em Phys. Rev. D} {\bf 108} (2023), no.~3
  032002, [\href{http://arxiv.org/abs/2212.09153}{{\tt arXiv:2212.09153}}].

\bibitem{Bhattacharya:2016mcc}
B.~Bhattacharya, A.~Datta, J.-P. Gu\'evin, D.~London, and R.~Watanabe, {\it
  {Simultaneous Explanation of the $R_K$ and $R_{D^{(*)}}$ Puzzles: a Model
  Analysis}},  {\em JHEP} {\bf 01} (2017) 015,
  [\href{http://arxiv.org/abs/1609.09078}{{\tt arXiv:1609.09078}}].

\bibitem{Kumar:2018kmr}
J.~Kumar, D.~London, and R.~Watanabe, {\it {Combined Explanations of the $b \to
  s \mu^+ \mu^-$ and $b \to c \tau^- {\bar\nu}$ Anomalies: a General Model
  Analysis}},  {\em Phys. Rev. D} {\bf 99} (2019), no.~1 015007,
  [\href{http://arxiv.org/abs/1806.07403}{{\tt arXiv:1806.07403}}].

\bibitem{Bigi:2016mdz}
D.~Bigi and P.~Gambino, {\it {Revisiting $B\to D \ell \nu$}},  {\em Phys. Rev.
  D} {\bf 94} (2016), no.~9 094008,
  [\href{http://arxiv.org/abs/1606.08030}{{\tt arXiv:1606.08030}}].

\bibitem{Bigi:2017njr}
D.~Bigi, P.~Gambino, and S.~Schacht, {\it {A fresh look at the determination of
  $|V_{cb}|$ from $B\to D^{*} \ell \nu$}},  {\em Phys. Lett. B} {\bf 769}
  (2017) 441--445, [\href{http://arxiv.org/abs/1703.06124}{{\tt
  arXiv:1703.06124}}].

\bibitem{Bigi:2017jbd}
D.~Bigi, P.~Gambino, and S.~Schacht, {\it {$R(D^*)$, $|V_{cb}|$, and the Heavy
  Quark Symmetry relations between form factors}},  {\em JHEP} {\bf 11} (2017)
  061, [\href{http://arxiv.org/abs/1707.09509}{{\tt arXiv:1707.09509}}].

\bibitem{Grinstein:2023njq}
B.~Grinstein, X.~Lu, L.~Merlo, and P.~Qu\'\i{}lez, {\it {Hilbert series for
  covariants and their applications to minimal flavor violation}},  {\em JHEP}
  {\bf 2024} (2024) 154, [\href{http://arxiv.org/abs/2312.13349}{{\tt
  arXiv:2312.13349}}].

\bibitem{Hou:2024vyw}
B.-F. Hou, X.-Q. Li, M.~Shen, Y.-D. Yang, and X.-B. Yuan, {\it {Deciphering the
  Belle II data on $B\to K\nu \overline{\nu}$ decay in the (dark) SMEFT with
  minimal flavour violation}},  {\em JHEP} {\bf 06} (2024) 172,
  [\href{http://arxiv.org/abs/2402.19208}{{\tt arXiv:2402.19208}}].

\bibitem{Fuentes-Martin:2019mun}
J.~Fuentes-Mart\'\i{}n, G.~Isidori, J.~Pag\`es, and K.~Yamamoto, {\it {With or
  without U(2)? Probing non-standard flavor and helicity structures in
  semileptonic B decays}},  {\em Phys. Lett. B} {\bf 800} (2020) 135080,
  [\href{http://arxiv.org/abs/1909.02519}{{\tt arXiv:1909.02519}}].

\end{thebibliography}\endgroup

\end{document}